\documentclass[12pt]{report}
\usepackage{a4              % A4 paper 
           ,graphics 	    % for including fancy graphics
           ,doublespace     % for double spacing
           ,epsfig
           ,subfigure
           ,float
            ,latexsym
            ,axodraw
           }
\makeindex 

\begin{document} 
\parskip 4mm 
\parindent 0mm
% The begining bit:
% The order and most of the layout of the preamble is set down 
% in University regulations. 
% 
% ************************** 
% ******* Title page ******* 
% ************************** 
\pagestyle{empty} 
\begin{center} 
    {\LARGE UNIVERSITY OF SOUTHAMPTON} \\
\vspace{4cm} 
    {\Huge{\bf The Derivative Expansion of the }} \vspace{12pt} \\ 
    {\Huge{\bf Exact Renormalization Group  }} \vspace{12pt} \\ 
    {\Huge{\bf ~  }} \vspace{12pt} \\ 
\vspace{2cm} 
    by \\
\vspace{2cm} 
    {\LARGE Michael Duncan Turner} \\
\vspace{2.5cm} 
    A thesis submitted for the degree of \\
\bigskip 
    Doctor of Philosophy \\
\vspace{1cm} 
\bigskip 
    Department of Physics \\
\bigskip 
August, 1996
\end{center}

% ************************** 
% ******* Dedication ******* 
% ************************** 

\newpage 
\pagestyle{empty} 
\begin{center} 
\vspace*{8cm} 
\hspace{-1cm}
\emph{"Nobody is completely worthless ---}\\
\hspace{1cm}
\emph{they can always serve as a bad example."}\\
\vspace{0.5cm}
\hspace{5cm} Seen on a t-shirt.
\end{center} 
% 
% ************************ 
% ******* Abstract ******* 
% ************************ 
\setstretch{1.5}
\newpage 
\pagestyle{empty} 
\begin{center} 
      {\Large UNIVERSITY OF SOUTHAMPTON}  \\
\bigskip 
      \underline{\large ABSTRACT} \\
\bigskip 
      {\Large FACULTY OF SCIENCE} \\
\bigskip 
      {\Large PHYSICS} \\
\bigskip 
      \underline{\large Doctor of Philosophy} \\
\bigskip 
      {\Large The Derivative Expansion of the Exact Renormalization Group} \\
\bigskip 
      {\large Michael Duncan Turner} \\
\end{center} 
We formulate a method of performing non-perturbative calculations in
quantum field theory, based upon a derivative expansion of the exact
renormalization group. We then proceed to apply this method to the
calculation of critical exponents for three dimensional $O(N)$
symmetric theory. Finally we discuss how the new approximation scheme
manages to reproduce  some exactly known solutions in critical phenomena.
%
% ***************************** 
% ******* Now the lists ******* 
% ***************************** 
\newpage 
% This is removed for debugging purposes.  With the little roman 
% numerals for the preamble pages, we have multiple pages with the same 
% 'number'.  Commenting this out makes thing easier in the debugging 
% stages. 
\typeout{final: Roman numbering in introduction}
\pagenumbering{roman} 
\pagestyle{plain} 
\tableofcontents 
\newpage 
\listoffigures 
\newpage 
\listoftables 
%
% *********************** 
% ******* Preface ******* 
% *********************** 
\newpage 
\chapter*{Preface} 
\addcontentsline{toc}{chapter} 
               {\protect\numberline{Preface\hspace{-27pt}}} 
%Preface

Work in chapter one is introductory and may be found in any of the
references cited. Chapters two, three and four are original and
carried out under the supervison of Tim Morris.
%
% ********************************** 
% ******** Acknowledgments  ******** 
% ********************************** 
\newpage 
\chapter*{Acknowledgments} 
\addcontentsline{toc}{chapter} 
                {\protect\numberline{Acknowledgments\hspace{-96pt}}} 

I would like to thank my parents and family for all their support --- I would
never have got through the past six years without them. Two other
people deserve a  special mention. Firstly, I would like to thank
Amanda Cambridge for all her love during 
my time at Southampton and for being there
when I wanted to give up (and putting up  with my all too frequent sulks).
Secondly, I would like to thank my 
supervisor,Tim Morris, for his help, comments and his undying
enthusiasm for the subject, especially when the going was tough. I
would never have completed this research 
without any of the above people.

Several other people deserve a mention: Andrew and Kevin for  their
computing help, Terry for his advice and comments, Treeve for being a
drinking partner over my three years here, the staff at the
Southampton HPC centre  and, finally, the Beer and
Darts Association for keeping me off the streets on a Friday evening.

I acknowledge the support of PPARC through a studentship.

% Go back to proper numbering:
\typeout{final: Numbering introductory pages sequentially}
\pagenumbering{arabic}

% Change rules for floats:
\renewcommand{\floatpagefraction}{0.8}
\renewcommand{\topfraction}{0.8}
\renewcommand{\bottomfraction}{0.8}
\renewcommand{\textfraction}{0.2}

% Include all the chapters:
\newcommand{\noi}{\noindent}

\chapter{Introduction}

It goes without saying that an efficient method of performing accurate
non-perturbative calculations in quantum field theory would be
extremely useful. The method should be able to produce a sequence of
approximations which can to be seen to converge. To be really useful,
the approximation scheme should also be applicable when there is no
identifiable small parameter to control the approximation, and hence
allow us to reach the areas of greatest interest in theoretical
physics, eg the strong hadronic physics. In this thesis we outline a
promising approximation scheme based upon the exact renormalization
group.  We start this introduction with a discussion of effective
field theories. We then move on to  discuss how we came to decide our
method of approximation was a sensible one and discuss some of the
problems posed by critical phenomena.

\newpage
\section{Effective Lagrangians}

The standard model of elementary particle physics has enjoyed
considerable success. It can neatly account for most of the phenomena
that are observed today to a high degree of accuracy. It is based upon
the gauge group $SU(3)_c \otimes SU(2)_L \otimes U(1)_Y$. The
$SU(3)_c$ deals with the strong interactions that bind the hadrons
together , whilst $SU(2)_L
\otimes U(1)$ deal with the electroweak sector. The Lagrangian density
of the theory is,

\begin{equation}
{\cal L}_{SM}={\cal L}_{kin} + {\cal L}_{yuk} - V_{Higgs}+ {\cal
L}_{\theta}
\end{equation}

where ${\cal L}_{kin}$ represents the kinetic terms of the Lagrangian
density and ${\cal L}_{yuk}$ are the Yukawa couplings which ultimately
give the particles their masses. The ${\cal L}_{\theta}$ term reflects
the non-trivial vacuum topology of a four dimensional non-abelian
gauge field theory. Perhaps the area of biggest speculation is that 
represented by $V_{Higgs}$. It has long been postulated that particles
gain their masses when the Higgs field , $H$, gains a vacuum
expectation value, $<H> = \nu$.  When this occurs the $SU(3)_c \otimes
SU(2)_L \otimes U(1)_Y$ is spontaneously broken down to $SU(3)_C
\otimes U(1)_{EM}$ resulting in both fermionic masses (via the Yukawa
couplings to the Higgs) and generating masses for the $W^{\pm}$ and
$Z^0$. This method of spontaneously breaking the symmetry has long
been supported, although other competitors have existed (eg top quark
condensates~\cite{a:topcond}, techni-colour, extended
techni-colour~\cite{a:techni1,a:techni2} etc).

Although the standard model has enjoyed considerable success there
have been innumerable attempts to go beyond this theory, that is to
extend it.  One of the most promising early attempts was made by
Georgi and Glashow~\cite{a:georgiGUTS} who, based upon arguments of
aesthetics and technical considerations, embedded the standard model
gauge group in a larger Lie group $SU(5)$ and hence unified the three
gauge groups at a very high energy scale $\Lambda$, giving birth to
the concept of grand unified theories. Since then considerable effort
has been placed into investigating such theories. For example,
supersymmetry~\cite{a:supsym} was introduced to cure what is known as
the hierarchy~\cite{hierarch} problem (where the low energy $SU(2)_L
\times U(1)_Y$ breaking doublets receive radiative contributions to
their mass of the order of the GUT scale). Other authors, interested in
the lack of right-handed neutrinos, have investigated left-right
symmetric Pati-Salam~\cite{a:patisalam} models based upon the gauge
group $SU(4) \otimes SU(2)_L \otimes SU(2)_R$ (and at the same time
introduced a fourth 'colour'!). Others interested in gravity have
introduced super-gravity models and super-string models.

The key point is as follows:-- the extended theories always assume
that the new theories only become important at high energies. To be
more precise, if the extended theory is based upon a gauge group
${\cal G}$, at some energy scale $\Lambda$, then at some lower energy
scales the gauge group is broken down to the standard model gauge
group. Two important points now arise. Firstly the low energy physics
depends on the high energy physics only in a limited
way~\cite{a:weineff,a:halleff,a:apple}. That is, we are largely blind
to any fundamental theory (ie a theory valid for all energy scales),
if such a thing exists, and if it does we can only see small 'windows'
of its effects on low energy phenomena. For example, it is usual for a
grand unified theory~(GUT) to predict proton decay. They also predict
that the proton has a very long life time and is to all intents and
purposes stable, eg the simplest $SU(5)$ GUT predicts a lifetime of
about \mbox{$3 \times 10^{30}$ years.} Proton decay, despite being
extremely rare, is one of the few windows open to us to look at the
physics at the GUT scale. Of course this extremely long lifetime is
important for our existence --- if the proton decayed at short
intervals then it would be unlikely that we would see a stable,
evolving universe around us today.  The weak effect of high energy
physics on low energy phenomena is also important from a practical
view point. If someone decides to investigate a new extension to the
standard model, then it must be checked whether the predictions lie
within the regions allowed by experimental results. For example,
extended techni-colour was largely rejected by it predicting values
for the so called S,T and U parameters that lie outside the bounds
dictated by experiment. However, to compute results in these extensions
requires the computation of a huge number of Feynman integrals, which
leads to an arduous task. It is much simpler to realize that at the
end of the day all that is required is to compute the values of a
number of parameters in an effective theory, and thus to reproduce the
result in a far more efficient manner. The point is that effective
theories represent our best chance to systematically organize the
experimental data and hence increase any chances of producing a
worthwhile result.

The second problem is a calculational one. We now have a problem with
several mass scales:-- in addition to the low-energy physics we now
have a new large energy scale $\Lambda$. This vastly complicates the
calculations. For example, within perturbation theory beyond tree
level, internal propagators will have several mass scales involved and
this leads to complicated Feynman parameter
integrals~\mbox{\cite{a:effect,b:georgi}.}

Perhaps a better way to proceed is to introduce the concept of an
effective Lagrangian. Instead of taking a theory to be valid over all
energy scales, we instead assume that the theory is valid only over a
limited range of energies, without requiring a detailed knowledge of
the physics outside its range of validity. For example, the physics of
soft pions scattering strongly below the chiral-symmetry breaking
scale is described well by chiral Lagrangians and the standard model
below the scale of grand unification. It appears that to some extent
all the theories that we know are described by effective field
theories up to the scale where some new physics occur.

\newpage
We will no longer assume that the theory is valid for all energies and
introduce an overall momentum cutoff $\Lambda_0$. Below $\Lambda_0$
the physics will be described by a very general Lagrangian with an
infinite set of couplings. Using 'naturalness' these couplings can be
expected to be of order unity at the scale $\Lambda_0$. That is we
will write,
\begin{equation}
{\cal L}_{eff}={\cal L}_0 + \frac{1}{\Lambda_0}{\cal L}_1
+\frac{1}{\Lambda_0^2}{\cal L}_2 + \cdots
\label{e:Leff}
\end{equation}
where ${\cal L}_0$ contains operators of canonical dimension $\leq D$,
where $D$ is the dimension of space-time, and ${\cal L}_n$ (for $n >
0$) contains operators of canonical dimension $n+D$. That is, ${\cal
L}_0$ contains what are usually referred to as the renormalizable
operators and the ${\cal L}_n$ (for $n>0$) contain the non-renormalizable
operators.

It may seem that we have lost all predictive power as our theory now
contains an infinite number of non-renormalizable interactions, but
this is not the case~\cite{b:georgi}:

\begin{itemize}
\item if we know the underlying theory at high energy then we can
calculate all the non-renormalizable interactions.

\item Looking at expression (\ref{e:Leff}) we see that it looks like
the effects of the non-renormalizable interactions are heavily
suppressed, unless the interactions lead to a divergence of a
sufficiently high degree, which then overwhelms the suppression
factor. For example, consider the six point coupling shown in
figure~(1.1a). If we consider the two loop graph in figure~(1.1b) then
we see that this will give a contribution of order $\Lambda_0^4$,
which will cancel out the $1/\Lambda_0^2$ factor and lead to a term of
order $\Lambda_0^2$. It again looks like we have lost all predictive
power. However, the two loop graph will only alter the coefficient of
the two point function (ie the mass term), which is a renormalizable
term. Hence we can renormalize the two point function to remove this
extra divergence. This argument can be shown to
generalize~\cite{a:pol}.
\end{itemize}

\begin{center}
\begin{picture}(300,140)(0,0)
\Vertex(50,90){5}
\Line(50,90)(85,90)
\Line(50,90)(15,90)
\Line(50,90)(75,115)
\Line(50,90)(25,65)
\Line(50,90)(75,65)
\Line(50,90)(25,115)
\Vertex(250,90){5}
\Line(250,90)(285,90)
\Line(250,90)(215,90)
\CArc(250,107.5)(17.5,0,360)
\CArc(250,72.5)(17.5,0,360)
\Text(50,10)[]{(a)}
\Text(250,10)[]{(b)}
\end{picture}
\end{center}
Figure 1.1: Six point interactions in a scalar field theory.\\
\setcounter{figure}{1}

The fact that we can still calculate in an effective theory can be
seen in one of the simplest of theories, quantum electro-dynamics. QED
has managed to make spectacularly accurate predictions at energy
scales from $eV$ up to a few $MeV$. For example, the anomalous
magnetic moment has been calculated to several loops in the
\emph{effective} field theory of QED. We say effective because
corrections from the 'full' theory will be more important at higher
energies, including the effects of QCD and electro-weak theory through
loop contributions, plus contributions from any theory defined at a
higher energy scale still. The suppression of these higher energy
scale theories still allows QED to make accurate low energy
predictions.

We will look at the second point above more closely. We will define
what is known as effective Lagrangian
flow~\cite{a:wk,a:pol,a:ball/thorne}. If we wish to look at the
physics at some energy scale $E \ll \Lambda_0$, instead of using the
full 'bare' Lagrangian with cut off $\Lambda_0$, we could lower the
cutoff to a lower scale $\Lambda \sim E$. To do this we will need to
allow the couplings to flow so that the physics (ie the partition
function) remains fixed for all the low energy processes. We will show
later that the action, $S_{\Lambda}=\int d^Dx {\cal L}$, will evolve
according to an equation of the general form (see \cite{a:timappr} and
references therein),
\begin{equation}
\Lambda \frac{\partial S[\phi]}{\partial\Lambda} = \tau[S]
\label{e:rg1}
\end{equation}
where $S_{\Lambda}$ is known as the Wilsonian effective action.

\begin{figure}[h!]
{\resizebox{\textwidth}{!}{
\rotatebox{0}{\includegraphics{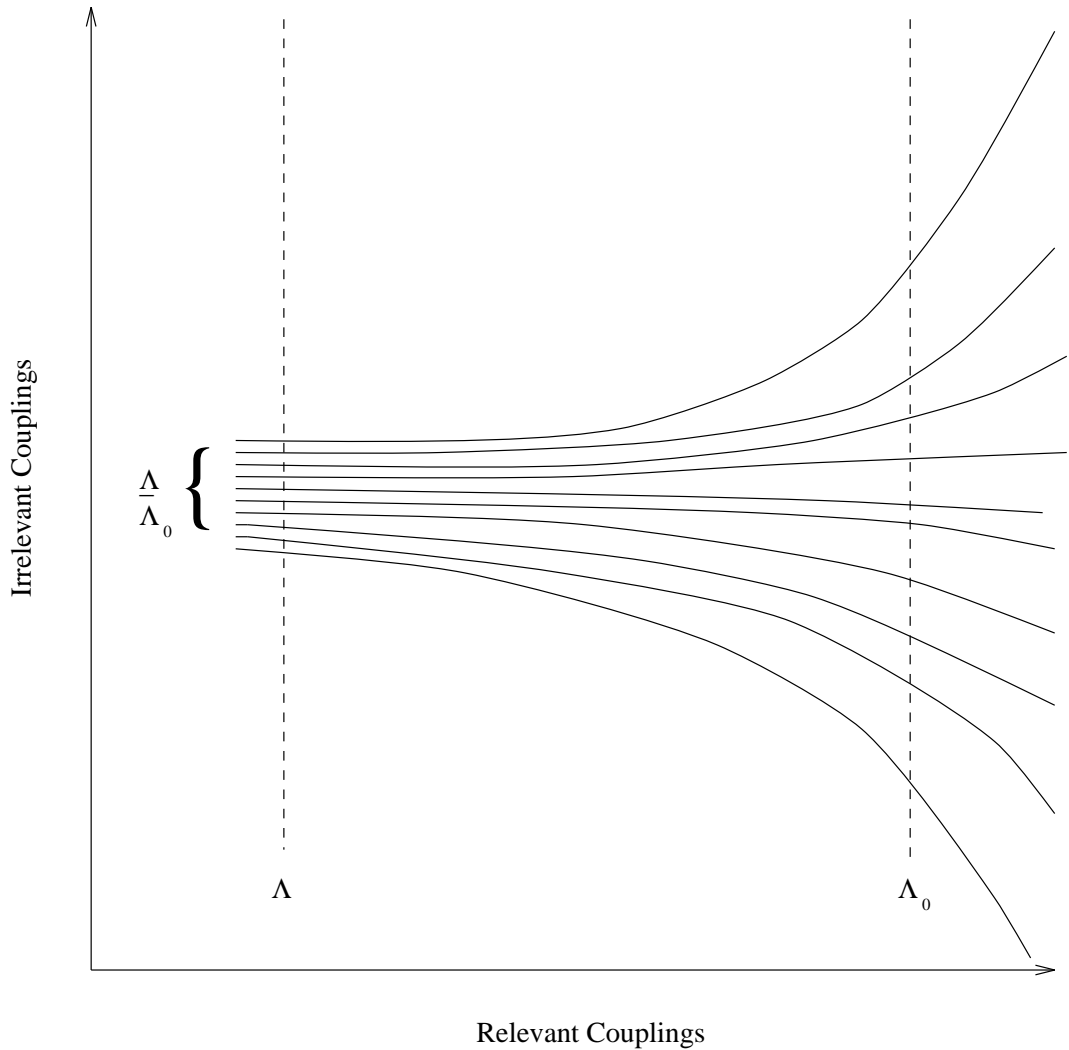}}}}
\caption{ The flow of the irrelevant couplings against the relevant
couplings as the cutoff is lowered from $\Lambda_0$ to $\Lambda$. The
couplings flow to a manifold of thickness $\Lambda / \Lambda_0$ for
$\Lambda \ll \Lambda_0$.}
\label{couplingflow}
\end{figure}

We no longer need to use the full theory at these low energy scales.
Instead we use the Lagrangian defined at $\Lambda$ to calculate
physical quantities at the scale $E$. We define the effective theory
to be renormalizable if we can calculate physical processes up to
errors of order $E / \Lambda_0$, once we have determined a finite
number of coupling constants at some scale $\Lambda \sim
E~$\cite{a:pol,a:ball/thorne,a:warr}. We call these couplings the
relevant couplings and the other couplings the irrelevant couplings.
We can now describe the low energy theory to a given order of accuracy
without any recourse to the full theory or an infinite number of
parameters. The flow of the couplings in the perturbative case is
shown in figure 
(\ref{couplingflow}). Each trajectory or coupling flow corresponds to
a particular choice of bare couplings at scale $\Lambda_0$. As we
evolve the couplings down to the scale $\Lambda$ we flow to a
sub-manifold of the coupling constant space with a dimension equal to
the number of relevant operators. This manifold has a thickness of
order $\Lambda / \Lambda_0$ (as we may expect considering that the
effective theory is renormalizable, as defined above). We see that
once the relevant couplings are known we automatically know the
irrelevant ones to an accuracy of order $\Lambda / \Lambda_0$ at the
scale $\Lambda$, cf figure (\ref{couplingflow}).

Looking at figure~(\ref{couplingflow}) more closely we see that at low
energy scales we have what is known as a 'self-similar' evolution ---
once we know the Lagrangian at low energy scale $\Lambda$, then the
Lagrangian at a lower scale $\Lambda'$ will look very much like the
one defined at $\Lambda$, except for a change in the values of the
relevant couplings. Now suppose we let the overall cutoff
$\Lambda_{0}$ tend to infinity, $\Lambda_{0}
\rightarrow \infty$. We will 
no longer have a self-similar evolution, but will now have an 'exactly
self-similar' evolution. That is lowering the cutoff will no longer
affect the values of the irrelevant couplings (expressed in terms of
the relevant couplings), and that the flow of the irrelevant couplings
has reached a fixed point.  This concept is known as the concept of
universality. The analysis of fixed points and their classification is
central to the understanding of what possible theories could be
relevant to the description of phenomena at present energies and is
hence central to the understanding of what phenomena can be understood
in terms of quantum field theory. Indeed, as Weinberg conjectured, any
quantum theory that is Lorentz covariant, unitary and satisfies
cluster decomposition is bound to look like a quantum \emph{field}
theory at sufficiently low energies~\cite{a:weinunitary,p:wein} . The
illustration that this is true was based upon an extensive analysis
involving the renormalization group and fixed point analysis
\cite{a:ball/thorne}.

This is a convenient point to mention that so far all four dimensional
field theories have been based around the Gaussian fixed
point~\cite{a:wk}.  This consists only of the part quadratic in the
fields and as such describes free, non-interacting field theories. As
it is free it has no anomalous dimension.  It is around this point
that perturbative proofs of renormalizabilty and perturbation theory
are based. It should be realized that we say \emph{based}. The actual
fixed point describes a non-interacting theory, whereas if we stay
close to the fixed point, as in perturbation theory, we can have a
theory defined in perturbation theory with an interaction. We can
perform the analysis of whether an operator is relevant or irrelevant
according to the dimension of the operator.  Operators of dimension
less than $D$ (the dimension of space-time) are referred to as
relevant, those of dimension greater than $D$ are known as irrelevant.
However, it is usual to classify the terms in a Lagrangian due to their
canonical dimension: we say that operators of canonical dimension
less than $D$ are relevant and those of canonical dimension greater
than $D$ are irrelevant. Of course it is well known that the dimension
of true importance is the canonical dimension plus an 'anomalous'
dimension due to quantum corrections.  Hence, it may be possible for
relevant operators, according to their canonical dimension, to become
irrelevant, and vice-versa.  Assuming that the anomalous dimension is
small, we see that relevant operators correspond to power counting
renormalizable interactions and irrelevant terms to power counting
non-renormalizable terms.  Therefore perturbative renormalizabilty
amounts to the statement that quantum corrections grow at most
logarithmically, so the anomalous dimensions remain
small~\cite{a:pol,a:ball/thorne}.  This is the case when we consider
the Gaussian fixed point, provided we don't tune the irrelevant
couplings to unnaturally large values.  This is the theoretical basis
upon which power counting renormalizabilty is based.

Some operators have zero canonical dimension, eg $\int d^Dx \,
\phi(x)^4$ in $D=4$. Operators with zero dimension are known as
marginal and perturbation theory with these interactions hence become
important in their classification. That is, we can use perturbation
theory with these operators to calculate $\beta$ function of
these operators.  For example, it can be
shown that $\int d^Dx \,
\phi(x)^4$ in $D=4$ is in fact an irrelevant interaction~\cite{a:hh}, 
so it is not possible to define an interacting scalar $\phi^4$ in four
dimensions with $\Lambda_0 = \infty$. (In fact $\Lambda_0$ must remain
less than some finite value.) This is a tremendously
important point as $\phi^4$ theories play an important role in the
Higgs sector of the standard model and if we have no interaction how
can we possibly hope to break the $SU(2)_L \otimes U(1)_Y$ symmetry of
the electro-weak sector?  Searches for non-Gaussian fixed points have
been taking place for a long time (see~\cite{a:wk} for an early
review), but have so far proved futile\footnote{Seiberg and Witten
have recently managed to find a non-trivial fixed points in $N=2$
supersymmetric theories \cite{a:Seiberg}}.  Of course if the cutoff is
kept finite then a quick glance at figure~(\ref{couplingflow}) will
show that it is possible to have an interacting effective $\phi^4$
theory in four dimensions. The key point is that the irrelevant
couplings are suppressed by a function of $\Lambda_0$, which vanishes
as $\Lambda_0 \rightarrow \infty$, but  can play a
r\^ole if the overall cutoff is finite.

We have now come to a point where we should make the above ideas more
precise. We begin with a discussion of fixed points, before moving
onto the more detailed areas of deriving and approximating
renormalization group equations. We will find it convenient to use
critical phenomena as an example in later chapters, so we end this
introduction with a brief outline of the theory behind problems in
critical phenomena.

\section{Fixed points}

One may ask what happens to the action as we continuously lower the
cutoff. The simplest possibility is that it flows into a fixed point
of the transformation. That is an action, $S^*$, such that continued
lowering of the cutoff leaves it unchanged. Such an action will
satisfy,
\begin{equation}
\tau[S^*]=0
\label{e:fp}
\end{equation}

Other possibilities exist, such as limit cycles and turbulent
behaviour, although they are less interesting from a physical point of
view (see \cite{a:wk} and references therein).

One may further enquire about the stability of these fixed points. To
do this we linearize about the fixed point by writing,
\begin{equation}
S[\phi]=S^*[\phi]+\delta S_{\Lambda}[\phi]
\end{equation}
This then yields,
\begin{equation}
\Lambda \frac{\partial}{\partial \Lambda} \delta S_{\Lambda}[\phi] = - L(\delta
S_{\Lambda})
\label{e:linear1}
\end{equation}
where $L$ is a linear operator acting upon $\delta S_{\Lambda}$.  We
will assume that $L$ will have a discrete set of eigenvalues
$\lambda_i$ corresponding to a set of eigenoperators
$\mathcal{O}_i$. In this case $\delta S_{\Lambda}$ can be expanded as
a series in the integrated $\mathcal{O}_i$'s
\begin{equation}
\delta S_{\Lambda}=\sum h_i(\Lambda) \mathcal{O}_i
\end{equation}
Equation (\ref{e:linear1}) then yields
\begin{equation}
\Lambda \frac{\partial}{\partial \Lambda} h_i(\Lambda)=  - \lambda_i h_i(\Lambda)
\label{e:linear2}
\end{equation}
Solution of equation (\ref{e:linear1}) will yield the values of the
eigenvalues $\lambda_i$.

We define the \emph{critical surface} as the surface, at
$\Lambda=\Lambda_0$, containing all points which will eventually flow
into a fixed point. We can classify the operators according to their
eigenvalue:-

If $\lambda>0$ we say the operator is relevant. The operator will flow
out of the fixed point as $\Lambda$
is lowered.

If $\lambda<0$ we say the operator is irrelevant. Successive
transformations on these operators force the operator into the fixed
fixed point.

If $\lambda=0$ we say the operator is marginal. We can't decide what
happens to the operator in the linearized approximation and have to go
the quadratic approximation to gain further information~\cite{a:hh}.

It is possible for a renormalization group equation to have several
fixed points. In this case we define the stability of a fixed point by
the number of relevant eigenvalues it has.

Given the above definitions we see that the critical surface is
locally spanned by the set of irrelevant operators.

This analysis is more extensively reviewed in \cite{b:zinn} and
\cite{b:amit}.

\section{Renormalization Group Equations and Approximations}

So far we have discussed renormalization group equations only in a
general sense. Eventually we will wish to build an approximation
scheme based upon the renormalization group. In this section we will
make the ideas of the first section more concrete by deriving an
equation for a scalar field theory. We will then proceed to discuss
how to approximate these equations in a sensible efficient manner.

\subsection{Why use the RG?}

Perhaps the first question to ask is why should we use the
renormalization group at all? For example why not use a scheme based
upon improving the ladder ansatz in Dyson-Schwinger equations? The
answer is quite simple -- we wish to preserve renormalizabilty. It was
pointed out in \cite{a:timappr} that Dyson-Schwinger equations quickly
run into problems with renormalizablity once we go beyond the ladder
ansatz. This illustrates an important point --- in any scheme
involving truncations perturbative renormalizabilty is not guaranteed.

We see that renormalizabilty is an important attribute to consider in
any approximation scheme and must be preserved by the scheme.
Polchinski \cite{a:pol} applied the renormalization group to provide a
conceptually elegant proof of the perturbative renormalizabilty of
$\phi^4$ theory in four dimensions. It is only necessary to show that
the irrelevant operators at zero coupling remain so at small coupling,
but this must be true because the right hand side of the flow equation
is a smooth function of the coupling.  Hence, it becomes clear, both
intuitively and in detail, that truncations of the flow equations are
perturbatively renormalizable. The renormalization group seems a safe
place to start the search for an approximation scheme.

\subsection{The Legendre Effective Action}

We could derive an equation for the Wilsonian effective action
$S_{\Lambda}[\phi]$. However it makes more sense to derive one for the
Legendre effective action with an infra-red cutoff $\Lambda$
\cite{a:timappr,a:marco,a:salm}. There are two main reasons for this:-

\begin{enumerate}
\item It will be shown that the flow equations for the (Wilsonian)
effective action $S_{\Lambda}$ have a tree like structure with one
particle irreducible bits linked by full infra-red cutoff
propagators~\cite{a:pol,a:timappr,a:timmom}. It will also be shown
that we will wish to use a momentum expansion in our approximation
scheme.  This tree like structure must be preserved by any momentum
expansion.  A momentum expansion corresponds to expanding the vertices
of the effective action in the scale of external momenta, regarding
this as small compared to $\Lambda$. In the sharp cutoff limit this
will cause all tree like terms with internal propagators to vanish,
as, by momentum conservation, the momenta flowing through this
internal propagator will be of the same scale as that of the external
momentum.  Noting that this internal propagator is also furnished with
a sharp infra-red cutoff $\Lambda$ in the sharp cutoff limit, we see
that the tree structure will be
destroyed~\cite{a:timappr,a:timmom}. This would be too much of a
mutilation of the theory as all tree level corrections to the theory
would be discarded as well as any loop diagram with more than one
vertex. If instead we apply a momentum expansion to the one particle
irreducible parts of $S_{\Lambda}$ then this problem is avoided.

\item It has long been known that we need to preserve a field
re-scaling invariance if we wish to calculate the anomalous dimension
$\eta$~\cite{a:golreparam,a:reparam1,a:reparam2,a:timderiv}. ie
\begin{equation}
\phi \rightarrow \frac{\phi}{\lambda}
\label{e:resc}
\end{equation}
should be preserved as an invariance of the approximation scheme.

If we use a momentum expansion of the Legendre flow equations then we
know that such a re-parameterization invariance is preserved with
certain choices of cutoff function, whereas there is no known one for
the Wilsonian effective action $S_{\Lambda}$, when a momentum expansion
is used \cite{timpriv}.

\end{enumerate}

The theory relating these two different types of action has been
extensively developed in \cite{a:timappr}.

\subsection{Deriving an RG equation}

Our starting point will be the following effective action
\begin{equation}
\exp{W_{\Lambda}[J]}=\int
D\phi
\;\exp{(-\frac{1}{2}\phi_.C^{-1}\;_.\phi+J_.\phi+S_{\Lambda_0}[\phi])}
\end{equation}
We assume the above has some internal symmetry, eg an $O(N)$ symmetry.
[Notation. It is perhaps a convenient point to explain the notation
that will be used. We will used a condensed notation whenever
possible. Hence $J_.\phi=\sum_{a=1}^{N}{\int d^Dx \phi^a(x)J^a(x)}$,
where $a$ is any internal symmetry index. Similarly propagators are
regarded as matrices in position/momentum space and any internal index
space.]  We assume the above is regulated by an overall UV cutoff
$\Lambda_0$, and that $C^{-1}$ is a smooth infra-red regulating cutoff
function of width $\varepsilon$. We assume it has the following
property,
$C^{-1}(q^2)=(\frac{1}{\theta_{\varepsilon}(q,\Lambda)}-1)q^2$.
$\theta_{\varepsilon}$ is a smooth regularization of the Heaviside
$\theta$ function, of width $2\varepsilon$ satisfying
$0<\theta_{\varepsilon}(q,\Lambda)<1$ for all positive $\Lambda$ and
$q$, and $\theta_{\varepsilon}(q,\Lambda)
\rightarrow \theta(q-\Lambda)$ as $\varepsilon \rightarrow 0$.

\begin{figure}[h!]
{\resizebox{\textwidth}{!}{
\rotatebox{90}{\includegraphics{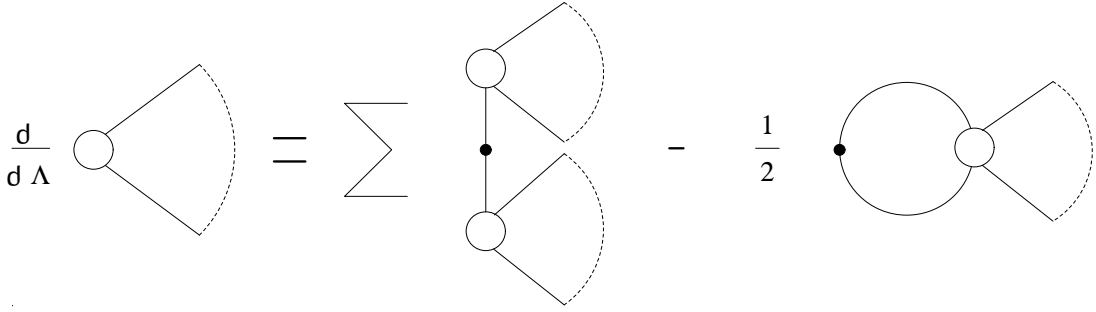}}}}
\caption{The flow equations for the vertices of the generating
functional $W_{\Lambda}$. The vertices are shown as open circles. In a
sharp cutoff the black dots will represent a delta function
restricting momenta to $q=\Lambda$. The sum occurs over all possible
disjoint partitions of momenta. }
\label{f:rg}
\end{figure}

If we differentiate with respect to $\Lambda$ we get
\begin{eqnarray}
\frac{\partial W_{\Lambda}[J]}{\partial\Lambda} & = &\int D\phi \; \phi_.
\frac{\partial C^{-1}}{\partial\Lambda}\;_.\phi \;
\exp{-(\frac{1}{2}\phi_.C^{-1}
\;_.\phi+J_.\phi+S_{\Lambda_0}[\phi])} \nonumber \\
 & = &- \left(\frac{\delta}{\delta J}.\frac{\partial
C^{-1}}{\partial\Lambda}.\frac{\delta}{\delta J}\right)
\exp{({W_{\Lambda}[J]})} 
\end{eqnarray}
Expanding the above then gives
\begin{equation}
\frac{\partial W_{\Lambda}[J]}{\partial \Lambda}=- \frac{1}{2} \left(\frac{\delta
W_{\Lambda}[J]}{\delta J}.\frac{\partial
C^{-1}}{\partial\Lambda}\;.\frac{\delta W_{\Lambda}[J]}{\delta J}
+tr[\frac{\partial C^{-1}}{\partial
\Lambda}\frac{\delta^2 W_{\Lambda}[J]}{\delta J \delta J}]\right)
\label{e:Weqn}
\end{equation}
Again in the above we have suppressed all internal indices. The first
term in the above represents the tree like terms to which we objected
before. The equations are best appreciated graphically as shown in
figure (\ref{f:rg}). This shows how the n-point functions
$W_n(\mathbf{p_1},\ldots,\mathbf{p_n})$ of $W_{\Lambda}[J]$ evolve.

\vspace{5mm}

We can now transform this to an equation for $\Gamma_{\Lambda}$, the
Legendre effective action, with an infra-red cutoff $\Lambda$.  We
define $\Gamma_{\Lambda}$, the generator of the one particle
irreducible parts of the vertices, by
\begin{equation}
\Gamma_{\Lambda}[\phi^c]+\frac{1}{2} \phi^c_.C^{-1}_.\phi^c =
-W_{\Lambda}[J]+J_.\phi^c
\label{e:legtran}
\end{equation}
where $\phi^c=\frac{\delta W_{\Lambda}}{\delta J}$ is the classical
field. Then using (derived in the standard way from~(\ref{e:legtran}))
\begin{equation}
\frac{\delta^2 W_{\Lambda}}{\delta J \delta
J}=\left(C^{-1}+\frac{\delta^2 \Gamma_{\Lambda}}{\delta \phi^c \delta
\phi^c}\right)^{-1} 
\end{equation}
we can re-write the equation as
\begin{equation}
\frac{\partial \Gamma_{\Lambda}[\phi^c]}{\partial \Lambda} =
\frac{1}{2} tr \left[\frac{\partial C^{-1}}{\partial
\Lambda}\;_.\left(C^{-1}+\frac{\delta^2 \Gamma_{\Lambda}}{\delta \phi^c
\delta \phi^c}\right)^{-1}\right]
\label{e:Legrge}
\end{equation}
Again all internal symmetry indices have been suppressed.  This will
be our starting point for all the work from now on

Note that as $\epsilon \rightarrow 0$ the $\frac{\partial
C^{-1}}{\partial \Lambda}$ will restrict momentum to the shell
$q=\Lambda$. Therefore, any other $C^{-1}(q)$ will contain
$\theta(0)$'s, which are ambiguous \cite{a:timappr,a:timmom}. Extreme
care must be taken with this limit.

\section{Sharp vs. Smooth}

It has come to the point where we have to decide what type of cutoff
to use. In this section we will briefly describe the pros and cons of
sharp \cite{a:hh,a:ncs2,a:HazenNager} and smooth
\cite{a:timderiv,a:wet,a:ellwanger} cutoffs.

\subsection{Sharp Cutoffs}

The first problem that arises with using a sharp cutoff is how to take
the sharp cutoff limit as $\varepsilon \rightarrow 0$. We mentioned
above that any other term containing a $C^{-1}(q)$ becomes ambiguous
in this limit. To circumvent this problem we separate the problematic
terms by writing
\begin{equation}
\frac{\delta^2 \Gamma_{\Lambda}}{\delta \phi \delta
\phi}(p,p')=\gamma(p,\Lambda) (2\pi)^D\delta(p+p')+\hat{\Gamma}[\phi](p,p',\Lambda)
\end{equation}
so that $\hat{\Gamma}[0]=0$, and drop the field independent vacuum
energy term. We can then write the subtracted equation as
\begin{equation}
\frac{\partial \Gamma_{\Lambda}}{\partial
\Lambda}=-\frac{1}{2}tr\left[
\frac{\partial C^{-1}}{\partial \Lambda}\;
.\left(C^{-1}+\gamma\right)^{-2} .\hat{\Gamma}.
\left(1+\{C^{-1}+
\gamma\}.\hat{\Gamma} \right)^{-1} \right]
\label{e:nearlysharp}
\end{equation}

The sharp cutoff limit can now be taken. This is not so
straightforward 
as it seems as we need to be able to deal with the
$\theta(0)$'s that occur when we do take this limit. The answer turns
out to be by no means as simple as using the usual physics convention
of setting $\theta(0)=\frac{1}{2}$! To take the limit we use the
following lemma:-

Let $f(\theta_{\epsilon},\Lambda)$ be any function whose dependence on
the second argument remains continuous at $\Lambda=p$ in the limit
$\epsilon \rightarrow 0$. Then
\begin{equation}
\lim_{\epsilon \rightarrow 0} \delta_{\epsilon}(p,\Lambda)
f(\theta_{\epsilon}(p,\Lambda),\Lambda) \rightarrow \delta(\Lambda-p)
\int^1_0 dt f(t,p) 
\label{e:limit}
\end{equation}
where $ \delta_{\varepsilon}(p,\Lambda)= - \frac{\partial}{\partial
\Lambda}
\theta_{\varepsilon}(p,\Lambda)$.
This is easily proved by noting that
\begin{equation}
\delta_{\epsilon}(p,\Lambda)f(\theta_{\epsilon}(p,\Lambda),\Lambda) =
\left . \left  \{ \frac{\partial}{\partial \Lambda}
\int^1_{\theta_{\epsilon}(p,\Lambda)} dt f(t,\Lambda') \right \} \right
|_{\Lambda ' = \Lambda}
\end{equation}
and that the integral is a representation of a step function with
height $\int_0^1 dt f(t,\Lambda')$. (This now yields
$\delta_{\epsilon}(p,\Lambda) \theta_{\epsilon}(p,\Lambda)^2
\rightarrow \frac{1}{3}\delta(p-\Lambda)$, which doesn't equal the
$\frac{1}{2} \delta(p-\Lambda)$ which some may have na\"{\i}vely
guessed.)

Using (\ref{e:limit}), we may now take the sharp cutoff limit in
equation (\ref{e:nearlysharp}) to yield,
\begin{equation}
\frac{\partial \Gamma_{\Lambda}}{\partial \Lambda} = - \frac{1}{2}
\int \frac{d^Dq}{(2 \pi)^D} \frac{\delta(q-\Lambda)}{\gamma(q,\Lambda)}
\left[\hat{\Gamma}.\left(1+G.\hat{\Gamma}\right)^{-1}\right](\mathbf{q},\mathbf{q'})
\label{e:RGsh}
\end{equation}
where we have defined,
\begin{equation}
\lim_{\varepsilon \rightarrow o}
\frac{1}{C^{-1}(p,\Lambda)+\gamma(p,\Lambda)} = G(p,\Lambda) =
\frac{\theta(p-\Lambda)}{\gamma(p,\Lambda)}
\end{equation}
which represents a sharply infra-red cutoff, full two point Greens
function.

Notice that this equation displays field re-parameterization
invariance under $\hat{\Gamma} \rightarrow \lambda^2\hat{\Gamma}, \gamma
\rightarrow \lambda^2 \gamma$ and $G \rightarrow G/\lambda^2$.

We now have a sharply cut-off RG equation. However, we quickly realize
that any attempt to solve any of the above equations by a direct
numerical approach will quickly grind to a halt due to the sheer
complexity of the problem \cite{a:timappr}. We are therefore forced to
choose some approximation scheme.

As a first attempt we could try expanding in powers of field and
truncating at some maximum order \cite{a:timtrunc,a:t+wett,a:alford},
ie we write (assuming a $Z_2$ symmetry),
\begin{equation}
\Gamma=\sum_{i=0}^{M} a_i \phi^{2 i}
\label{e:truncexp}
\end{equation}
This expression is then substituted into equation (\ref{e:RGsh}) and
the equation expanded up to a maximum power of $\phi$. This results in
relationships between the coefficients which can easily be
solved. This approach has been extensively investigated. However few
people at first recognized certain problems with this scheme.

Firstly as pointed out in \cite{a:timtrunc} the truncations at first
seem to converge to an answer but stop converging after a certain
value of $M$.  The reason for this is that there are in fact only a
finite number of true solutions \cite{a:hh,a:timlat,a:timhalp},
together with an infinite number of solutions with singular behaviour
for some real value of $\phi$. Very bad solutions will have singular
field dependence close to the origin causing the coefficients of
$\phi^{2m}$ in (\ref{e:truncexp}) to diverge with m. Of course, the
truncation for which the coefficient of the $2 M + 2$ vertex vanishes
will therefore better approximate the the Taylor series of a
non-singular solution.  At first increasing the value of $M$ will
improve the approximation, by forcing the singularities further away
form the origin. However, even the non-singular solutions have
singularities, but for complex $\phi$, at a radius of, say ,
$|\phi|=r$. Therefore the truncations can not be expected to converge
to better results than would be obtained with 'moderately bad'
solutions with singular behaviour at or beyond the value of $r$.  Also
spurious solutions are generated and no completely reliable method can
be found to reject these \cite{a:timtrunc}.

Taking the above into account we could try some sort of momentum
expansion. It should be noted that (\ref{e:RGsh}) contains $\theta$
functions, which do not have an analytic expansion. For example,
\begin{equation}
\theta( |\mathbf{p} + \mathbf{q} | - 1) = \theta(\mathbf{\hat{p}_.q} + p/2)
= \theta(\mathbf{q_.\hat{p}}) + \sum^{\infty}_{m=1} \frac{(p/2)^m}{m!}
\delta^{(m-1)}(\mathbf{q_.\hat{p}})
\end{equation}
where we have defined $\mathbf{\hat{p}}=\mathbf{p}/|\mathbf{p}|$ and
$\delta^{(m-1)}$ is the $m^{\scriptsize{th}}$ derivative of the
$\delta$-function with respect to its argument.  The above expansion
means we cannot expand in powers of $p^2$ but are forced to use a
non-analytic expansion in powers of momentum scale,
$\rho=\sqrt{p_.p}$, instead
\cite{a:timappr,a:timmom}.

To lowest order in the approximation we discard all external momentum
dependence and write
\begin{equation}
\Gamma_{\Lambda}=\int d^Dx \left\{ \frac{1}{2}\left(\partial_{\mu} \phi \right)^2
+ V(\phi) \right\}
\end{equation}

Such an ansatz yields the following equation,
\begin{equation}
\frac{\partial}{\partial \Lambda}V(\phi) =\ln\left(1+\frac{\partial^2 V(\phi)}
{\partial \phi \partial \phi} \right)
\end{equation}

This forms what is known as the \emph{local potential approximation}
and has been extensively investigated by several authors
\cite{a:WandH, a:hh, a:ncs1,a:ncs2,a:alford}. It produces
good results, although it does fail to take account of any wave
function renormalization and other momentum dependent effects.

The problems arise when we try to go beyond this level of
approximation. To get any further, we will eventally be forced to
calculate certain averages, which appear only to be calculable, in a
closed form, in certain truncations~\cite{a:timmom}. As we already know
that truncations do not work we are forced to look down another
avenue.

\subsection{Smooth Cutoffs}

We have rejected sharp cutoffs and we know that truncations don't
work.  We also know that any attempt to use a momentum expansion (that
is a derivative expansion) must preserve a re-parameterization
invariance.  We also know that we need to keep all powers of the
fields involved as we cannot use truncations in the fields. To satisfy
these requirements we will employ a derivative expansion:-
\begin{equation}
\Gamma[\phi]=\int d^Dx \, V(\phi) + \frac{1}{2}\left(\partial_{\mu} \phi
\right)^2 K(\phi) + \cdots
\label{e:derexp}
\end{equation}
This is substituted into~(\ref{e:Legrge}) and the right hand side
expanded up to a maximum number of derivatives.

 The momentum expansion corresponds to a local derivative expansion in
the effective Lagrangian, with radius of convergence $p \approx
\varepsilon$ (this arises from the expansion of terms such
$\theta_{\varepsilon}(|\mathbf{q} + \mathbf{p}|, \Lambda)$ with $q
\approx \Lambda$). Since these expansions are substituted back into
the flow equations and averaged over $p \approx \Lambda$ we must have
$\varepsilon > \Lambda$ for convergence. Obviously the cutoff must
have $\Lambda > \varepsilon$, as otherwise there would be no
suppression of low momenta. Hence we see that $\varepsilon \approx
\Lambda$ and typically the expansion will converge only slowly, if at all. To maximize the
rate of convergence it clearly follows that we should choose the width
to be as large as possible, as then we will have convergence as well
as suppression of low momentum modes. We see that we are forced to use
a cutoff that has a width of at least
$\Lambda$~\cite{a:timappr,a:timderiv}.

To preserve a re-parameterization invariance we will use a power law
additive cutoff
\begin{equation}
C(q,\Lambda)=\frac{q^{2 k}}{\Lambda^{2 k}}
\end{equation}
for $k$ a non-negative integer. The re-parameterization invariance
manifests itself in the form of a scaling symmetry with a set of
(non-physical) scaling dimensions. That is if we choose
\begin{eqnarray}
[q_{\mu}]=[\partial_{\mu}]=1, & [\phi]=k+D/2, & [\Gamma_{\Lambda}]= 0,
\end{eqnarray}
as the non-physical scaling dimensions, then the scaling symmetry of
equation (\ref{e:Legrge}) becomes apparent.  We already know that we
need to choose a cutoff width of at least $\Lambda$ and that
convergence will be quicker the 'wider' the cutoff function is. As we
have the following identity,
\begin{eqnarray}
\theta_{\varepsilon}(q,\Lambda) & = & \frac{q^2 C(q,\Lambda)}{1+q^2
C(q,\Lambda)} 
\\
& = & \frac{q^{2(k+1)}}{\Lambda^{2k}+q^{2(k+1)}}
\end{eqnarray}
we see that it is beneficial (ie results in the widest cutoff width)
if we take $k$ as small as possible. As we also need the cutoff to
regularize the theory it can be shown that $k > D/2 -1 $. Hence we
chose $k$ to be the smallest integer greater than $D/2-1$. For $D=3$
this means we take $k=1$.

The study of the above approximation scheme will form the major part
of this thesis. We will apply it to some simple problems in critical
phenomena and show that it does indeed work. We leave further
development of the scheme to a later chapter and we will review some
of the concepts in critical phenomena required for later chapters.

\newpage
\section{Critical Phenomena}

In this section we only be concerned with the problems posed by second
order phase transitions. These are transitions where there is a
continuous change in the properties of the system from one state to
another. eg the continuous appearance of a magnetization as a
ferromagnetic material is cooled. (As opposed to first order
transitions where there is a discontinuous change, see~\cite{b:zinn}
for details).  Such phase transitions are well known, even in everyday
life. For example, consider water boiling in a kettle -- the liquid is
changing from a liquid to a vapour. At a certain critical temperature
this phase transition is second order. In fact, if we look at
this  transition more closely we see the phenomena
known as critical opalescence, where the fluid takes on a milky
appearance at the transition point. This happens exactly at the
transition point and is due to regions the size of microns fluctuating
coherently on a large scale. This illustrates two difficulties with
critical phenomena,
\begin{itemize}
\item we need to consider all length scales.
\item we need to consider long ranges.
\end{itemize}

This makes theoretical work extremely difficult; we find it nearly
impossible to deal with problems involving just three degrees of
freedom, let alone one involving maybe hundreds of thousands.

There are other problems that we would like to understand in second
order transitions. For example, we would like to know why seemingly
separate physical systems display very similar critical behaviour. eg
uniaxial ferromagnets and fluids near their critical points display
similar types of behaviour, even though they are totally different
physical systems.  This phenomenon is known as universality.

In the next section we describe some simple phenomenology of phase
transitions and then briefly outline a theoretical background.  There
are many references to the subject but the books by Zinn-Justin
\cite{b:zinn} and Amit \cite{b:amit}, and the article by Weinberg
\cite{a:wein} are particularly good introductions.

\subsection{Scaling}

For convenience we will mostly use the language of ferromagnetic
systems.

The basic quantity of interest in statistical mechanics is the free
energy, $F$. This is defined by:
\begin{equation}
F= - k T \log Z
\end{equation}
where Z is the partition function of the theory. For example, consider
a classical spin model, so that the partition function takes the
form
\begin{equation}
Z = \int \prod_i \, dS_i \, \rho(S_i) \exp - \beta {\cal H}(S)
\end{equation}
where $\beta=1/k T$, $k$ is the Boltzmann constant, $T$ is the
temperature, $\rho(S_i)$ is a spin weighting factor describing the
local microscopic properties of the system, and ${\cal H}$ is the
Hamiltonian given by,
\begin{equation}
\sum_{i j} V_{i j} S_i S_j + \sum_i H_i S_i
\label{e:ising}
\end{equation}
The variable $H_i$ in the above represents an externally applied
magnetic field.  Such models have achieved great success in describing
simple ferromagnetic systems. The statistical average of a quantity
$X$ is given by
\begin{equation}
<X> = \frac{1}{Z}\int \prod_i dS_i \, \rho(S_i) X_i \exp - \beta {\cal H}(S)
\end{equation}
Using this definition we define the magnetization of a system to be
the average of the spins
\begin{equation}
M(H,T)=V^{-1} \sum_n a^D < S_n >
\end{equation}
where $a$ is the lattice spacing.
In what is known as the thermodynamic limit, where $N \rightarrow
\infty$, we can describe the magnetization as
\begin{equation}
M(H,T)=-\frac{\partial}{\partial H} F(H,T)
\end{equation}
This can easily be seen by realizing that taking a derivative with
respect to $H$ brings down a $S_i$ into the sum. It can happen, that
even in the presence of zero external magnetic field, below a certain
temperature $T_c$ there is a non-zero value for the
magnetization. When this occurs there is said to be a spontaneous
magnetization and the temperature at which this first occurs is called
the critical temperature.

There are other quantities of interest. The susceptibility is defined
as
\begin{equation}
\chi(H,T)=\frac{\partial M(H,T)}{\partial H} = \frac{1}{V} \sum_{n\,  m} a^{2D}
<(S_n-<S>) (S_m - <S>)>
\label{e:suscept}
\end{equation}
and this represents the response of a system to a small applied
magnetic field. Also of interest is the specific heat which is defined
as
\begin{equation}
C = -T\frac{\partial^2 F}{\partial T^2}
\end{equation} 

Various physical quantities diverge as we approach a second order
critical point. For example, the correlation function of two
fluctuating fields behaves as
\begin{equation}
g_{n m} = <(S_n -<S>)( S_m - <S>)>
\sim  \exp{(-|x_n - x_m|/{\xi})} \, x^{\tau} \; \ \mbox{as $x \rightarrow \infty$ }
\label{e:corr1} 
\end{equation}
We call $\xi$ the correlation length.  As we approach the critical
temperature the correlation length diverges as follows
\begin{equation}
\xi \sim \begin{array}{cc}
     (T-T_c)^{-\nu} & T>T_c \\ (T_c-T)^{-\nu'} & T<T_c \end{array}
\end{equation}
We call $\nu$ a critical exponent.  Other exponents can also be
 defined. For example, the susceptibility diverges as
\begin{equation}
\chi \sim \begin{array}{cc}
     (T-T_c)^{-\gamma} & T>T_c \\ (T_c-T)^{-\gamma'} & T<T_c
     \end{array}
\end{equation}
and the specific heat acts as
\begin{equation}
C \sim \begin{array}{cc} (T-T_c)^{-\alpha} & T>T_c \\
     (T_c-T)^{-\alpha'} & T<T_c \end{array}
\end{equation}
Below the critical temperature a spontaneous magnetization appears
which behaves as
\begin{equation}
M \sim (T_c-T)^{\beta}
\end{equation}
If we are exactly at the critical point, $T=T_c$, then the correlation
function will act like
\begin{equation}
<(S_n - <S>) (S_m -<S>)> \sim \frac{1}{|x_n - x_m|^{D-2+\eta}}
\; \ \mbox{for $a \ll |x_n-x_m| \ll \infty$}
\label{e:corr}
\end{equation}
Notice that we have lost all
functional dependence on a fundamental length scale --- this reflects
the fact that at criticality the system is scale invariant, and is
also consistent with the appearance of long range order.  The exponent
$\eta$ is known as the anomalous dimension.

\newpage
Various relationships exist between these exponents defined at the
fixed point\footnote{The exponents discussed so far are defined by the
behaviour at the fixed point. In addition to these there is also an
infinite set of exponents characterizing corrections to the scaling
laws, when the system is close to the fixed point}, known as scaling
relations. These mean that you only need to know two of the above
critical exponents to calculate the rest. For example,
\begin{eqnarray}
2\beta + \gamma & = & 2-\alpha \\
\gamma    &    =         & \nu ( 2- \eta) \label{e:fisher} 
\end{eqnarray}
are the two of the best known.

These relations are easily reproduced. As an example consider the
relation (\ref{e:fisher}). To prove this, first consider the correlator
of two spins (\ref{e:corr}). We see from equation (\ref{e:corr}), that
when we are close to the transition point that the correlator has
dimension $-D+2-\eta$. Hence, if we consider $\frac{a^{2D}}{V}\sum_{i
j} g_{i j}$ then this will have scaling dimension $2-\eta$. Therefore,
as the correlation length sets the basic scale of the system at
criticality, so that at criticality the system loses all dependence on
all other length scales, we can determine the behaviour of the
magnetization, specific heat etc, in terms of the behaviour of
$\xi$. This hypothesis is known as the scaling hypothesis~\cite{a:widom}. Using the
scaling hypothesis, we see that $\sum_{i j} g_{i j}$ satisfies,
\begin{equation}
\sum_{i j} g_{i j} \sim \xi^{2-\eta} \sim (T-T_c)^{\nu (-2 + \eta)}
\end{equation}
near the critical point.  However, a quick glance at equation
(\ref{e:suscept}) reveals that $\chi = \frac{a^{2D}}{V} \sum_{i j}
g_{i j}$. Therefore, we must have
\begin{equation}
\chi \sim (T-T_C)^{- \gamma} \sim (T-T_c)^{\nu ( -2 + \eta)}
\end{equation}
This then forces equation (\ref{e:fisher}) to hold, ie $\gamma =
\nu (2 - \eta)$. This is known as Fisher's scaling relation.
It is also found that the indices are symmetrical about the transition
point so $\alpha=\alpha'$, $\nu=\nu'$ etc. The reader is referred to
the extensive literature on the subject for further details.

\subsection{Critical Phenomena and the RG}

In this section we will indicate how to relate the above theory to the
renormalization group.  We know that near criticality the correlation
length is very large and we need to consider a large number of degrees
of freedom. Hence, if we wish to perform calculations near the phase
transition and gain a theoretical understanding of the problem, we
need to find some method of reducing the number of degrees of
freedom. The renormalization group does such a thing. If we
continually integrate out the fluctuations with the shortest
wavelength we gradually reduce the degrees of freedom, making the
problem more tractable. The key point relating this to the critical
point is that at criticality all wavelengths are equally
important. This is because the correlation length diverges and we
therefore lose all dependence on any fundamental length scale (cf
equation~(\ref{e:corr})). This also means that as we continually
integrate degrees of freedom out of the theory we essentially see the
same picture --- ie we are at a fixed point of the transformation.

To calculate the critical exponents we need to consider the behaviour
near the fixed points. To aid this discussion we will re-write the
equations in terms of dimensionless quantities. Suppose we are near a
fixed point, but not on the critical surface. Equation
(\ref{e:linear2}) shows that
\begin{equation}
h_i \sim \Lambda^{ -\lambda_i}
\end{equation}
As the $h_i$'s are now dimensionless we see that we have
\begin{equation}
h_i= \left ( \frac{\mu}{\Lambda} \right )^{ \lambda_i}
\end{equation}
Then, as the flow of the action will be dominated by the largest
relevant eigenvalue, we can write,
\begin{equation}
\delta S_{\Lambda} \sim  \left ( \frac{\mu}{\Lambda} \right )^{\lambda_L}{\cal O}_L 
\end{equation}
where $\Lambda_L$ is the largest eigenvalue and ${\cal O}_L$ is the
operator associated with it. We see that the integrated operator
${\cal O}_L$ has dimension $ - \lambda_L$ and is associated with a
coupling $\mu^{\lambda_L}$ of dimension $\lambda_L$.  Now consider the
situation at $\Lambda=\Lambda_0$. There will be an operator
corresponding to the deviation from criticality. This will be the
dominant operator in determining the statistical state of the
system. Therefore, we  associate this operator with ${\cal O}_L$:
\begin{equation}
\delta S_{\Lambda_0} = (T-T_c) {\cal O}_L +O((T-T_c)^2)
\end{equation}
As we know that the coefficient of ${\cal O}_L$ has dimension
$\lambda_L$, we see that the scaling dimension of the temperature 
difference is also $\lambda_L$. Therefore, as $[\xi]=-1$,
we must, 
have
\begin{equation}
\xi \sim (T-T_c)^{-1/\lambda_L}
\end{equation}
or, $\nu = 1/\lambda_L$.

The value of $\nu$ is therefore determined by the linearization
procedure. The anomalous dimension is a property of the fixed point
itself --- in the full un-approximated equations the value of $\eta$
is determined by the field re-parameterization invariance.  The
scaling symmetry of the equations turns the renormalization group
equations into a non-linear eigenvalue for $\eta$, with only a few
values of $\eta$ leading to acceptable fixed point solutions. By
preserving the re-parameterization invariance in an approximation
scheme we can still determine $\eta$ without recourse to any
non-physical arguments or parameters. We can now determine two
critical exponents and therefore determine the others.  Corrections to
the scaling behaviour are given by the subleading eigenvalues. For
example, the leading correction to scaling exponent, denoted by
$\omega$, is defined as being equal to minus the least negative
eigenvalue. For example, close to the critical temperature we have the
following scaling relation,
\begin{equation}
\xi \sim |T-T_c|^{-\nu} + \cdots
\end{equation}
However, there are also corrections to this expression, so further
away from the 
critical temperature we have,
\begin{equation}
\xi \sim |T-T_c|^{-\nu} + a_{\xi} |T-T_c|^{(\omega-1)\nu} +\cdots
\end{equation}

\subsection{Critical Phenomena and Field Theory}

We have now reached the stage where we should relate physical systems
to field theoretical models. It should have been realized by now that
there are two equivalent descriptions of critical phenomena. In the
above we have sometimes used the language of classical spin like
systems and sometimes we have found it more convenient to use a
field-theoretical like language. There is in fact an intrinsic link
between the two descriptions and we will make an attempt to briefly
describe the relation between the descriptions.  Consider the
classical spin model defined above by equation
(\ref{e:ising}). If we assume that the lattice is hyper-cubic and that
we only have nearest neighbour couplings then we can write
\begin{equation}
\sum_{i j} V_{i j} S_i S_j = K \sum_n \left ( \sum_{\mu} (S_{n+\mu}
-S_n )^2 - 2 d S_n^2 \right )
\end{equation}
where $\mu $ runs over the directions of the lattice. In fact, if we
consider a model where the spin is constrained by $S^2=\pm 1$ (known
as the Ising model), we can describe this using a spin weighting
function of (taking an appropriate choice of $\lambda$ and $\kappa$),
\begin{equation}
\rho(S_i) \propto \exp  - ( \kappa S_n^2 + \lambda S_n^4)
\label{e:spinweight}
\end{equation}
We can then write the Hamiltonian as
\begin{equation}
{\cal H} = \sum_n \left ( K(\beta) \sum_{\mu} (S_{n+\mu} -S_n)^2
+\mu(\beta) S_n^2 +\lambda(\beta) S_n^4 + H_n S_n \right )
\label{e:isingH}
\end{equation}
[It is perhaps worth pointing out that other versions of the
spin-weighting function (\ref{e:spinweight}), with terms of higher
powers in $\phi$ could also be used. However it turns out that such
terms are irrelevant, in the sense defined above, and have no
influence on the critical theory.]

The above expression (\ref{e:isingH}) for the Ising Hamiltonian should
look familiar --- it looks like a lattice regularized $\phi^4$ theory
\cite{b:rothe}. To see this take a massive $\phi^4$ in $D$ dimensions,
\begin{equation}
S(J) = \int d^Dx \left (\frac{1}{2} (\partial_{\mu} \phi)^2 +
\frac{1}{2} m^2\phi^2 + \frac{\lambda}{4 !} \phi^4 \right )
\label{e:lat}
\end{equation}
and place it on a hyper-cubic lattice by writing,
\begin{eqnarray}
x & \mapsto & x_n = a n_{\mu} \hat{e}_{\mu} \\
\phi(x) & \mapsto & \phi_n = \phi(x_n) \\
D \phi & \mapsto & \prod_n d \phi_n \\
\partial_{\mu} \phi & \mapsto & \frac{1}{a} \left ( \phi(x_n +
\hat{e}_{\mu}) - \phi(x_n) \right ) = \frac{1}{a} ( \phi_{n+\mu} -
\phi_n) 
\end{eqnarray}
where $a$ is the lattice spacing and $\hat{e}_{\mu}$ is a set of $D$
orthonormal vectors. Under these transformations the lattice
regularized theory is
\begin{equation}
S(\phi,J) = \sum_n a^D \left \{ \frac{1}{a^2} \sum_{\mu} \frac{1}{2}
\left ( \phi_{n+ \mu} - \phi_n \right )^2 + m^2 \phi_n^2 + \lambda
\phi_n^4 +J_n \phi_n \right \}
\end{equation}
and the partition function becomes,
\begin{equation}
Z(J) = \prod_n \int d \phi_n \exp S(\phi, J)
\end{equation}
where the functional measure $D\phi$ is replaced by a finite
dimensional integral $\prod_n d\phi_n$.

The correspondence between (\ref{e:lat}) and (\ref{e:isingH}) should
now be clear. So we see that there is a direct link between lattice
theories and the Ising model. However, what we are really interested
in is the critical theory. To clearly see the link we will have to
show explicitly what happens at the phase transition and consider the
ground states of the Ising model.

\begin{figure}[t]
\centering
\subfigure[$\mu >
0$]{\epsfig{figure=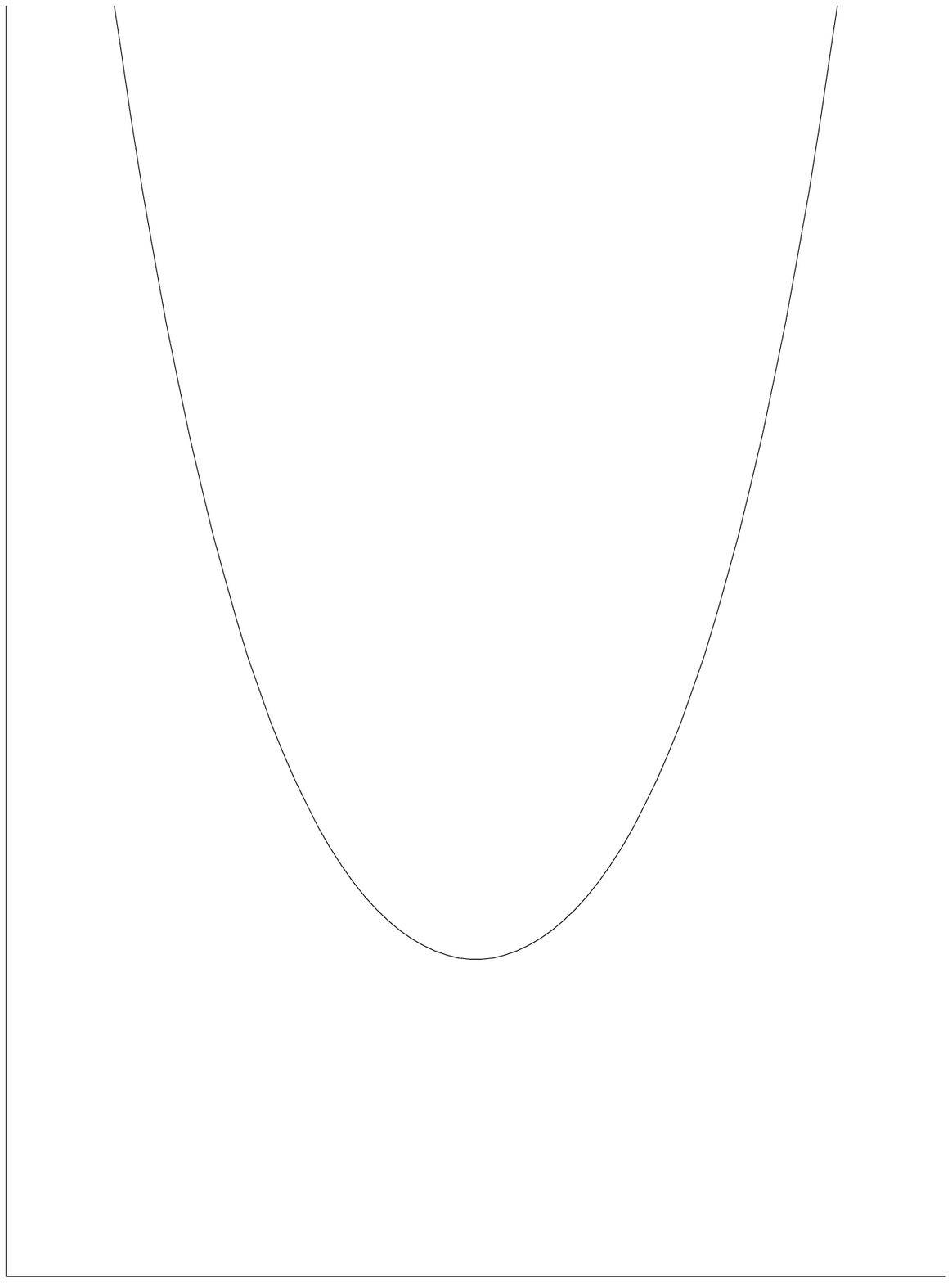,width=0.46\textwidth,height=0.46
\textwidth}} 
\subfigure[$\mu <
0$]{\epsfig{figure=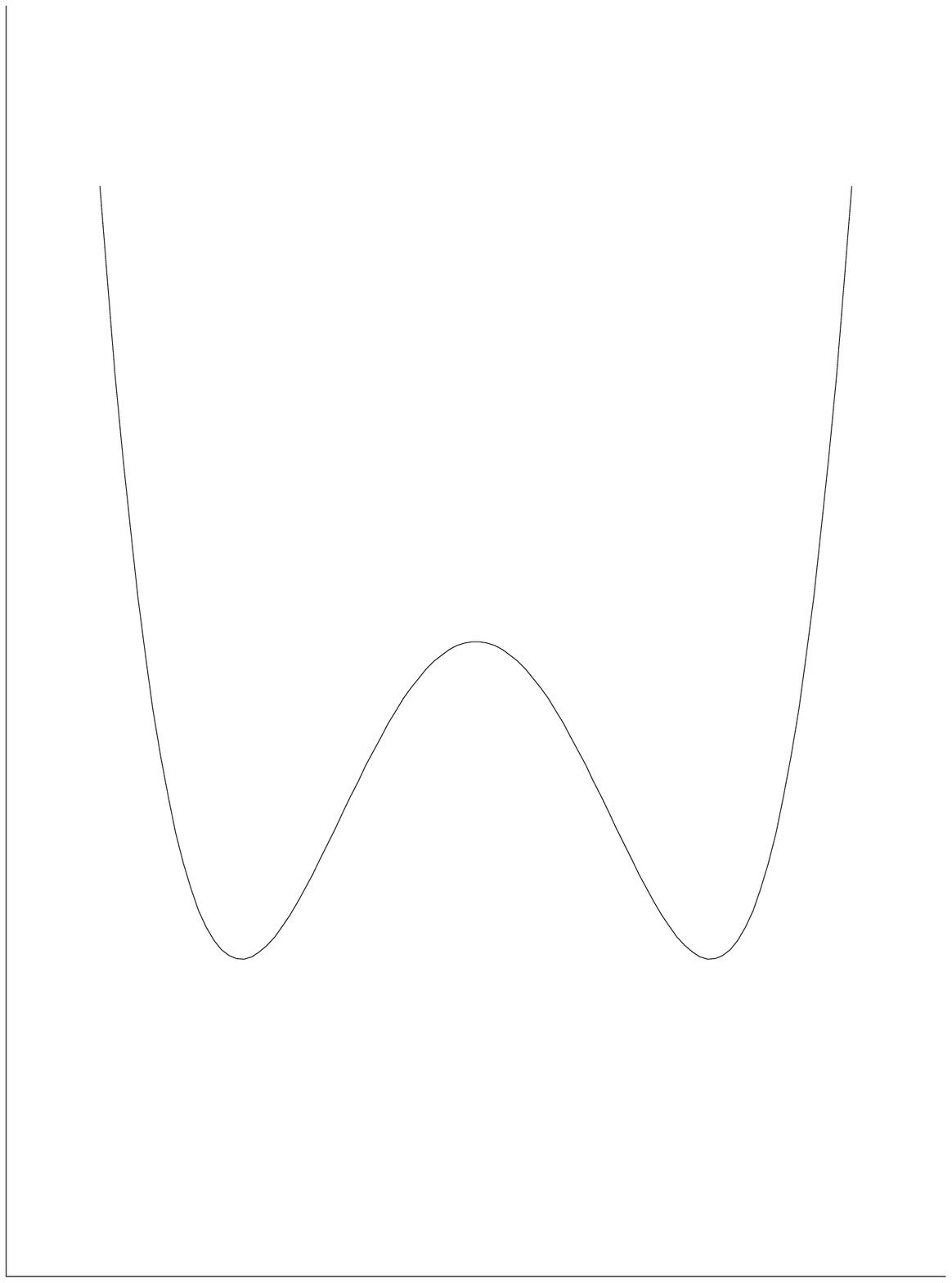,width=0.46 \textwidth,height=0.46
\textwidth}} 
\caption{The potential, $\mu(\beta) S^2 + \lambda(\beta)
S^4$, for the unbroken case $\mu > 0$ (a), and the broken case $\mu <
0$, (b).}
\label{pot}
\end{figure}

Consider the potential of the model $\mu(\beta) S^2 + \lambda(\beta)
S^4$, where for the moment we will ignore quantum corrections (ie we
will use what is known as mean field theory \cite{b:zinn}).  For
$\mu(\beta) > 0$ we have a potential like figure (\ref{pot}a). We see
we have a ground state in which $S_n=0$ and
\begin{equation}
<S> = \frac{1}{V} \sum_n S_n = 0 \ \ \mbox{ for $\mu({\beta}) > 0$}
\end{equation}
If $\mu(\beta)<0$ then we have a potential like figure (\ref{pot}b).
Now the ground state corresponds to the $S_n$ aligned with $S_n=
\sqrt{\frac{-\mu} {2 \lambda}}$, so that 
\begin{equation}
<S> = \sqrt{ \frac{-\mu} {2 \lambda}} \ \ \mbox{ for $\mu({\beta}) <
0$}
\end{equation}
We see that the phase transition corresponds to the point where
$\mu=0$. The temperature that corresponds to this is called the
critical temperature.

Now let us make the correspondence between the critical Ising model
and and the critical lattice theory. To make direct comparisons we
first move to non-dimensional variables. That is we write
\begin{eqnarray}
\lambda=g^2 a^{D-4}, & \phi'=g a^{\frac{D}{2} -1 } \phi, &  \mbox{and}
\ \ 
J'=g a^{D/2+1} J
\end{eqnarray}
the action then becomes
\begin{equation}
S(\phi',J') = -\frac{1}{g^2}\sum_n \left \{ \frac{1}{2} \sum_{\mu}
(\phi'_{n+\mu} -
\phi'_n)^2 +\frac{1}{2} m^2 a^2 \phi'^2_n +\frac{1}{4!} \phi'^4_n
+J'\phi'_n \right \}
\label{e:lat2}
\end{equation}

We see comparing equations (\ref{e:isingH}) and (\ref{e:lat2}) that we
have the following correspondence,
\begin{equation}
m^2 a^2 \sim \mu(\beta)
\end{equation}
We know that as we approach the critical temperature $\mu(\beta)
\rightarrow 0$. This forces us to fine tune $a$ to zero as well, as
the physical mass of the system $m$ must be held fixed. Hence, we see
that as we approach the fixed point we must also approach the
continuum limit.

We should really have expected this on heuristic arguments:- we know
that at the critical point the Ising model is scale invariant.
Therefore to preserve this invariance we cannot have any massive
parameter in the field theoretical model and so there cannot be any
cutoff involved. We can also use this argument to show that the
critical theory corresponds to a massless theory --- any mass
involvement in the critical theory would break scale invariance.
We see that the critical theories correspond to massless, renormalized
theories.

We now have a definite link between field theories and the statistical
mechanics of critical phenomena. There is one final missing link in
our puzzle --- how to relate the theory to physical systems. This is
not as complex as may first be thought. For most simple systems it is
possible to find local observables whose values depend upon the phase
that they are in. For example, in the above the spin played the role
of the order parameter, differentiating between the phase with a
spontaneous magnetization and the one without. The relation of a
physical theory and a theoretical analysis boils down to determining
what are the order parameters and what are their internal
symmetries. For example, the order parameter for the helium superfluid
transition is complex having a continuous symmetry corresponding to
multiplication by a phase. This can be shown to be equivalent to a
theory of two real scalar fields. Therefore, we may use a field theory
consisting of a real scalar field with a global $O(2)$ symmetry.

We will primarily concentrate on systems described by $O(N)$ symmetric
scalar field theories with the following action
\begin{equation}
S=\int d^Dx \left\{ \frac{1}{2} \left (\partial_{\mu}
\mathbf{\Phi^a(x)} \right )^2 + \frac{m^2}{2} \mathbf{\Phi^a(x) \Phi^a(x)}
+\frac{g}{4 !} \left({\mathbf{\Phi^a(x) \Phi^a(x)}} \right)^2 \right
\}
\end{equation}
As usual the index a refers to an internal symmetry under the $O(N)$
group.  These models correspond to different physical systems
according to the value of $N$ (see \cite{b:zinn} for details on how to
derive these),
\begin{description}
\item[N=0] critical behaviour of polymers. This is only formally
defined in the limit $N \rightarrow 0$, as first noted by de Gennes
\cite{a:deGennes}. 
\item[N=1] the liquid-vapour phase transition \cite{a:Weigel}, the
alloy order-disorder transition, uniaxial ferromagnets. We see that
here we have no internal symmetries involved. eg the liquid-vapour
transition can be modelled by particles living on a lattice: allowing
the occupation of each lattice site to be either be 0 or 1 leads to a
link with the Ising model~\cite{a:YangLee}.
\item[N=2] $\mathit{He}^2$ superfluid phase transition, planar ferromagnets.
The first example was discussed above. The planar ferro-magnet will
clearly have a symmetry in a plane. This then leads to the choice of a
two component field with an $O(2)$ symmetry between the two
components.
\item[N=3] ferro magnetic phase transitions. Now we have a true three
dimensional symmetry so a three field order parameter is chosen, with
an $O(3)$ symmetry between the components.
\item[N=4] It has been postulated that this corresponds to the chiral
phase transition for two flavours of quarks \cite{a:wilzcek}.
\end{description}

The fact that one model for a particular value of $N$ can descibe
several physical systems is the theoretical basis of universality. It
should also be noted that it doesn't matter what the initial value of
any coupling is, or the lattice spacing or even upon the shape of the
lattice: once we have adjusted the temperature (or pressure, density
etc,..) so that we are on the critical surface, we know that
renormalization group transformations will take us into a fixed point,
which will yield the same critical behaviour. In fact we can take this
further and ask why do different materials have the same critical
behaviour. eg in critical binary fluids, which correspond to $N=1$,
$D=3$, a mixture of aniline and cyclohexane shows the same critical
behaviour as a mixture of triethlamine and water~\cite{a:critbin}.
This is because the critical behaviour is independent of the
underlying structure of the physical system. At distances of order of
the lattice spacing the detailed fine structure, such as the lattice
spacing and the lattice structure, will be important. However when we
use the renormalization group to decrease the amount of
'magnification' these details are washed out, as the details at the
scale of the lattice will be averaged out \cite{a:wilsciam}.

\subsection{Other methods of investigating critical behaviour.}

Before going any further we briefly consider other methods of
investigating critical phenomena and calculating exponents.

Original predictions of critical exponents were made using \emph{mean
field theory} in which a saddle point expansion is made about the
classical minimum of the field equations. Exact predictions for the
exponents are made. In fact for the $O(N)$ models descibed above the
predictions are independent of the value of $N$. The main problems
occur when we try to go beyond the tree level expansion. Below four
dimensions infra-red divergences plague an expansion making it useless
\cite{b:zinn,b:amit}.

To go beyond mean field theory Wilson and Fisher developed what is
known as the \emph{$\varepsilon$ expansion}. The dimension of space is
taken as $4-\varepsilon$ and a double expansion is made in the
coupling constant $g$ of the theory and $\varepsilon$. Thus, for
example, a $n$-point function may be expanded as
\begin{equation}
\Gamma^n(p_1,\ldots,p_n)=\sum_{r s} \Gamma_{r s}^{(n)}g^r \varepsilon^s
\end{equation}

 This coupled with a renormalization group analysis has produced some
spectacular results.  This method avoids the infra-red divergences by
moving from a gaussian fixed point to the non-trivial, infra-red
stable Wilson-Fisher fixed point \cite{b:zinn}. The $\varepsilon$
expansion has enjoyed considerable success and has made accurate
predictions of the exponents. However it does have some drawbacks. For
example the series it produces is only asymptotic and requires Borel
resummation \cite{b:zinn}.

The $N$ vector model has also been studied on lattices with nearest
neighbour interactions. The results for the exponents come from
analysis using high temperature series by different types of ratio
method, Pad\'e approximants or differential approximants. For large
$N$, the $1/N$ expansion has also been used. However the large $N$
models are unphysical and are only studied out of theoretical interest
(the main one being that the $N=\infty$ model is exactly solvable).

There are various other methods that have been used to calculate
critical exponents which deserve a mention. The real space
renormalization group method~\cite{b:realspace} consists of
considering spins on a lattice and performing blocking transformations
to reduce the number of degrees of freedom, and then truncating the
resulting expression at a certain number of operators.  Similar in
style are Monte-Carlo renormalization group
calculations~\cite{a:mcrg1,a:mcrg2,b:mcrg3}.  Here the path integral
is expanded in a certain set of operators and a blocking
transformation performed. A Monte-Carlo calculation is then performed
to calculate what values the couplings associated with these operators
should have afterwards.  Finally there is perturbation in fixed
dimension~\cite{b:zinn}.  Developed by Parisi, this consists of
calculating the $\beta$ function in perturbation theory at a fixed
dimension. The resulting expression can then be re-summed to provide
information about the zero's of the $\beta$ function and hence
information about the critical exponents.

\chapter{The Results at Leading Order}

In the introduction we outlined why an approximation scheme based upon
a derivative expansion of the renormalization group equations is
sensible approach to analytic  non-perturbative calculations.
In this chapter we will report the results of applying the derivative
expansion at the leading order in the approximation  to a theory with
a global $O(N)$ symmetry, and 
calculate the critical exponents $\eta$, $\nu$ and $\omega$.

\section{Deriving the Leading Order Equation}
Our starting
point will be the flow equation for  the Legendre effective action,
\begin{equation}
\frac{\partial \Gamma_{\Lambda}[\phi]}{\partial \Lambda} =
  \frac{1}{2} tr \left[\frac{\partial C^{-1}}{\partial
\Lambda}\;_.\left(C^{-1} \delta^{a b} +\frac{\delta^2
\Gamma_{\Lambda}}{\delta \phi^a 
\delta \phi^b}\right)^{-1}\right]
\label{e:Legrge:indices}
\end{equation}
where we have dropped the superscript $c$ on the $\phi$'s, denoting
the classical field,  
and re-introduced the internal symmetry indices.  The trace represents
a sum over the internal spin indices and an integration over
momentum/position space. As in \cite{a:timappr} it is convenient to
write the trace as an integral over momentum space and factor out the
$D$-dimensional solid angle:
\begin{equation}
\frac{\partial \Gamma_{\Lambda}[\phi]}{\partial \Lambda} =
- \frac{\Omega}{2} \mathrm{tr} \int^{\infty}_0 dq \, \frac{q^{D-1}}{C(q,\Lambda)}
\frac{\partial C(q,\Lambda)}{\partial \Lambda} \left \langle \left [
\delta^{a b} + C.\frac{\delta^2 \Gamma_{\Lambda}}{\delta \phi^a
\delta \phi^b} \right ]^{-1} (\mathbf{q},\mathbf{-q}) \right \rangle 
\label{e:rgint}
\end{equation}
where $\Omega=2/[\Gamma(D/2)(4\pi)^{D/2}]$ is the solid angle of a
$D-1$-dimensional sphere divided by $(2 \pi)^D$, the brackets $\langle
\cdots \rangle$ represent an average over all directions of the
momentum $\mathbf{q}$, and the trace now represents only the trace over
the spin indices, $a$ and $b$.

It is not yet clear how the anomalous dimension will be determined. At
a fixed point we know the field $\phi$ scales anomalously as 
$\phi \sim \Lambda^{D_{\phi}}$, where $D_{\phi} = \frac{1}{2}
(D-2+\eta)$ and  $\eta$ is the anomalous scaling dimension. Considering the
defining expression for $\Gamma_{\Lambda}$, $\Gamma_{\Lambda}[\phi]
= - \frac{1}{2} \phi.C^{-1}.\phi - W_{\Lambda}[J] + J_.\phi$,
dimensional analysis 
then shows that  we require $C$ to behave as follows,
\begin{equation} 
C(q,\Lambda) \rightarrow \Lambda^{\eta -2 } \tilde{C}(q^2/\Lambda^2)
\end{equation}
for some $\tilde{C}$, if we are to have
$\Gamma_{\Lambda}$ independent of $\Lambda$ as we approach the fixed
point. From now on we will write $C(q,\Lambda)$ as $\Lambda^{\eta -2 }
\tilde{C}(q^2/\Lambda^2)$ and drop the tilde on the scaled $C$. As we are only interested in the
behaviour near or at fixed points this is a sensible definition and we can take $\eta$ to be a
constant. This
is another reason for choosing an additive cutoff. A 
multiplicative cutoff would not scale anomalously, as is required if
we wish to reach non-Gaussian fixed points \cite{a:timderiv}.
From this discussion we  see that it will be particularly convenient to
re-write the equations in terms of dimensionless variables,
\begin{eqnarray}
\mathbf{q} & \rightarrow & \Lambda \mathbf{q} \nonumber \\
\phi(\Lambda \mathbf{q}) & \rightarrow & \Lambda^{D-D_{\phi}} \phi(\mathbf{q})
\label{e:rescaling}
\end{eqnarray}
We will also re-write the equations in terms of $t=\log (\Lambda_0 /
\Lambda)$ and rescale the fields and the effective action to absorb
the factor of $\Omega/2$ in (\ref{e:rgint}),
\begin{eqnarray*}
\Gamma & \rightarrow & \frac{\Omega}{2 \zeta} \Gamma \\
\phi^a & \rightarrow & \sqrt{\frac{\Omega}{2 \zeta}} \phi^a
\end{eqnarray*}
where we have dropped the subscript on $\Gamma$, and $\zeta$ is a
normalization factor to be chosen for later convenience. Upon doing
this we get,
\begin{eqnarray}
\lefteqn{(\frac{\partial}{\partial t} +D_{\phi} \Delta_{\phi}
+\Delta_{\partial} -D) \Gamma[\phi] =}  \label{e:rgscaled}  \\
& & -\zeta \, \mathrm{tr}\int^{\infty}_0 dq \, q^{D-1} \left (\frac{q}{C(q^2)}
\frac{\partial C(q^2)}{\partial q} + 2 - \eta \right )   
 \left \langle
\left [  \delta^{a b} + C_.\frac{\delta^2 \Gamma}{\delta \phi^a \delta
\phi^b} \right ]^{-1} (\mathbf{q},-\mathbf{q}) \right \rangle \nonumber
\end{eqnarray}

In the above  $\Delta_{\phi} = \phi.\frac{\delta}{\delta \phi}$ is the
field counting operator: it counts the number of occurrences of the
field $\phi$ in a given vertex, and arises due to the scaling of the
field in (\ref{e:rescaling}). $\Delta_{\partial}$ is the momentum
counting operator plus the dimension of space $D$ and arises through
the rescaling of the momenta in equation (\ref{e:rescaling}). It can
be represented as
\begin{equation}
\Delta_{\partial} = D + \int \frac{d^D p}{(2  \pi)^D}
\phi(\mathbf{p}) p^{\mu} \frac{\partial}{\partial p^{\mu}}
\frac{\delta}{\delta \phi(\mathbf{p})}
\end{equation}
Operating on a given vertex it counts the total number of derivatives
acting on the fields $\phi$.

Equation (\ref{e:rgscaled}) will be the starting point for all the work
from now onwards. Notice that for the first time $\eta$ explicitly
appears in the equation.

We write  $\Gamma[\phi]$ as the space-time integral of an effective Lagrangian expanded in
powers of derivatives,
\begin{equation}
\Gamma[\phi] = \int d^Dx \left \{ V(\phi^2,t) + \frac{1}{2} K(\phi^2,t)
\left (\partial_{\mu} 
\phi^a \right)^2  + \frac{1}{2} Z(\phi^2,t) \left ( \phi^a \partial_{\mu}
\phi^a \right)^2 + \cdots \right \}
\label{e:gammalo}
\end{equation}
Each linearly independent (under integration by parts) combination
of differentiated fields will 
carry its own  general (t-dependent) coefficient. The global $O(N)$
symmetry forces us to choose the coefficient 
functions to be functions of $\phi^2$.
We will require that the fixed point solutions for $V$,$K$,$Z$ etc will be non-singular for all
$\phi^2$, that perturbations about these solutions grow no faster
than a power, and that $K(0) \ne 0$.

The rescaling symmetry is made explicit by choosing the following
(non-physical) scaling 
dimensions, as follows from (\ref{e:gammalo}) and the definition of $\Gamma$.
\begin{eqnarray}
 [q^{\mu}]   =  1 &  [\partial^{\mu}]  =  1   & [\phi^a]  =  k+D/2
\label{e:scalsym}   \\
\mbox{} [V]=D           & [K]= -2(k+1)               & [Z]  = -2(2 k + 2) - D
\nonumber 
\end{eqnarray}
where $k$ is the exponent in the cutoff function, $C(q^2)=q^{2k}$.
The expansion is performed by substituting equation (\ref{e:gammalo})
into equation (\ref{e:rgscaled}) and expanding the right hand side
up to a maximum number of derivatives (in this chapter this will be no
derivatives). The angular average in (\ref{e:rgscaled}) can be easily
computed by translating them into invariant tensors, eg $\langle
q^{\mu} q^{\nu} \rangle = q^2 \delta^{\mu \nu} /D$ etc. From now
 on we
will specialize to $D=3$.

To lowest order in the expansion we drop all the derivatives from the
right hand side of (\ref{e:rgscaled}). The coefficient functions
$K$,$Z$, etc are then determined by linear equations given by the
vanishing of the left hand side of (\ref{e:rgscaled}). Consider the
equations for $K$,
\begin{equation}
\frac{\partial}{\partial t}K(\phi^2,t) + (1+\eta) \phi^2 K'(\phi^2,t) + \eta K(\phi^2,t) = 0
\end{equation}
We see that at fixed points, where $\partial/\partial t=0$, this then predicts 
\begin{equation}
K(\phi^2)  \propto (\phi^2)^{-\eta/(1+\eta)}
\end{equation}
where $'= \frac{\delta}{\delta \phi^2}$.
To avoid $K(\phi^2)$ being singular and ensure that $K(0) \ne 0$
we see we must 
have $\eta=0$. Therefore   $K(\phi^2)$ must be a constant. Using the scaling
symmetry we can set this constant to be $1$.

If we now consider the equation for $Z$ we get
\begin{equation}
\frac{\partial}{\partial t}Z(\phi^2,t) + \phi^2 Z'(\phi^2,t) + Z(\phi^2,t)=  0
\end{equation}
Hence, we see that at fixed points that $Z$ will satisfy,
\begin{equation}
Z \propto (\phi^2)^{-1}
\end{equation}
Again to avoid non-singular behaviour  we set the constant of
proportionality to zero and get $Z(\phi^2) \equiv 0$. In general,
considering the form of a general term in the expansion we see that a
similar conclusion must hold for all other terms. A general term, $H$,  will
satisfy
\begin{equation}
\frac{\partial}{\partial t}H(\phi^2,t) +  (
\phi^2 H'(\phi^2,t) +n_{\phi} H(\phi^2,t))
+ n_{\partial} H(\phi^2,t) - 3 H(\phi^2,t) =0
\end{equation}
 where $n_{\phi}$ is the number of $\phi$'s occurring in the term
multiplying $H$, divided by two,  and $n_{\partial}$ is the number of
derivatives 
occurring in the term multiplying $H$. Hence we see that at a fixed
point, where $\eta$ must be zero, we have
\begin{equation} 
H \propto (\phi^2)^{- (n_{\phi} + n_{\partial} -3)}
\end{equation}
We see that for terms of higher order in the expansion, where
$n_{\phi} > 2$ and $ n_{\partial} > 2$, that $H$ will be singular
unless the constant of proportionality is set to zero. Therefore, at
leading order, the derivative expansion must reproduce the local
potential expansion at fixed points:
\begin{equation}
\Gamma[\phi] = \int d^Dx \, V(\phi^2) + \frac{1}{2} (\partial_{\mu} \phi^a)^2
\end{equation}

The next step is to calculate the trace of the following expression,
\begin{equation}
\left [ \delta^{a b} +C.\frac{\delta^2 \Gamma[\phi]}{\delta \phi^a
\delta \phi^b} \right ]^{-1}(\mathbf{q},\mathbf{-q}) = \left [ \delta^{a b} + 2 q^2 \left \{
\delta^{ab} V' + 2  \phi^a
\phi^b V'' + q^2 \delta^{ab} \right \} \right ]^{-1}
\label{e:inverse}
\end{equation}
Working in momentum space and integrating by parts
where necessary we have,
\begin{equation}
\frac{\delta^2 \Gamma[\phi]}{\delta \phi^a \delta \phi^b} = 2
\delta^{ab} V' +4 \phi^a \phi^b V'' + q^2 \delta^{ab}
\end{equation}

The taking of the trace over the spin indices of (\ref{e:inverse})
is easily accomplished by noting that an operator
$A \delta^{a b} + B \phi^a \phi^b$  has eigenvalues $A$ and
$A+\phi^2 B$, with the eigenvalue $A$ having multiplicity $N-1$, where
$N$ is the dimension of the internal symmetry space.  Hence, provided
that $A \ne 0$ and $A + \phi^2 B \ne 0$, then $[A
\delta^{a b } + B \phi^a \phi^b]^{-1}$ has 
eigenvalues $1/A$, with multiplicity $N-1$, and $1/(A+\phi^2 B)$.
Hence we must have $\mathrm{tr} [A\delta^{a b } + B \phi^a \phi^b]^{-1} =
(N-1)/A + 1/(A+B \phi^2$). Therefore, taking the trace of equation
(\ref{e:inverse}) gives
\begin{equation}
\frac{N-1}{1+2 V' q^2 +q^4} + \frac{1}{1+(2V' + 4 \phi^2 V'') q^2 +q^4}
\end{equation}

The leading order equation has now become
\begin{eqnarray}
\lefteqn{\frac{\partial V(\phi^2,t)}{\partial t} + \phi^2 V'(\phi^2,t) -
3 V(\phi^2,t)   =}  \\ 
& & - 4 \zeta \int dq \, q^2
\frac{N-1}{1+2 V'(\phi^2,t) q^2 +q^4} + \frac{1}{1+(2V'(\phi^2,t) + 4 \phi^2
V''(\phi^2,t)) 
q^2 +q^4} \nonumber
\end{eqnarray}
Performing the $q$-integrals, doing the trivial angular average and setting $\zeta$ to $1/2\pi$ yields
the equation at leading order,
\begin{eqnarray}
\frac{\partial V(\phi^2,t)}{\partial t} +  \phi^2 V'(\phi^2,t)
- 3 V(\phi^2,t)& = & 
 - \frac{1}{\sqrt{2 + 2 V'(\phi^2,t) + 4
\phi^2 V''(\phi^2,t)}} \nonumber \\
& & -  \frac{N-1}{\sqrt{2 + 2 V'(\phi^2,t)}}
\label{e:lo}
\end{eqnarray}
In the rest of this chapter we will report the results of the study of
this equation.

\section{Fixed Point Solutions}

We  are now ready to start the search for fixed point solutions of
(\ref{e:lo}), that is solutions with $\partial / \partial t =0$. At
first sight it may seem that (\ref{e:lo}) has infinitely many solutions
parameterized by the value of $V''(0)$. In fact this is not the case,
as only finitely many solutions do not end in a singularity~\cite{a:timhalp,a:hh}. Of course
this is sensible on physical grounds, as the fixed points correspond
to massless continuum limits (ie second order phase transitions) with the
prescribed field content. To see that we should only expect to see a discrete set of solutions we
must consider the boundary conditions supplied to (\ref{e:lo}).

The only requirement we have imposed upon $V(\phi^2)$ so far is that it is
non-singular for all $\phi^2$. Considering the form of (\ref{e:lo}),
we 
see that this implies that either $V(\phi^2)$ is trivial (the Gaussian
fixed point), or that $V(\phi^2)$ must satisfy the following for large $\phi^2$,
\begin{equation}
A_v \;(\phi^2)^3 + \frac{\sqrt{30}}{120}\frac{(\sqrt{5} (N-1) + 1)}{\phi^2
\sqrt{A_v}} +O((\phi^2)^{-2})
\label{e:loasy}
\end{equation}
for  some constant $A_v$.

If we consider the  perturbations  about this solution we see that at
large  $\phi^2$ 
the perturbation must behave as  a combination of   $\exp( \frac{1}{8}[30 A_v]^{3/2}
(\phi^2)^{3/2})$  and
$(\phi^2)^3$. The second perturbation merely alters the value of $A_v$,
whilst we disallow the first  as it is incompatible with
(\ref{e:loasy}). We see that the solution space of fixed 
point solutions, defined for all 
$\phi^2$ and satisfying (\ref{e:loasy}), divides up into an
\emph{isolated} one parameter set, 
parameterized by the value of $A_v$.

The only other possibility that makes the two sides of (\ref{e:lo})
balance is that $V(\phi^2)$ is
only defined for $\phi^2 < \phi_c^2$ and ends at a singularity, at
$\phi^2=\phi_c^2$  as follows,
\begin{equation}
\frac{1}{\phi_c^2} \left ( \frac{27}{16} \right )^{2/3} \left ( \phi_c^2 -
\phi^2 \right )^{4/3} 
\label{e:losing}
\end{equation}
where we have suppressed non-singular and lower order-singular parts.
Notice that if either $V$ or $V'$ where to diverge then it would not
be possible for both sides of  equation (\ref{e:lo}) to balance.
We will disregard these non-physical singular solutions. 

The imposition of the $O(N)$ symmetry forces $V$ to be a function of
$\phi^2$, so it is symmetric under the $Z_2$ transformation $\phi^a
\leftrightarrow   -\phi^a$. To ensure this, we need  $V(\phi^2)$ to exist at
the origin and satisfy (\ref{e:lo}) there.  Setting $\phi^2=0$ in
equation~(\ref{e:lo}) gives us our final boundary condition,
\begin{equation}
-3 V(0) = - \frac{N}{\sqrt{2 + 2 V'(0)}}
\label{e:bc0lo}
\end{equation}
We now have a second order differential equation with two boundary
conditions. Thus we expect at most a discrete set of acceptable solutions.
In fact we only find two:- the Gaussian fixed point and an
approximation to the Wilson-Fisher fixed point \cite{a:wk}.

\subsection{The Gaussian Fixed Point}

It can be quickly seen that imposing $V'(\phi^2)=0$ and $V''(\phi^2)=0$
that a trivial fixed point solution exists,
\begin{equation}
V(\phi^2) = \frac{N}{3 \sqrt{2}}
\end{equation}
This is the fixed point mentioned in the introduction.

\subsection{The Wilson-Fisher Fixed Point}

\begin{figure}[H]
\centering
\epsfig{figure=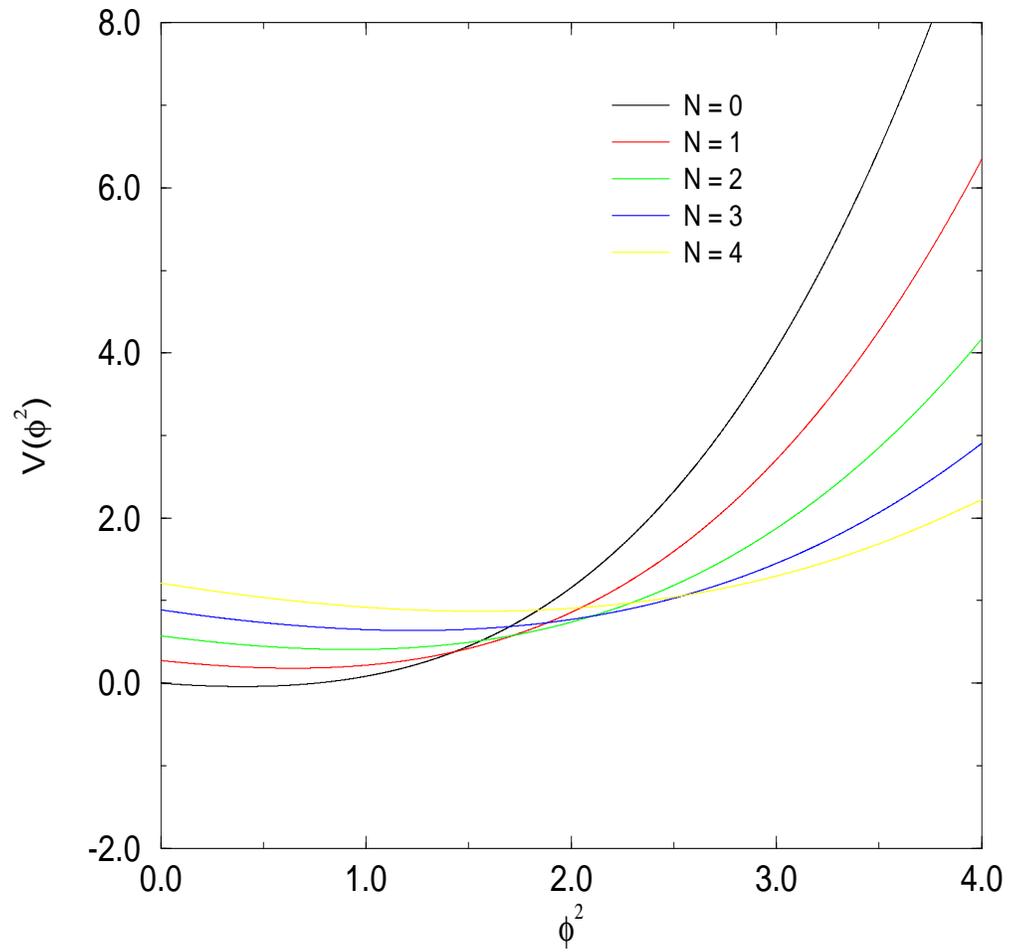,width = \textwidth,height=\textwidth}
\caption{The potential at leading order in the derivative expansion
for \mbox{$D=3$}, $O(N)$ symmetric field theory, for \mbox{$N=0,1,2,3,4$}.}
\label{lo}
\end{figure}

A more interesting case is when the solution is non-trivial.  We only
find one example of a non-trivial point, which we deem to be an
approximation to the Wilson-Fisher fixed point. To find this fixed
point we need to rely on numerical methods as outlined in appendix~A.
 We will impose (\ref{e:bc0lo}) as a condition at the origin and
force non-trivial behaviour by imposing (\ref{e:loasy}) as $\phi^2
\rightarrow \infty$. The results of these calculations are shown in
figure (\ref{lo}) for various values of $N$.

\newpage

\section{Critical Exponents at Leading Order}

The next step in the program is to calculate the critical exponents
for $\nu$~and~$\omega$ 
corresponding to the fixed points. 
To calculate the critical exponents it is necessary to linearize about
this fixed point potential, $V^*(\phi^2)$. We will write $V(\phi^2,t) = V^*(\phi^2) + \delta 
V(\phi^2)$, where $\delta V(\phi^2,t)$ is given by $ \varepsilon
\mathrm{e}^{\lambda t} v(\phi^2)$, with $\varepsilon \ll 1$, and expand
to first order in $\varepsilon$. Upon doing
this we have
\begin{equation}
(\lambda -3 ) v(\phi^2) + v'(\phi^2) \phi^2 =
\frac{N-1}{(2+2V'(\phi^2))^{3/2}} v'(\phi^2) + \frac{v'(\phi^2) + 2
\phi^2 v''(\phi^2)}{ (2+ 2V'(\phi^2) + 4 \phi^2 V''(\phi^2))^{3/2}}
\label{e:loolin}
\end{equation}

\subsection{Critical Exponents for the Gaussian Fixed Point}
\label{gaussian} 
For the Gaussian fixed point we will expect to find  perturbations
with eigenvalues given by the classical scaling dimension of the
coupling (cf
Introduction), eg an operator corresponding to the mass  with
dimension 2, an operator corresponding to the $\phi^4$ interaction
with dimension $1$ etc. In the case of the Gaussian fixed point equation
(\ref{e:loolin}) becomes,
\begin{equation}
(\lambda -3 ) v(\phi^2) + v'(\phi^2) \phi^2 = \frac{N v'(\phi^2)}{2
\sqrt{2}} + \frac{2 \phi^2 v''(\phi^2)}{2 \sqrt{2}}
\end{equation}
The general solution to this equation is 
\begin{equation}
A _{[\lambda-3]}F(\sqrt{2} \phi^2)_{[N/2]} + B 2^{-\lambda/2 +3/2}
(\phi^2)^{3-\lambda}_{[\lambda-3,-2+\lambda - 
N/2]}F(-\frac{1}{\sqrt{2} \phi^2})_{[]}
\end{equation}
where $ _{[a_1,a_2,\ldots,a_i]}F(x)_{[b_1,b_2,\ldots,b_j]}$ represents
Barnes' extended hypergeometric function,defined by,
\begin{equation}
_{[a_1,\ldots,a_i]}F(x)_{[b_1,\ldots,b_j]}= \sum_{k=0}^{\infty} \frac
{\left ( \prod_{m=1}^i  \frac{\Gamma(a_m+k)}{\Gamma(a_m)} \right )}
{\left ( \prod_{n=1}^j \frac{\Gamma(b_n +k)}{\Gamma(b_n)} \right )} x^k
\end{equation}
If any of the  $a_i= - k$, for $k$ a non-positive integer, then the
series stops after $k$ terms. 
If we wish the solutions to
be bounded by polynomials then we are forced to set the coefficient
$B$ in the above to zero and chose $\lambda-3=k$, for some negative  integer
$k$. This then forces $\lambda=2,1,0,-1,\ldots$, and we have  the
expected results. We  see that the at the Gaussian fixed
point the scaling dimensions are given by the canonical dimensions
\cite{a:hh}, that $\eta=0$ and $\nu=1/2$.

\subsection{Critical Exponents at the Wilson-Fisher Fixed Point}

To calculate the exponents at the non-trivial fixed point we will
need to think about the boundary conditions. 
By linearity of the perturbation we can choose $v(0)=1$. We will again
ensure that the perturbation exists at the origin by imposing the
equation as a boundary condition at $\phi^2=0$. That is,
\begin{equation}
(\lambda -3 ) v(0)  =
\frac{N}{(2+2V'(0))^{3/2}} v'(0)
\label{e:lolinbc1}
\end{equation}

The case $N=0$ is treated slightly differently. If we were to impose
(\ref{e:lolinbc1}) then we either have $\lambda=3$ or $v(0)=0$. We
discard the case of $\lambda=3$ as this corresponds to the
uninteresting case of the vacuum energy operator. We must therefore
have $v(0)=0$. The other boundary condition again comes from
linearity, which we take to be $v'(0)=1$.

We can also
consider the solutions for large $\phi^2$. This time we will see
that the perturbation will be a linear combination of $(\phi^2)^{3 -
\lambda}$ and $\exp( \frac{1}{8}[30 Av]^{3/2} (\phi^2)^{3/2})$.
Once more we  will
enforce a coefficient of zero on the latter perturbation, as we
require the perturbations to grow no 
faster than a power in $\phi^2$. The imposition of
this condition will  ensure that the solutions satisfying 
$(\phi^2)^{3  -\lambda}$ as $\phi \rightarrow \infty$ form an isolated
one parameter set.

Upon imposing these two conditions we will have a second order
eigenvalue problem  for $\lambda$ with three  boundary conditions.
Therefore, we  expect 
to see a  discrete number of solutions, which is indeed the  case.
We find only one positive eigenvalue, which yields $\nu$ through
$\nu=1/\lambda$. The least negative  eigenvalue yields the first
correction to scaling through $\omega=-\lambda$. The results are
summarized in table~(\ref{t:lo}).

\begin{table} [h]
\renewcommand{\arraystretch}{1.5}
\hspace*{\fill}
\begin{tabular}{|c||c|c||c|c|}     \hline
$N$
&\multicolumn{2}{c||}{$\nu$}
&\multicolumn{2}{c||}{$\omega$}
\\
\hline \hline
&
&$0.5880^a$
&
&$0.80(4)^a$
\\
0
&$0.5961$
&$0.5880(15)^b$
&$0.6175$
&$0.82(4)^b$
\\
&
&$0.592(3)^c$
&
&
\\ \hline 

&
&$0.6300(15)^a$
&
&$0.79(3)^a$
\\
1
&$0.6604$
&$0.6310(15)^{b}$
&$0.6285$
&$0.81(4)^b$
\\

&
&$0.6305(15)^{c}$
&
&
\\ \hline

&
&$0.6695(20)^{a}$
&
&$0.78(25)^a$
\\
2
&$0.7253$
&$0.671(5)^{b}$
&$0.6621$
&$0.80(4)^a$
\\

&
&$0.672(7)^{c}$
&
&
\\ \hline

&
&$0.705(3)^{a}$
&
&$0.78(2)^a$

\\
3
&$0.7811$
&$0.710(7)^{b}$
&0.7068
&$0.79(4)^b$
\\

&
&$0.715(20)^{c}$
&
&
\\ \hline
4

&$0.8240$
&$0.7479(90)^e$
&$0.7510$
&
\\  \hline
10
&$0.9380$
&$0.877^{d}$
&$0.8910$
&$0.78^f$
\\ \hline
20
&$0.9577$
&$0.942^{d}$
&$0.9469$
&$0.89^f$
\\ \hline
100
&$0.9938$
&$0.989^{d}$
&$0.9911$
&$0.98^f$
\\ \hline
\end{tabular}
\hspace*{\fill}
\renewcommand{\arraystretch}{1}
\caption[Critical exponents at leading order in the derivative expansion]
{
Critical exponents of the three-dimensional theory for various values of $N$.
For comparison we list results obtained with other methods: \\
a) From summed perturbation series in fixed dimension 3 at six-loop
order \cite{b:zinn}. \\
b) From the $\varepsilon$-expansion at order $\varepsilon^5$ \cite{b:zinn}. \\
c) From lattice calculations \cite{b:zinn}. \\
d) From the $1/N$-expansion at order $1/N^2$ \cite{a:nu:1overN2}.\\
e) From a recent lattice study \cite{a:KanN=4}.\\
f) From the $1/N$ expansion at order $1/N$ \cite{a:om:1overN}.}
\label{t:lo}
\end{table}

We see that there is  quite an impressive agreement between the
results gained from other methods and those gained at this simplistic
level of approximation. At first there is a
gradual decrease in the accuracy of the approximation for $\nu$  and a
slight improvement in $\omega$, as $N$ increases. This
has been noticed before by several authors and should 
therefore not be entirely surprising~\cite{a:alford}.  The results are
better in the large $N$ regime, but this should be expected as
the local potential approximation effectively becomes exact at
$N=\infty$. We will 
discuss the special case of $N=\infty$ in a later chapter. The results
compare well with those obtained by others using different forms of
cutoff  \cite{a:t+wett,a:wet2,a:wet3,a:ballrubbish} and those obtained
using a sharp cutoff \cite{a:hh,a:alford,a:timtrunc}.
Using a different form of smooth cutoff will not generally preserve
a re-parameterization invariance of the equations  and so  other
authors are forced to  
appeal to some 'heuristic' argument to set a value for $\eta$.

\begin{figure}[p]
\epsfig{figure=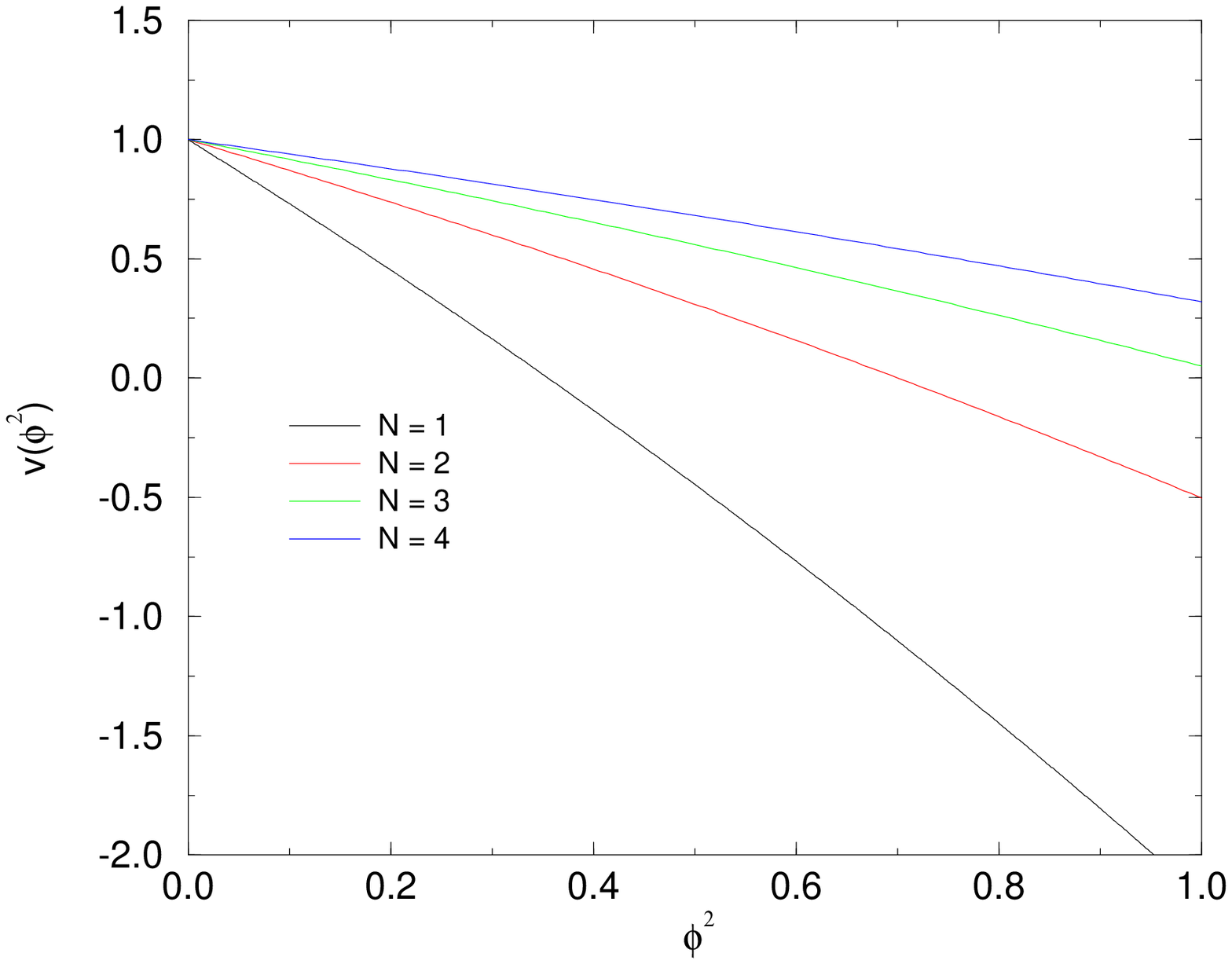,width=\textwidth,height=\textwidth}
\caption{The perturbation corresponding to $\nu$ at leading order in
the derivative expansion, 
for three dimensional $O(N)$ symmetric field theory, for $N=1,2,3,4$.}
\label{lonu}
\end{figure}

\begin{figure}[p]
\epsfig{figure=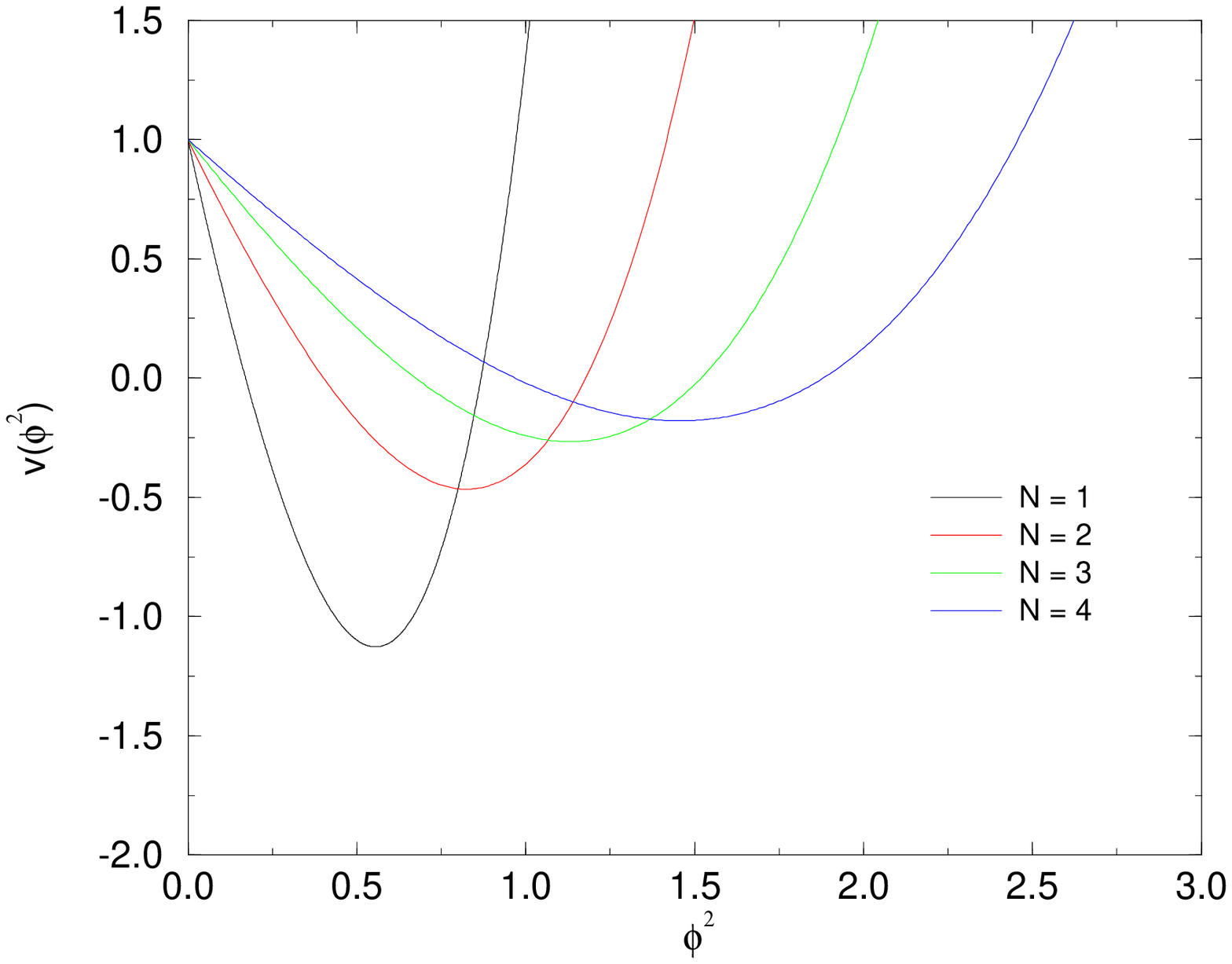,width=\textwidth,height=\textwidth}
\caption{The perturbation corresponding to $\omega$ leading order in
the derivative expansion 
for three dimensional $O(N)$ symmetric field theory, for $N=1,2,3,4$.}
\label{loom}
\end{figure}

\begin{figure}[p]
\epsfig{figure=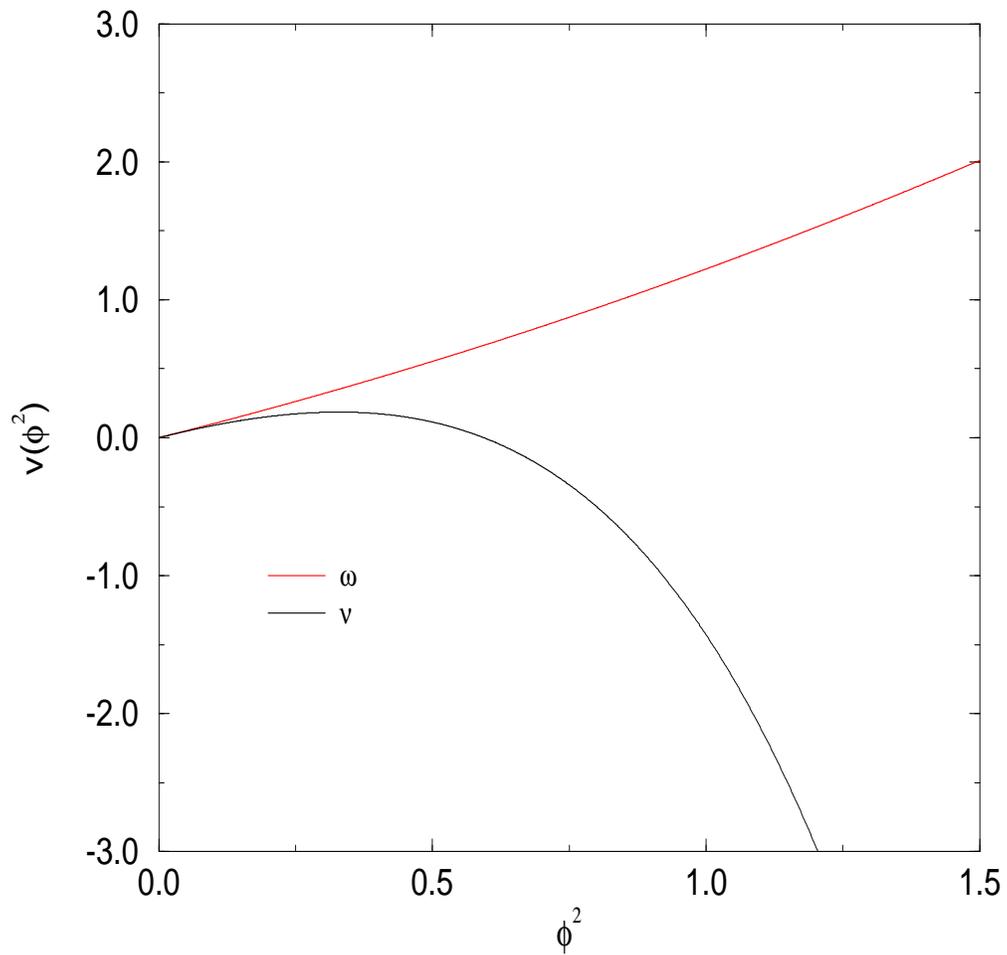,width=\textwidth,height=\textwidth}
\caption{The perturbations corresponding to $\nu$ and  $\omega$ at leading order in
the derivative expansion 
for three dimensional $O(N)$ symmetric field theory, for $N=0$. The
normalization condition is now given by $v'(0)=1$.}
\label{lon0}
\end{figure}

%\begin{figure}[h]
%\centering
%\subfigure[$\nu$]{\epsfig{figure=nu.ps,height= 0.4 \textheight}}
%\subfigure[$\omega$]{\epsfig{figure=om.ps,height= 0.4  \textheight}} 
%\caption{The solutions for the perturbations corresponding to $\nu$
%and  $\omega$ at leading order in 
%the derivative expansion 
%for three dimensional, $O(N)$ symmetric field theory, for $N=1,2,3,4$.}
%\end{figure}

\chapter{The Results at Second Order}

In the previous chapter we showed how to obtain the results coming
from the leading order of the approximation. In this chapter we will
report the results of taking the approximation to the next order and
including the effects of the terms involving two derivatives on the
right hand side of the flow equation. This will lead to a vast
increase in the complexity of the equations and the boundary
conditions supplied to them. We will again specialize to the case
$D=3$.

\section{The Second Order Equations}

Our starting point will again be equation (\ref{e:rgscaled}),
reproduced here,
\begin{eqnarray}
\lefteqn{(\frac{\partial}{\partial t} +D_{\phi} \Delta_{\phi}
+\Delta_{\partial} -D) \Gamma[\phi] =} \label{e:rgscaled2} \\ & &
 -\frac{1}{2 \pi} \mathrm{tr}\int^{\infty}_0 dq q^{D-1} \left
 (\frac{q}{C(q^2)}
\frac{\partial C(q^2)}{\partial q} + 2 - \eta \right )   
 \left \langle
\left [  \delta^{a b} + C_.\frac{\delta^2 \Gamma}{\delta \phi^a \delta
\phi^b} \right ]^{-1} (\mathbf{q},-\mathbf{q}) \right \rangle \nonumber
\end{eqnarray}
 We will again take $\Gamma[\phi]$ to be expanded as in equation
 (\ref{e:gammalo}). We will also force the coefficient functions to be
 non-singular for all $\phi^2$ at fixed points, and that the
 perturbations about these fixed points grow no faster than a power of
 $\phi^2$. The equations at the second order in the expansion are
 calculated by substituting expression (\ref{e:gammalo}) into equation
 (\ref{e:rgscaled2}) and dropping all terms with more than two
 derivatives from the right hand side. We will first consider what
 happens to the coefficient functions multiplying terms with more than
 two derivatives in them. We see that a general term $H(\phi^2,t)$
 will satisfy,
\begin{equation}
\frac{\partial}{\partial t} H(\phi^2,t) + (1+\eta)(\phi^2 H'(\phi^2,t) + n_{\phi} H(\phi^2,t) ) +
n_{\partial} H(\phi^2,t) - 3 H(\phi^2,t)=0
\end{equation}
where $n_{\phi}$ denotes the number of $\phi^a$'s in the term
multiplying $H$ divided by 2 and $n_{\partial}$ is the number of
derivatives in the term multiplying $H$. We can no longer force that
$\eta$ be zero at fixed points, and this leads to the following
expression for $H$ at fixed points (setting D=3),
\begin{equation}
H \propto \phi^{-\frac{(1+\eta)n_{\phi} + n_\partial -3}{1+\eta}}
\end{equation}
Hence,  as $n_{\phi}>2$ and $n_{\partial}>2$, we see that $H$ will
be singular at the origin, under mild assumptions about $\eta$, eg
$\eta>0$. We are therefore forced to set all the terms with more than
two derivatives in them equal to zero. Hence at the second order of
the derivative expansion we parameterize $\Gamma$ as follows,
\begin{equation}
\Gamma[\phi^2] = \int d^D x \left \{ V(\phi^2,t) + \frac{1}{2} K(\phi^2,t) \left ( \partial_{\mu}
\phi^a \right )^2 + \frac{1}{2} Z(\phi^2,t) \left ( \phi^a \partial_{\mu} \phi^a \right)^2 \right \}
\end{equation}
A little bit of thought shows that this conclusion holds at all orders
in the approximation: if we substitute an expression into the equation
that is of higher order than the order we are working at, then we are
forced to set the higher order terms to zero~\cite{a:timderiv}.

The next step is to compute the inverse in equation
(\ref{e:rgscaled2}). This is not as straightforward as in the leading
order case and a more involved technique is required. The steps that
are required will depend upon the value of $N$: the case of $N=1$ is
slightly different than the more general case. We discuss the more
general case first before briefly reporting the results at $N=1$.

We will regard $[\delta^{a b}+C.\delta^2
\Gamma / \delta \phi^a \delta \phi^b]^{-1}$ as a differential operator:
\begin{eqnarray}
\left [\delta^{ab}+C.\frac{\delta^2 \Gamma}{\delta \phi^a \phi^b}
\right ]^{-1}\!\!\!(\mathbf{q},\mathbf{-q}) & = & \int d^Dx \, d^Dy \,
\mathrm{e}^{-i \mathbf{q.x}} \left[ 
\delta^{ab} + C.\frac{\delta^2 \Gamma}{\delta \phi^a \delta \phi^b} \right ]^{-1}\!\!\!(\mathbf{x},\mathbf{y}) \, \mathrm{e}^{i \mathbf{q.y}} \nonumber \\
& \equiv & \int d^Dx \, Q ^{ab}\\ \mbox{where} \ \ \ \ \ \ \ \ \ \ \ \
 \ \ Q^{ab} & = & \mathrm{e}^{-i \mathbf{q.x}} \left[
\delta^{a b} + C.\frac{\delta^2 \Gamma}{\delta \phi^a \delta \phi^b} \right
]^{-1}\!\!\!\mathrm{e}^{i \mathbf{q.x}}
\label{e:Q}
\end{eqnarray}
$Q^{ab}$ is a function of $\mathbf{q}$ and $\phi(\mathbf{x})$ and its
derivatives evaluated at $\mathbf{x}$.  To calculate $Q^{ab}$ we will
regard $C$ and $\frac{\delta^2 \Gamma}{\delta \phi^a \delta
\phi^b}$ as differential operators. Noting that we are performing a derivative expansion,  we should 
expect $Q^{ab} \approx (\delta^{ab} + \nu^{ab})^{-1} + \cdots$, where
$\nu^{ab}$ is the expression obtained by dropping all terms containing
differentials of $\phi^a$ from $C(q^2)
\mathrm{e}^{-i \mathbf{q.x}} \frac{\delta^2 \Gamma}{\delta \phi^a \delta \phi^b} \mathrm{e}^{i
\mathbf{q.x}}$.
Hence noting that
\begin{equation}
Q^{ac} \left ( \delta^{cb} + \nu^{cb} \right ) =\left
(\delta^{ac}+\nu^{ac} \right) \left [(\delta^{cb} +\nu^{cd})^{-1} Q^{d
b} + (\delta^{cd} +\nu^{cd})^{-1} \nu^{de} Q^{eb} \right ]
\label{e:Q1}
\end{equation}
and that from (\ref{e:Q}) we have,
\begin{equation}
Q^{ab} = \delta^{ab} - \mathrm{e}^{-i \mathbf{q.x}} C(-\Box)
\frac{\delta^2 \Gamma}{\delta
\phi^a \delta \phi^c} \mathrm{e}^{i \mathbf{q.x}} Q^{cb}
\label{e:Q2}
\end{equation}
we must have $Q^{ab}$ satisfying the following expression,
\begin{equation}
Q^{ab} = (\delta^{ab} + \nu^{ab})^{-1} + (\delta^{ac} + \nu^{ac})^{-1}
\left \{
\nu^{cd}Q^{cb} - \mathrm{e}^{-i \mathbf{q.x}} C(-\Box) \frac{\delta^2 \Gamma}{\delta \phi^c
\delta \phi^d} \mathrm{e}^{i \mathbf{q.x}} Q^{d b} \right \}
\label{e:Qit}
\end{equation}
The derivative expansion is performed by iterating expression
(\ref{e:Qit}) to the required order, here up to two derivatives,
starting with $Q^{ab}=1/(\delta^{ab}+ \nu^{ab})$, and remembering that
$C(-\Box)=-\Box$. To do this we must calculate $\frac{\delta^2
\Gamma}{\delta \phi^a \delta \phi^b}$, which is found, after a long
but straightforward calculation, to be,
\newcommand{\fa}{\phi^a}
\newcommand{\fb}{\phi^b}
\newcommand{\p}{\partial_{\mu}}
\newcommand{\dd}{\delta^{ab}}
\newcommand{\nn}{\nonumber}
\begin{eqnarray}
\frac{\delta^2 \Gamma}{\delta \phi^a \delta \phi^b} &= & 4 \fa \fb V'  + 2 \dd V' \nn \\ 
&&- \dd K \Box - 2 (\Box \fa) \fb K' \nn \\ && - 2 \dd (\phi^c \p
\phi^c)^2 K' \p - 2 (\p \fa) (\p \fb) K' \nn \\ && - 2 (\p \fa) \fb K'
\p -4 (\p \fa) \fb (\phi^c \p \phi^c)^2 K' \nn \\ && + 2 \fa (\p \fb)
K' \p + \dd (\p \phi^c)^2 K' \nn \\ && + 2 \fa \fb (\p \phi^c)^2 K'
\nn \\ && - \dd (\p \phi^c)^2 Z - 2 \fa (\p \fb) Z \p \nn \\ && - 2
\fa \fb (\p \phi^c)^2 Z' - \dd (\phi^c \Box \phi^c) Z \nn \\ && - \fa
(\Box \fb) Z - \fa \fb Z \Box \nn \\ && - 2 \fa \fb (\phi^c \Box
\phi^c) Z' - \dd (\phi^c \p \phi^c)^2 Z' \nn \\ && - 2 \fa (\p \fb)
(\phi^c \p \phi^c)^2 Z' -2 \fa \fb (\phi^c \p \phi^c) Z' \p \nn \\ &&
- 2 \fa \fb (\phi^c \p \phi^c)^2 Z'
\label{e:Gpp}
\end{eqnarray} 

This expression is then used in (\ref{e:Qit}) to iterate $Q^{ab}$ to
two derivatives. This is an extremely long calculation and was
performed using the symbolic manipulation package Form
\cite{a:Form}. The only remaining step is  
to calculate the trace over the spin indices. This will require the
calculation of $(\dd +
\nu^{ab})^{-1}$, and the derivative of its inverses. This is easily done by noting that if $\dd +
\nu^{ab} = A
+ B \fa \fb$ then,
\begin{equation}
 (\dd + \nu^{ab})^{-1} = \frac{\dd}{A} - \frac{B}{A ( A + B \phi^2)}
\fa \fb
\end{equation}
and also by noting that if we have a matrix $A(x)$ then
$\frac{dA^{-1}}{dx} = - A^{-1}
\frac{dA}{dx} A^{-1}$. 
\newpage

This calculation then yields the following equation for $V$, near
fixed points,
\begin{eqnarray}
\lefteqn{\frac{\partial }{\partial t}V + (1 + \eta) \phi^2 V' - 3 V  =
 - \left ( 1 - \eta/4 \right ) \times } \\ & & \left [
\frac{N-1}{\sqrt{K} \left ( 2 V' + 2 \sqrt{K} \right )^{1/2}} +
\frac{1}{\sqrt{K + \phi^2 Z} \left (2 V' + 4 \phi^2 V'' +2
\sqrt{K + \phi^2 Z} \right )^{1/2}}  \right ] \nn 
\end{eqnarray}
we refer to the above equation as the $V$ equation. Similarly the
equations coming from the $(\partial_{\mu} \phi^a)^2$ and $(\phi^a
\partial_{\mu} \phi^a)^2$  parts of the action will be referred to as
the $K$ and $Z$ equations respectively.  The equations for $K$ and $Z$
are  a lot longer and are relegated to an appendix.  We will
perform one further transformation on the equations. To enable an
easier comparison with the $N=\infty$ equation we will scale the
equations as follows,
\begin{eqnarray}
\phi & = & N \tilde{\phi} \\
V & = & N \tilde{V} \\ K &=& \tilde{K} \\ Z &=& \tilde{Z}/N
\end{eqnarray}
This then yields the following equation for $V$,
\begin{eqnarray}
\lefteqn{\frac{\partial }{\partial t}V + (1 + \eta) \phi^2 V' - 3 V =
  - \left ( 1 - \eta/4 \right ) \times } \\ && \!\!\!\! \left [
\frac{1}{N \sqrt{K + \phi^2 Z} \left( 2 V' + 4 \phi^2 V'' +2
\sqrt{
K + \phi^2 Z} \right )^{1/2}} + \frac{1-1/N}{\sqrt{K} \left (2 V'
 + 2 \sqrt{K} \right )^{1/2}} \right ] \nn
\end{eqnarray}
where we have dropped the tildes on $V$, $K$ and $Z$.

\section{ The  Fixed Points Solutions}

To find the fixed points we will again need to consider the boundary
conditions supplied to the equations. We will again ensure that the
solutions exist at the origin by imposing the equation at the
origin. This will provide three conditions. We can also perform the
asymptotic analysis as in the previous chapter. Once more we find that
either the solution is the trivial Gaussian fixed point
($\eta=0,V=N/3\sqrt{2},K=1,Z=0$), that the solutions are only defined
for $\phi^2<\phi_c^2$, or that the solutions behave as follows as
$\phi^2 \rightarrow \infty$,
\begin{eqnarray}
V(\phi^2) & \sim & \mathit{A_v} \ (\phi^2)^{\frac{3}{1+\eta}} + \cdots
\label{e:2asy1}\\ K(\phi^2) & \sim & \mathit{A_k} \
(\phi^2)^{\frac{-\eta}{1+\eta}} + \cdots \\ Z(\phi^2) & \sim &
\mathit{A_z} \ (\phi^2)^{\frac{-(1+ 2 \eta)}{1+ \eta}} + \cdots
\label{e:2asy2}
\end{eqnarray}
We will be most interested in the non-trivial solutions defined for
all $\phi^2$.  We can also study the perturbations about these
solutions. Forcing  the perturbations to grow no faster than a
power, we see that the solutions
satisfying~(\ref{e:2asy1})~to~(\ref{e:2asy2}) will form a discrete
set, with the solutions parameterized by the values of $\mathit{A_v}$,
$\mathit{A_k}$ and $\mathit{A_z}$.

\begin{figure}[H]
\epsfig{figure=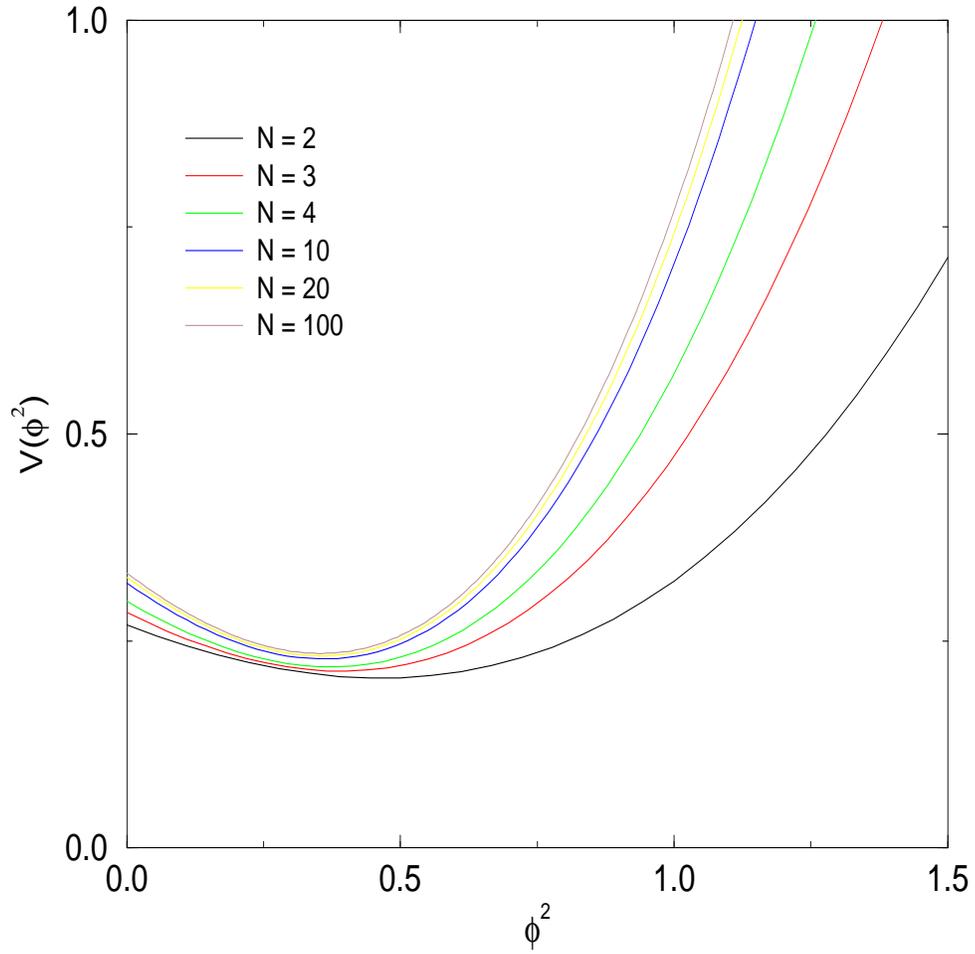,width=\textwidth,height=\textwidth}
\caption{The Legendre effective potential at second order in 
the derivative expansion, for three dimensional $O(N)$ symmetric field
theory, for $N=2, 3, 4, 10, 20, 100$.}
\label{V2nd}
\end{figure}

\begin{figure}[H]
\epsfig{figure=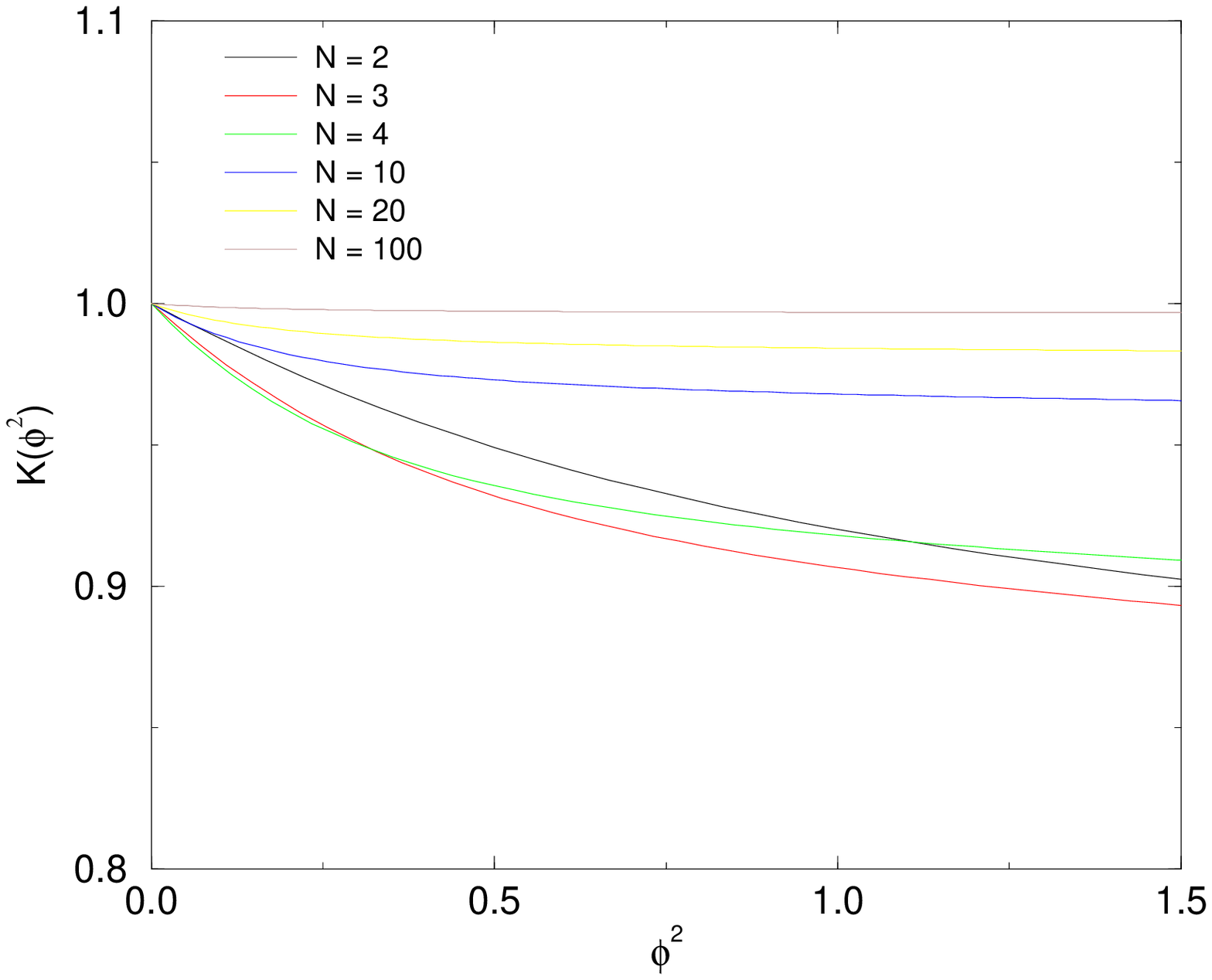,width=\textwidth,height=\textwidth}
\caption{The K component of the wave function renormalization of the Legendre effective action
at second order in the derivative expansion, for three dimensional
$O(N)$ symmetric field theory, for $N=2, 3, 4, 10, 20, 100$.}
\label{K2nd}
\end{figure}

\begin{figure}[H]
\epsfig{figure=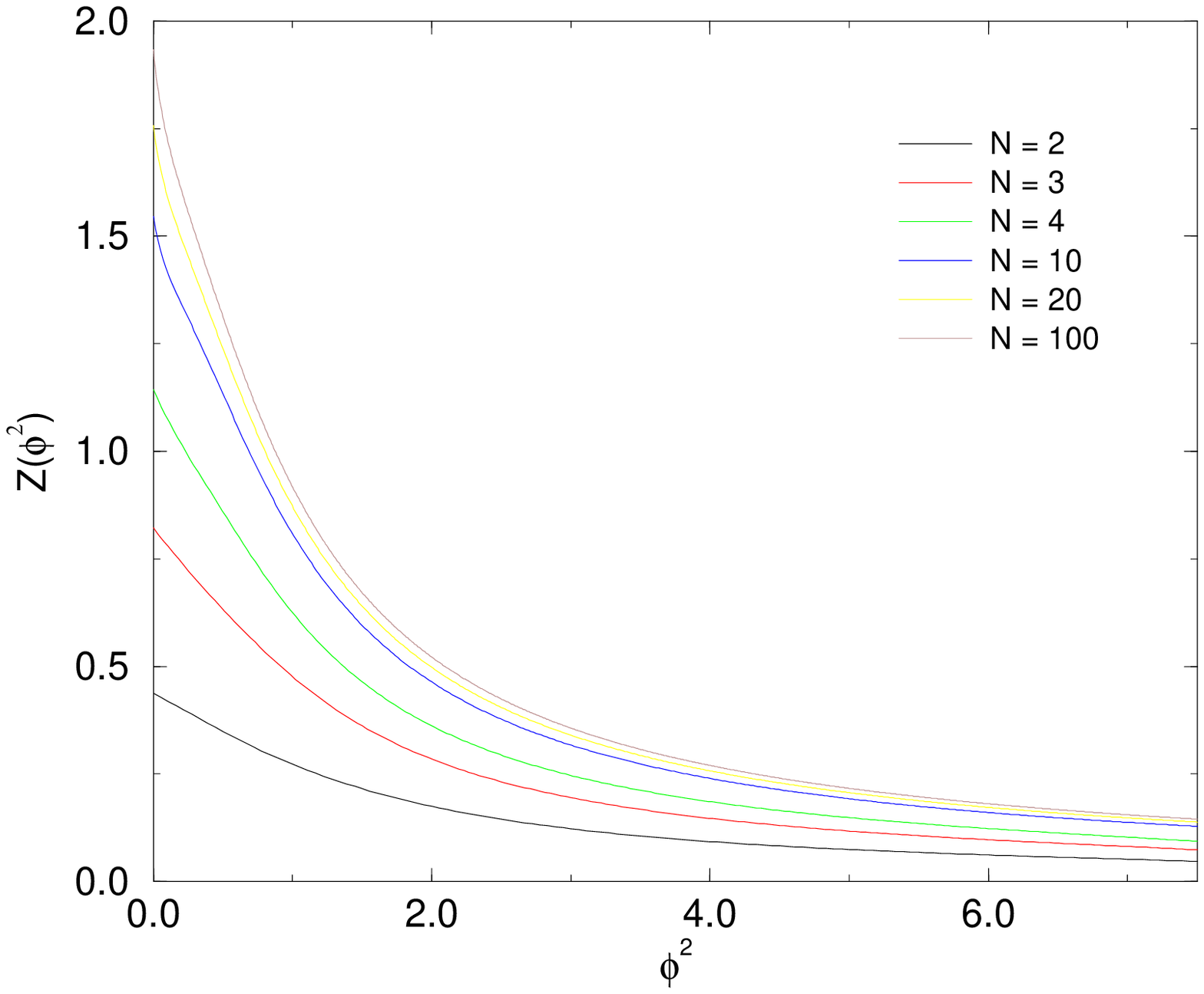,width=\textwidth,height=\textwidth}
\caption{The Z component of the wave function renormalization of the
Legendre effective action at second order in the derivative
expansion, for three dimensional $O(N)$ symmetric field theory, for
$N=2, 3, 4, 10, 20, 100$.}
\label{Z2nd}
\end{figure}

At the moment the value of $\eta$ as been left undetermined and it
appears that we have a free parameter. However we can use the scaling
symmetry to impose an extra condition and hence fix this extra
parameter. We will take \mbox{$K(0)=1$}, with other possible solutions
being reached by using the re-parameterization invariance. We now have
a seven parameter set with seven conditions imposed and hence expect
at most a discrete set of solutions. Again we only find two:- the
Gaussian point mentioned above and an approximation to the
Wilson-Fisher fixed point. The results for~$\eta$ are summarized in
table~(\ref{t:2crit}) and the results for the fixed point solutions
shown in figures~(\ref{V2nd}) to~(\ref{Z2nd}).

\pagebreak
\section{The Critical Exponents at Second Order}

To calculate the critical exponents we will again need to consider the
perturbations about the fixed point solutions. That is we will write,
\begin{eqnarray}
V(\phi^2,t) & = & V^*(\phi^2) + \delta V(\phi^2,t) \nn \\ & = &
 V^*(\phi^2) +\varepsilon \mathrm{e}^{\lambda t} v(\phi^2) \\
 K(\phi^2,t) & = & K^*(\phi^2) + \delta K(\phi^2,t) \nn \\ & = &
 K^*(\phi^2) + \varepsilon
\mathrm{e}^{\lambda t} k(\phi^2) \\
Z(\phi^2,t) & = & Z^*(\phi^2) + \delta Z(\phi^2,t) \nn \\ & = &
Z^*(\phi^2) +\varepsilon \mathrm{e}^{\lambda t} z(\phi^2)
\end{eqnarray} 
where $V^*(\phi^2)$, $K^*(\phi^2)$, $Z^*(\phi^2)$ are the fixed point
solutions calculated above, and expand the equations to first order in
$\varepsilon$.

To find the perturbations we will again need to consider the boundary
conditions. We will want the perturbations to exist for all
$\phi^2$. If we insist that $v$,$k$ and $z$ grow no faster that a
power at large $\phi^2$, then asymptotic analysis will show that $v$,
$k$ and $z$ will grow according to their scaling dimension,
\begin{eqnarray}
v(\phi^2) & \sim & \mathit{a_v} (\phi^2)^{\frac{3}{1 + \eta} -\lambda}
+ \cdots \\ k(\phi^2) & \sim & \mathit{a_k}
(\phi^2)^{\frac{-\eta}{1+\eta} - \lambda} + \cdots \\ z(\phi^2) & \sim
& \mathit{a_z}(\phi^2)^{-\frac{1+ 2 \eta}{1 + \eta} - \lambda} +
\cdots
\end{eqnarray}
The imposition of power law growth will again force $\mathit{a_v}$,
$\mathit{a_k}$ and $\mathit{a_z}$ to form an isolated three parameter
set.  The other three boundary conditions come from forcing the
equations to hold at the origin. Using linearity to set $v(0)=1$, we
will have a seven parameter set with seven boundary conditions
imposed. Hence we expect to find at most a discrete set of solutions,
which is what is found.  As before we find just one positive
eigenvalue, which yields $\nu$ through $\nu=1/\lambda$, and determine
the least negative eigenvalue, which gives the first correction to
scaling exponent through $\omega=-\lambda$. These values are shown in
table~(\ref{t:2crit}).

It is important to recognize that we will also find other solutions of
the equations which do not correspond to critical
indices~\cite{a:redun}. These solutions are known as redundant
perturbations and the eigenvalue corresponding to the solution depends
on the exact form of the renormalization group chosen, but as their
eigenvalue depends on the exact form of the equation chosen these
perturbations can not be physical. The redundant perturbation reflect
invariances of the equations. In general if we change variables
$\phi^a(x) = \tilde{\phi}^a(x) + \varepsilon \Phi^a[\tilde{\phi}]$
in~(\ref{e:rgscaled2}) (with $J.\phi$ replaced by $J.\tilde{\phi}$),
then this induces a change in the effective action of $\delta
\Gamma=
F. \frac{\delta \Gamma}{\delta \phi}$ with $F[\phi] = \varepsilon \exp
\left ( - W[J] \right )
\Phi[\delta/\delta J] \exp \left( W[J] \right )$ and a change in the
cutoff functional~\cite{a:timderiv,a:redun}. A general choice of $F$
that will leave~(\ref{e:gammalo}) invariant at this order is,
\begin{equation}
F[\phi] = \left \{ f(\phi^2(x)) \phi^a + \alpha x^{\mu}\p\phi^a(x)
\right \}
\end{equation}
 for any function $f$ and any constant $\alpha$. A redundant
 perturbation must then satisfy,
\begin{eqnarray}
v & \propto & 2 f \phi^2 V' - 3 \alpha V \\ 
k & \propto & 2 f \phi^2 K' +
2 f K - \alpha K \\
 z & \propto &  2 f \phi^2 Z' + 4 f' \phi^2 Z + 4 f
Z - \alpha Z + 2 f'K
\end{eqnarray}
In fact we expect and find only one redundant perturbation
corresponding to the re-parameterization
invariance~(\ref{e:scalsym}). This has eigenvalue
\mbox{$\lambda=0$}, \mbox{$f=5/2 \alpha$} and \mbox{$\alpha \ne 0$}, yielding
\mbox{$(v,k,z)=(-5\phi^2V'+ 3V,-5\phi^2 K' - 4K, -5 \phi^2 Z' -9Z)$},
where we have set $\alpha=-1$. An
obvious question is why no redundant perturbations are found at the
leading order. The answer is simple: the choice of $K \equiv 1$ breaks
the re-parameterization invariance, and we should not expect to see
any redundant perturbation. We should also not expect to see any
perturbation corresponding to $\phi$-translation
invariance~\cite{a:WandH} as the choice of $V,K$ and $Z$ being
functions of $\phi^2$ breaks this invariance.

The redundant perturbation also proves to be a convenient check on the
numerical accuracy of the equations for the perturbations. Once we
know the fixed point solutions we automatically know a solution to the
perturbation equation, which should have eigenvalue $\lambda=0$.  We
can then try to find our redundant solution using numerical
methods. The degree to which the known solution (from the fixed
points), and the solution found numerically agree, gives a bound on
the accuracy of the critical exponents.  Using numerical methods, we
find that the value for the eigenvalue corresponding to the redundant
perturbation typically lies in the range $0.0003$ (for $N=2$) to
$0.00005$ (for $N=100$), indicating that we should only trust any
numeric results to three figures. The graphs of the two types of
solutions also agree to a high degree of accuracy.

\section{The case of $N=1$}

\begin{figure}[t]
\centering
\subfigure[$V(\phi^2)$]{\epsfig{figure=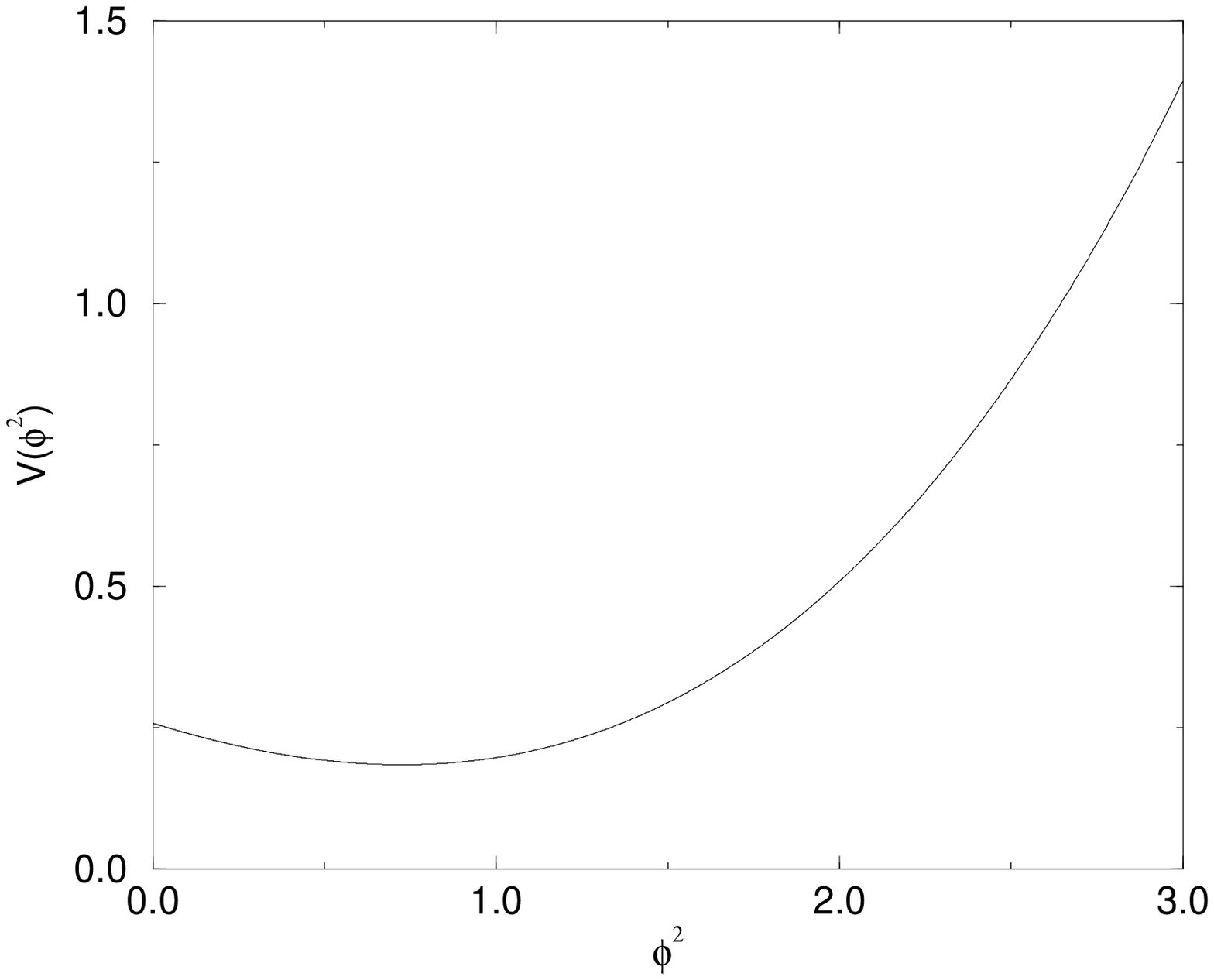,width=0.46\textwidth,height=0.46
\textwidth}} 
\subfigure[$\kappa(\phi^2)$]{\epsfig{figure=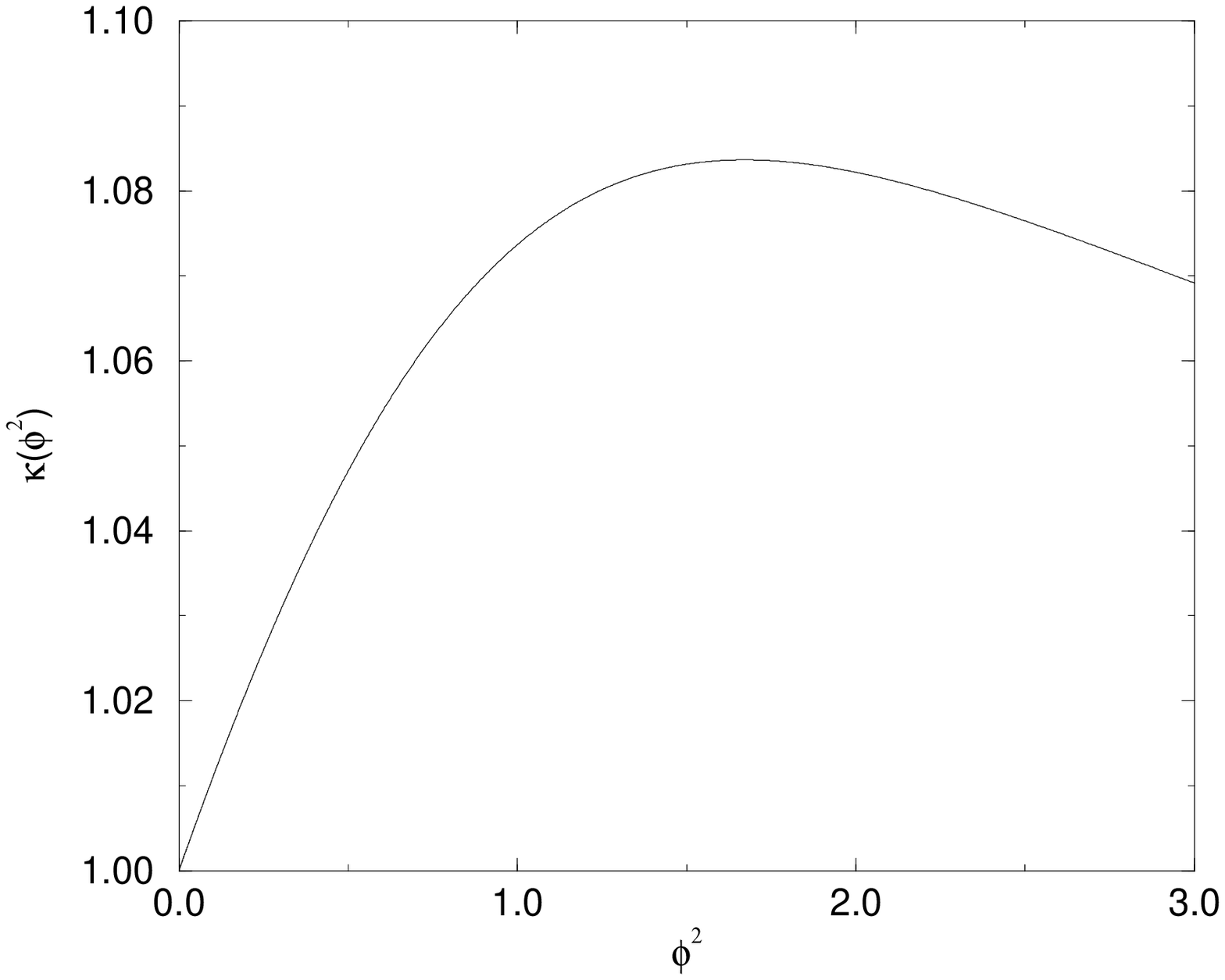,width=0.46 \textwidth,height=0.46
\textwidth}} 
\caption{The results of applying the derivative expansion at second
order for the case of $N=1$.}
\label{2ndN1}
\end{figure}

As mentioned above the case of $N=1$ is slightly different to the more
general case of $N \ne 1$. This should not be entirely surprising as
there is now no internal symmetry at all (other than the discrete
$Z_2$ symmetry). The derivative expansion now becomes,
\begin{equation}
\Gamma[\phi] = \int d^Dx \left \{ V(\phi^2)  +  \frac{1}{2} \left ( K(\phi^2) + \phi^2
Z(\phi^2) \right ) \left ( \partial_{\mu} \phi \right )^2 \right \}
\end{equation} 
We see that we should no longer consider the $Z$ and $K$ components of
wave function renormalization separately, but should consider instead
a single function \mbox{$\kappa=K + \phi^2 Z$}. The derivative
expansion at second order, at $N=1$, thus becomes,
\begin{equation}
\Gamma[\phi] = \int d^Dx \left \{ V(\phi^2) +  \frac{1}{2} \kappa(\phi^2)
\left ( \partial_{\mu} \phi  \right )^2 \right \} 
\end{equation}
We will no longer have three coupled second order equations but will
have only two coupled second order equations for $V$ and $\kappa$. The
results for this case can be found in~\cite{a:timderiv}.  It is an
important check on our equations that if we write \mbox{$\kappa = K +
\phi^2 Z$} and set $N=1$, then we get the same equations as found
in~\cite{a:timderiv}.  The results of this expansion are included in
figure~(\ref{2ndN1}) for completeness. These come from an independent
calculation.

It is interesting to ask what does happen to the $K$ and $Z$
 components as $N \rightarrow 1$. This can be done by taking the $K$
 equation and substituting $K=\kappa-\phi^2 Z$ into it. This then
 yields an equation which can be solved numerically using the known
 result for $\kappa$ and $V$ (from
\cite{a:timderiv} or our independent calculation). It is then found,
that  at $N=1$, $Z$
diverges as $1/\phi^2$ at the origin. Again this should not be
entirely surprising: we can write $Z= (\kappa -K)/\phi^2$, and should
therefore expect to see some divergence at the origin, provided $K(0)
\ne \kappa(0)$.  Numerically this is an important point.  As $N
\rightarrow 1$ we will expect that $Z$ will become steeper and steeper
at the origin before it eventually becomes divergent at $N=1$. This
divergent behavior may lead to numerical instability as we approach
the regime $N \approx 1$.  In fact we start to feel the effects of
this divergence as early as $N=2$, where it is already found to be
much harder to produce an accurate numerical solution than at $N=3$,
say.

\begin{table} [ht]
\renewcommand{\arraystretch}{1.5}
\hspace*{\fill}
\begin{tabular}{|c||c|c||c|c|c||c|c|c|}  \hline
$N$ &\multicolumn{2}{c||}{$\eta$} &\multicolumn{3}{c||}{$\nu$}
&\multicolumn{3}{c|}{$\omega$} \\ \hline & $O(\partial^2)$ & &
$O(\partial^2)$ & $O(\partial^0)$ & & $O(\partial^2)$ &
$O(\partial^0)$ & \\ \hline 1 & 0.054 &0.035 &0.62 &0.66 &0.63 &0.898
&0.63 &0.80 \\ \hline 2 & 0.044 &0.033 - 0.04 & 0.65 & 0.73 & 0.67 &
0.38 &0.66 & 0.79 \\ \hline 3 & 0.035 &0.033 - 0.04 & 0.745 & 0.78 &
0.71 & 0.33 & 0.71 & 0.78 \\ \hline 4 & 0.022 &0.025 & 0.816 & 0.8240&
0.75 & 0.42 & 0.75 & \\ \hline 10& 0.0054 &0.025 & 0.95 & 0.93 & 0.88
& 0.82 & 0.89 & 0.78 \\ \hline 20& 0.0021 & 0.013& 0.98 & 0.96 & 0.94
& 0.93 & 0.95 & 0.89 \\ \hline 100 & 0.00034 &0.003 & 0.998& 0.994&
0.989& 0.988 & 0.991 & 0.98 \\
\hline
\end{tabular}
\hspace*{\fill}
\renewcommand{\arraystretch}{1}
\caption[Critical exponents at second order in the derivative expansion]
{ Critical exponents of the three-dimensional theory for various
values of $N$.  We list the results gained at second order together
with those from the leading order and an indication of what should be
expected from other methods as summarized
in~\cite{b:zinn,a:nu:1overN2,a:om:1overN,a:KanN=4}. $\eta$ is
predicted to be zero for all $N$ at the leading order in the
approximation.}
\label{t:2crit}
\end{table}

\section{Discussion}

The results at second order are puzzling: for $N=2,3,4$ we have
reasonable accuracy for $\eta$ \mbox{and $\nu$}, with an improvement
upon the leading order results, but terrible predictions for $\omega$,
which is predicted to be worse than the leading order results.  On the
other hand the predictions at $N=10,20,100$ show the converse: we have
$\eta$ seriously underestimated, by a factor of about ten, and the
results for $\nu$ compare less favourably with other methods than
those predicted by the leading order of the approximation. However we
also see an improvement in the predictions for $\omega$ compared to
other methods.

These results are very puzzling as the only other comparable
calculations at this order in the derivative expansion
\cite{a:timderiv,a:tim2d}, performed at $N=1$, showed an  improvement for all
the calculated exponents when the change was made from the leading
order to the second order of the derivative expansion.  Other authors
who have tried to go beyond the leading order of the approximation
certainly do not report results for $\omega$ like those we have found
at $N=2,3,4$~\cite{a:ballrubbish}.  It would be easy to explain these
results away by calling upon numerical inaccuracy, especially at
$N=2,3,4$ where the known steep behaviour of $Z$ at the origin may
have an effect.  However, although it is true that the equations are
extremely difficult to solve (cf appendix A) we feel that this is
unlikely for several reasons. Firstly the results are insensitive to
changes in where we impose the asymptotic boundary condition, the
number of points in the solution mesh and other numerical factors,
whereas we might expect there to be random errors introduced if the
results were down to numerical roundoff or discretization error. For
example, it can be seen that the eigenvalues seem to flow continuously
with the value of N, whereas we would perhaps expect some random
fluctuations if we were encountering severe numerical
problems. Secondly, we already know that the known redundant
perturbation can be found to a reasonable degree of accuracy,
indicating that we should have a high degree of faith in our numerical
methods and in the numerical stability of the equations themselves. For
these two reasons it seems rather unlikely that these effects are due
to numerical problems.

\begin{figure}[t]
\centering
\subfigure[]{\epsfig{figure=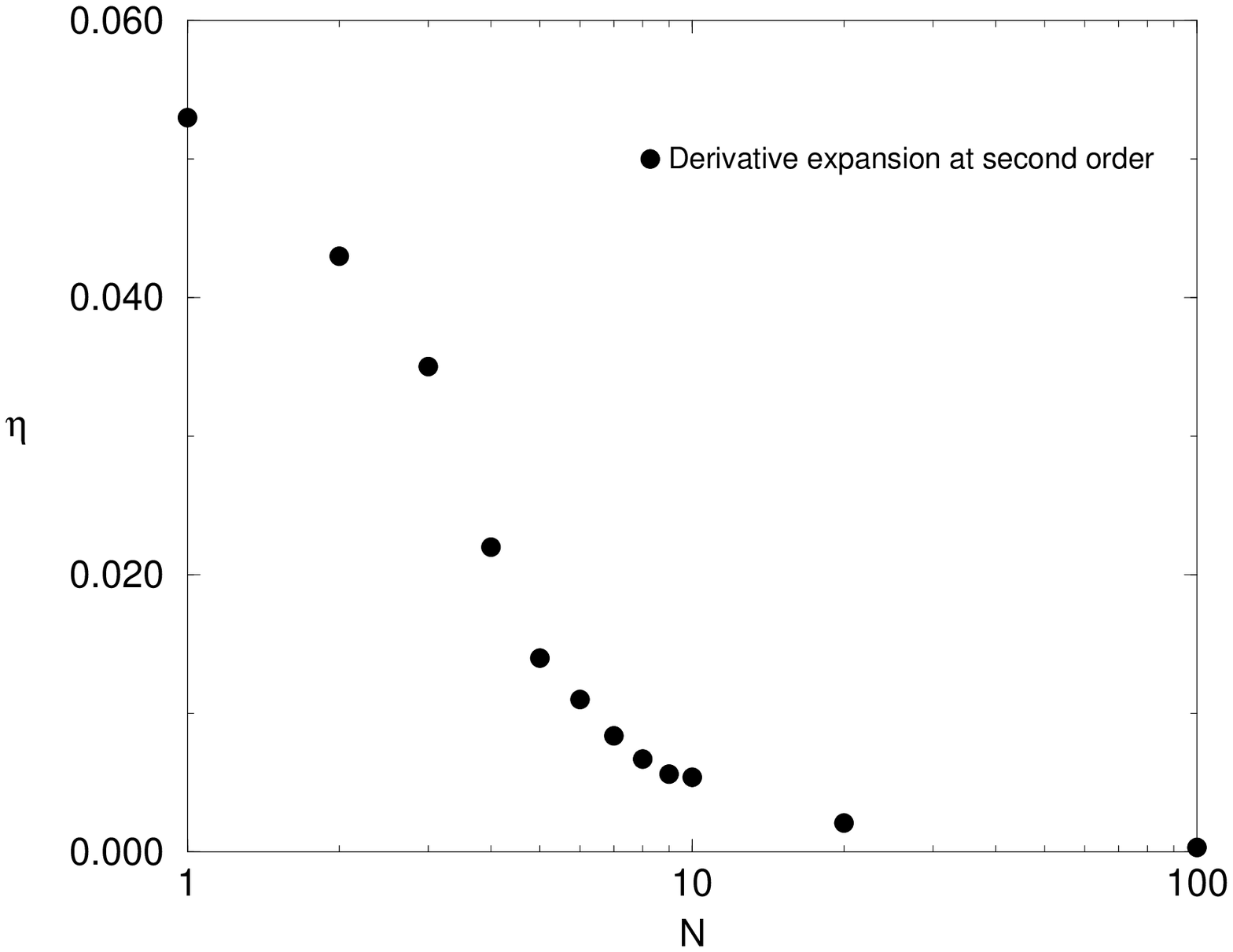,width=0.46\textwidth,height=0.46
\textwidth}} 
\subfigure[]{\epsfig{figure=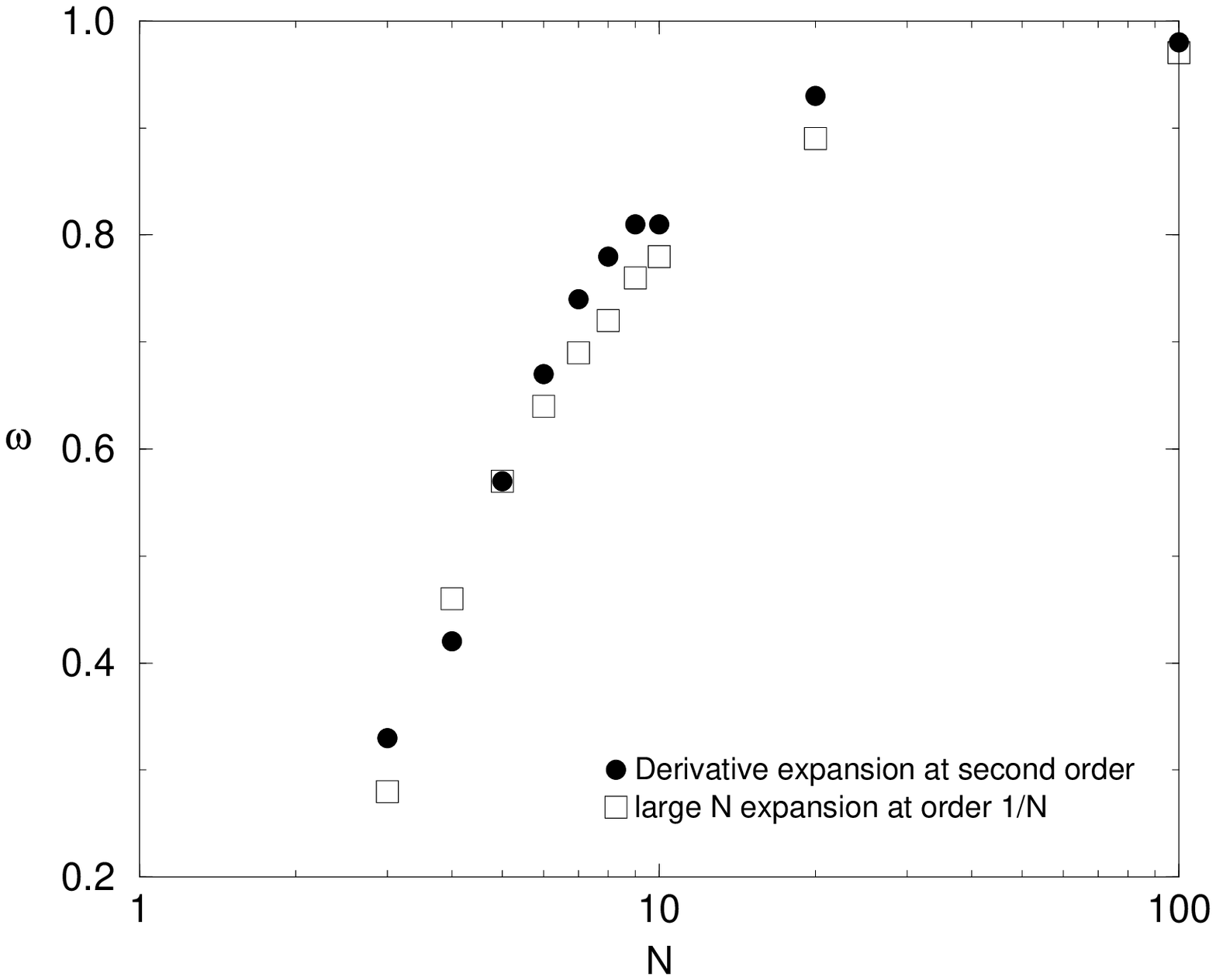,width=0.46 \textwidth,height=0.46
\textwidth}} 
\caption{The flow of $\eta$ and $\omega$ with the value of $N$.
Figure~b also shows what the large $N$ expansion predicts at order
$1/N$.}
\label{eigenflow}
\end{figure}

Quite why the results do not compare so well with those presented by
others is puzzling. The serious under-prediction of $\eta$ for large
$N$ could be attributed to the approximation scheme being 'too biased'
towards the $N=\infty$ results, which would predict $\eta=0$, at large
$N$. It is also interesting to note that the flow of $\omega$ with $N$
seems to follow the large $N$ results at order $1/N$ for 
$N=3,4,5,\ldots...$. The results of these two flows are shown in
figure~(\ref{eigenflow}).

Overall we must conclude that the scheme for intermediate $N$ does not
work as well as it does for $N=1$ or $N=\infty$. The $N=\infty$ result
is in fact exact and is well understood (cf the next chapter
and~\cite{a:WandH}). The reason why the encouraging results at
$N=1$~\cite{a:timderiv,a:tim2d} are not continued for $N \ne 1$ are
not yet understood. However, it should be emphasized that we find no
evidence for spurious results, such as those found by truncations: we
find only the fixed points and perturbations expected.

\begin{figure}[th]
\centering
\subfigure[$v(\phi^2)$]{\epsfig{figure=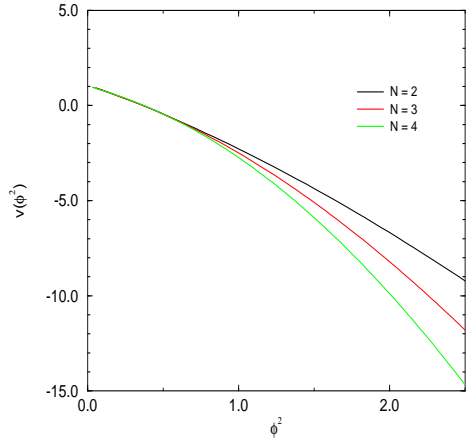,width=0.46\textwidth,
height=0.46\textwidth}}
\subfigure[$k(\phi^2)$]{\epsfig{figure=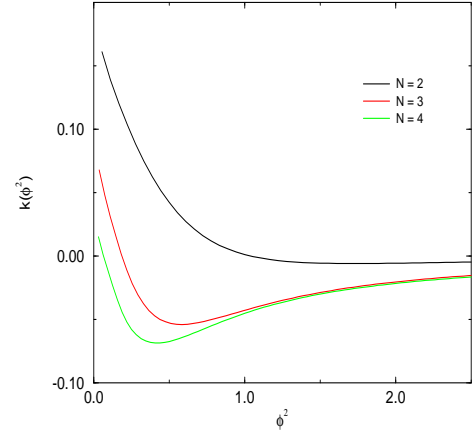,width=0.46
\textwidth,height=0.46\textwidth}}  
\subfigure[$z(\phi^2)$]{\epsfig{figure=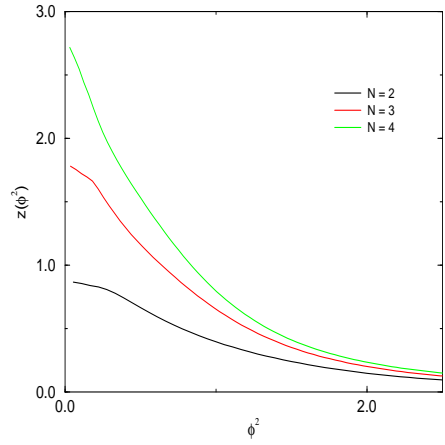,width=0.46
\textwidth,height=0.46\textwidth}}  
\caption{The $v$, $k$ and $z$ components of the perturbation
corresponding to $\nu$ at second order in the derivative expansion}
\label{2ndnu}
\end{figure}

\begin{figure}[th]
\centering
\subfigure[$v(\phi^2)$]{\epsfig{figure=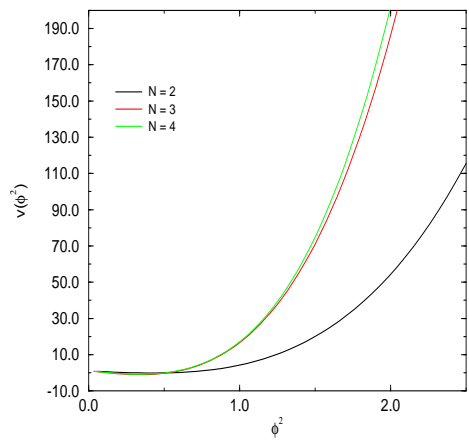,width=0.46\textwidth,
height=0.46\textwidth}}
\subfigure[$k(\phi^2)$]{\epsfig{figure=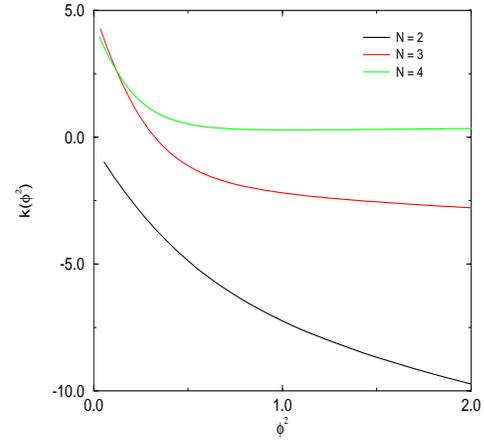,width=0.46
\textwidth,height=0.46\textwidth}}  
\subfigure[$z(\phi^2)$]{\epsfig{figure=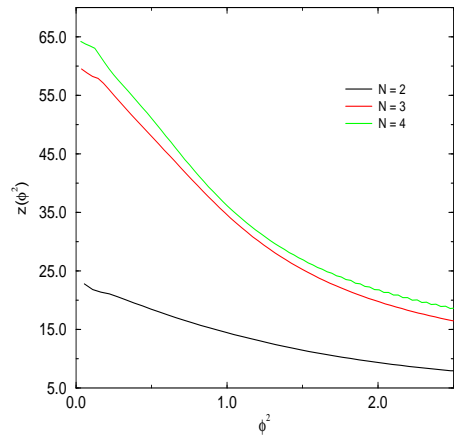,width=0.46
\textwidth,height=0.46\textwidth}}  
\caption{The $v$, $k$ and $z$ components of the perturbation
corresponding to $\omega$ at second order in the derivative expansion}
\label{2ndom}
\end{figure}

\chapter{Exactly Solvable Systems}

It has been well known that certain theoretical models can be exactly
solved. That is, have all their physical parameters determined without
any form of approximation. In this section we will discuss how the
derivative expansion reproduces the exact results found in two well
known cases:-- $N=\infty$ and $N=-2$. It is important that the
derivative expansion reproduces these two exactly known  results, for if
it didn't then the approximation would be too great  a mutilation of
the theory.

\section{Exact Results at $N=\infty$}

It was noted a long time ago that the large $N$ limit is
equivalent~\cite{a:stanley} to the exactly solvable spherical
model~\cite{a:joyce}. It was shown that in the large $N$-limit that
the local potential expansion effectively becomes
exact\footnote{Wegner pointed out that although the other terms exist
in the expansion, 
they do not affect the value of the critical
exponents~\cite{a:WandH}}. ie  we have, 
\begin{equation}
\Gamma[\phi] \equiv  \int d^Dx \, \frac{1}{2} \left (\partial_{\mu} \phi^a
\right )^2 + V(\phi^2) \ \ \ \ \mbox{at $N=\infty$} 
\end{equation}  
In this section we will show that the derivative expansion re-produces
this result and give an expression for $V(\phi^2)$ and for the
perturbations about the fixed point.

In the previous chapter we scaled the equations to provide a
comparison with the $N=\infty$ results. We now justify where these
scalings come from. 

We will consider a simple scalar $\Phi^4$ theory in $D$ dimensions,
\begin{eqnarray}
Z &=& \int D\Phi \exp \left \{ - S[\Phi] + \int d^Dx J.\Phi \right \} \\
S[\Phi] &=& \int d^Dx \, \frac{1}{2} (\partial_{\mu} \Phi)^2 + \frac{m^2}{2}
\Phi^2 +\frac{u}{4!} [\Phi^2]^2
\end{eqnarray}
We can perform the following transformation
\begin{equation}
\exp -\frac{u}{4!} \int d^Dx \,  [\Phi^2]^2  \sim \int D\lambda \exp
\left \{ \int 
d^Dx  \frac{N \lambda^2}{2} - \frac{\lambda}{2} \Phi^2
\sqrt{\frac{Nu}{3}} \right \}
\label{e:trans}
\end{equation}
and hence re-write the path integral as,
\begin{equation}
Z = \int D\lambda \exp  \left \{ \int d^Dx \, \left ( \frac{1}{2} N
\lambda^2 \right ) + \frac{N}{2}
\mathrm{tr}
\log \left ( { -\Delta + m^2 + \sqrt{\frac{Nu}{3}} \lambda}  \right )\right \}
\label{e:scaledaction}
\end{equation}
We see from this that we now have an explicit dependence on $N$ in the
(non-local) action. If we wish to perform a saddle point expansion in
$1/N$ then we must have a factor of $N$ outside the action, but none
in the $\mathrm{tr} \log$. From this we see that we must hold $Nu$
fixed, so $u$ is of order $1/N$, and we must have $\lambda$ of order one also.
Hence, counting the powers of $N$ in equation~(\ref{e:trans}) and
demanding that the powers of $N$ match up on both sides, we see that we
must have $\Phi^2$ of order $N$. 

From this we see that when we come to consider the flow equations we
should scale $\phi^2$ as $\phi^2= N 
\tilde{\phi^2}$. Hence,  demanding that $\Gamma$ scales as
$\Gamma= N \tilde{\Gamma}$ (as follows from (\ref{e:scaledaction}), we
have the following scalings,
\begin{eqnarray}
V & \sim & N \tilde{V}\\
K & \sim & \tilde{K}\\
Z & \sim & \tilde{Z}/N
\end{eqnarray}
where $\tilde{V}$,$\tilde{K}$ and $\tilde{Z}$ are of order one in the
$1/N$ expansion.
Upon doing this the flow equation for $V$ becomes (upon dropping the
tildes), 
\begin{eqnarray}
\lefteqn{\frac{\partial}{\partial t}V + (1 + \eta) \phi^2 V' - 3 V =
  - \left ( 1 - \eta/4 \right ) \times }\\
&& \left  [ \frac{1}{N \sqrt{K + \phi^2 Z} \left( 2 V'  + 4 \phi^2 V''  +2
\sqrt{
K + \phi^2 Z} \right )^{1/2}} 
 +   \frac{1-1/N}{\sqrt{K} \left (2 V'
+ 2 \sqrt{K} \right )^{1/2}}  \right ] \nonumber
\end{eqnarray}
plus some longer equations for $K$ and $Z$.
If we now take the $N=\infty$ limit we end up with the following sets
of equations,
\begin{eqnarray}
\frac{\partial }{\partial t}V + (1 + \eta) \phi^2 V' - 3 V &=&
  - \frac{\left ( 1 - \eta/4 \right)}{\sqrt{K} \left (2 V'
+\sqrt{K} \right )^{1/2}}    \label{e:Vinfin}\\
\frac{\partial}{\partial t}K + (1+\eta)\phi^2 K' +\eta K & = & -
\frac{(1-\eta/4)K'}{\sqrt{K} \left ( 2 
V' + 2\sqrt{K} \right)^{3/2}} \label{e:Kinfin}\\
\frac{\partial}{\partial t}Z + (1+\eta)\phi^2 Z' +(1+2\eta) Z & = &
\cdots 
\end{eqnarray}

We immediately notice something different:- the equations for $K$ and
$V$ have decoupled from the equation for $Z$ and the equations are now
\emph{first} order equations. This is the first simplification.

Now lets consider the fixed point solutions. The
next simplification comes by looking at the local potential  expansion
and noting that this will require that $K \equiv 1$. Hence we try
$K\equiv 1$ as a fixed point solution. Equation~(\ref{e:Kinfin}) then
yields, $\eta = 0$.

Upon setting $\eta=0$ we find that  the fixed
point potentials  will then satisfy,
\begin{equation}
 \phi^2 V' - 3 V =
  - \frac{1}{ \left (2 V'
+ 2 \right )^{1/2}} 
\label{e:potninfin}
\end{equation}

Although this is a first order equation it is still not in a
particularly useful form for solution.
If we differentiate the equation with respect to $\phi^2$ and set
\mbox{$W(\phi^2) =  V'(\phi^2)$} then we have,
\begin{equation}
-2 W + \phi^2 W' = \frac{W'}{(2+2 W)^{3/2}}
\end{equation}
This equation can the be re-arranged as partial differential equation
for $\phi^2$ in $W$ as follows,
\begin{equation}
2 W \frac{\partial \phi^2}{\partial W} - \phi^2 = - \frac{1}{(2+2W)^{3/2}}
\end{equation}
Using $1/\sqrt{W}$ as an integrating factor we  then have,
\begin{equation}
\frac{\partial \frac{\phi^2}{\sqrt{W}}}{\partial W} = - \frac{1}{2 W^{3/2} (2 +
2 W)^{3/2}}
\end{equation}
Performing the final integration then yields a final expression for
$\phi^2$,
\begin{equation}
\phi^2 = A \sqrt{W} + \frac{1}{2 \sqrt{2}} \sqrt{1+W} \left ( 1+
\frac{W}{1+W} \right)
\end{equation}
where $A$ is a constant of integration.
If we have $A \ne 0 $ then it can be shown that $\phi^2$ will reach a
maximum value for some non zero value of $W$. ie the solution is not defined
for all values of $\phi^2$. We are therefore  required to set $A=0$ and the
solution then becomes,
\begin{equation}
\phi^2 = \frac{1}{2 \sqrt{2}} \sqrt{1+W} \left ( 1+
\frac{W}{1+W} \right)
\end{equation}
This can be inverted to give the solution for $W$,
\begin{equation}
W=-1 + \left [ \frac{1}{\sqrt{2}} \phi^2 + \frac{1}{\sqrt{2}} \sqrt{ (\phi^2)^2
+1} \right ]^2
\end{equation}
Integrating with respect to $\phi^2$ then gives the solution for the
potential,
\begin{equation}
V= -\frac{1}{2} \phi^2 +\frac{(\phi^2)^3}{3} + \frac{((\phi^2)^2
+1)^{3/2}}{3}
\label{e:Vsol}
\end{equation}
It is important to stress the difference between the scaled and the
unscaled results in the large $N$ regime. The \emph{unscaled} $V$ tends to
infinity, $K$ tends to $1$ and $Z$ tends to
zero as $N \rightarrow \infty$. However when we consider the scaled
solutions we see a different pattern: $V$ now tends to our solution
(\ref{e:Vsol}), $K$ still tends to one, but so far $Z$ has been left
undetermined. 

\newpage
Performing the scaling upon the $Z$ equation and setting $K=1,K'=0$ we
see that $Z$ must satisfy the following equation,
\begin{eqnarray*}
\lefteqn{\phi^2 Z'+ Z =} \\
 & &- {{\it i}_{2, 0, 2}}\,{\it Z'} 
+ {\displaystyle \frac {32}{3}}{ \phi^2}\,{\it V''} 
{\displaystyle \frac {1024}{3}}\,{{\it i}_{4, 3, 2}}\,
{\it V''}^{4}\,{ (\phi^2)}^{2}\\
 & & +  \left( \! \,4\,{{\it i}_{6, 2, 3}} - 8\,{{\it i}_{4, 2, 3
}} + {{\it i}_{4, 2, 2}} + 4\,{{\it i}_{2, 2, 3}} - 5\,{{\it
i}_{2,2,2}} +   {{\it i}_{2, 2, 1}}\, \!  \right) \,{\it V'''}\\
& & + {\displaystyle \frac {64}{3}}{ \phi^2}
 \left( {\vrule height0.53em width0em depth0.53em}
 \right. \! \! 16\,{{\it i}_{5, 3, 2}}\,{Z}\,{ \phi^2} - 8\,{
{\it i}_{5, 2, 3}} - 11\,{{\it i}_{3, 2, 2}} 
 \mbox{} + 8\,{{\it i}_{3, 2, 3}} - 8\,{{\it i}_{3, 3, 2}} \\
& & + 8\,{{\it i}_{3, 3, 1}} + 8\,{{\it i}_{5, 3, 2}} \! 
\! \left. {\vrule height0.53em width0em depth0.53em} \right) 
{\it V''}^{3}\mbox{} \\
& & +  \left( {\vrule 
height0.79em width0em depth0.79em} \right. \! \! 
\mbox{} {\displaystyle \frac {256}{3}}\,{ \phi^2}\,{Z}\,
{{\it i}_{6, 3, 2}} - {\displaystyle \frac {256}{3}}\,{ \phi^2}
\,{Z}\,{{\it i}_{4, 3, 2}} + {\displaystyle \frac {256}{3}}\,{ 
\phi}^{2}\,{Z}\,{{\it i}_{4, 2, 3}} \\
 & & \mbox{} - {\displaystyle \frac {416}{3}}\,{ \phi^2}\,{Z}\,
{{\it i}_{4, 2, 2}} + {\displaystyle \frac {256}{3}}\,{ (\phi^2)}^{2
}\,{Z}^{2}\,{{\it i}_{6, 3, 2}} + {\displaystyle \frac {128}{3}}
\,{{\it i}_{6, 2, 3}} - {\displaystyle \frac {256}{3}}\,{{\it i
}_{4, 2, 3}} \\
 & & \mbox{} - 64\,{{\it i}_{4, 2, 2}} + {\displaystyle \frac {
128}{3}}\,{{\it i}_{2, 2, 3}} - {\displaystyle \frac {64}{3}}\,{
{\it i}_{2, 2, 2}} - {\displaystyle \frac {64}{3}}\,{{\it i}_{2
, 2, 1}} + {\displaystyle \frac {128}{3}}\,{{\it i}_{6, 0, 5}}
 \\
 & & \mbox{} + {\displaystyle \frac {64}{3}}\,{{\it i}_{6, 3, 2}
} + {\displaystyle \frac {128}{3}}\,{{\it i}_{4, 3, 1}} - 
{\displaystyle \frac {128}{3}}\,{{\it i}_{6, 1, 4}} + 
{\displaystyle \frac {256}{3}}\,{{\it i}_{4, 1, 4}} + 
{\displaystyle \frac {80}{3}}\,{{\it i}_{4, 1, 3}} \\
 & & \mbox{} + {\displaystyle \frac {128}{3}}\,{{\it i}_{2, 0, 5
}} + {\displaystyle \frac {16}{3}}\,{{\it i}_{2, 0, 3}} - 32\,{
{\it i}_{2, 0, 4}} + {\displaystyle \frac {64}{3}}\,{{\it i}_{2
, 3, 0}} + {\displaystyle \frac {64}{3}}\,{{\it i}_{2, 3, 2}} \\
 & & \mbox{} - {\displaystyle \frac {128}{3}}\,{{\it i}_{2, 1, 4
}} + {\displaystyle \frac {112}{3}}\,{{\it i}_{2, 1, 3}} + 
{\displaystyle \frac {16}{3}}\,{{\it i}_{2, 1, 2}} - 
{\displaystyle \frac {128}{3}}\,{{\it i}_{2, 3, 1}} - 
{\displaystyle \frac {32}{3}}\,{{\it i}_{4, 0, 4}} \\
 & & \mbox{} - {\displaystyle \frac {256}{3}}\,{{\it i}_{4, 0, 5
}} - {\displaystyle \frac {128}{3}}\,{{\it i}_{4, 3, 2}} \! 
\! \left. {\vrule height0.79em width0em depth0.79em} 
{\displaystyle 
\frac {256}{3}}\,{ \phi^2}\,{Z}\,{{\it i}_{4, 3, 1}} - 
{\displaystyle \frac {256}{3}}\,{ \phi^2}\,{Z}\,{{\it i}_{6, 2
, 3}} 
 \right) 
{\it V''}^{2}\mbox{} \\
&&+  \left( {\vrule 
height0.79em width0em depth0.79em} \right. \! \!
  \mbox{} {\displaystyle \frac {8}{3}}\,{ \phi}^{2}\,{\it Z'
}\,{{\it i}_{3, 2, 1}} + {\displaystyle \frac {8}{3}}\,{Z}\,{
{\it i}_{5, 2, 2}} - {\displaystyle \frac {40}{3}}\,{ \phi^2}
\,{\it Z'}\,{{\it i}_{3, 2, 2}} - {\displaystyle \frac {64}{3}}
\,{ \phi^2}\,{\it Z'}\,{{\it i}_{5, 2, 3}} \\
 & & \mbox{} - {\displaystyle \frac {40}{3}}\,{Z}\,{{\it i}_{3, 
2, 2}} + {\displaystyle \frac {8}{3}}\,{Z}\,{{\it i}_{3, 2, 1}}
 + {\displaystyle \frac {32}{3}}\,{Z}\,{{\it i}_{3, 2, 3}} - 
{\displaystyle \frac {64}{3}}\,{Z}\,{{\it i}_{5, 2, 3}} \\
 & & \mbox{} + {\displaystyle \frac {32}{3}}\,{ \phi^2}\,{\it 
Z'}\,{{\it i}_{7, 2, 3}} + 8\,{Z}\,{{\it i}_{3, 0, 3}} + 
{\displaystyle \frac {32}{3}}\,{Z}\,{{\it i}_{7, 2, 3}} + 
{\displaystyle \frac {32}{3}}\,{ \phi^2}\,{\it Z'}\,{{\it i}_{
3, 2, 3}}   \\
& & \mbox{} + \,{\displaystyle 
\frac {8}{3}}\,{ \phi^2}\,{\it Z'}\,{{\it i}_{5, 2, 2}} 
\! \! \left . {\vrule height0.79em width0em depth0.79em}
 \right)  
 {\it V''}
\label{e:Zifin}
\end{eqnarray*}
where $i_{j,k,m}$ represents the following integral,
\begin{displaymath}
\int d^{D}q  \frac{q^{2j}}{\left ( 1 + (2 V^{'} +4 \phi^2 V^{''}) q^2 + (1 +
\phi^2 Z) q^4 \right)^k \left(1 + 2 V^{'} q^2 +  q^4 \right )^m}
\end{displaymath}

\begin{figure}[th]
\subfigure[$V(\phi^2)$]{\epsfig{figure=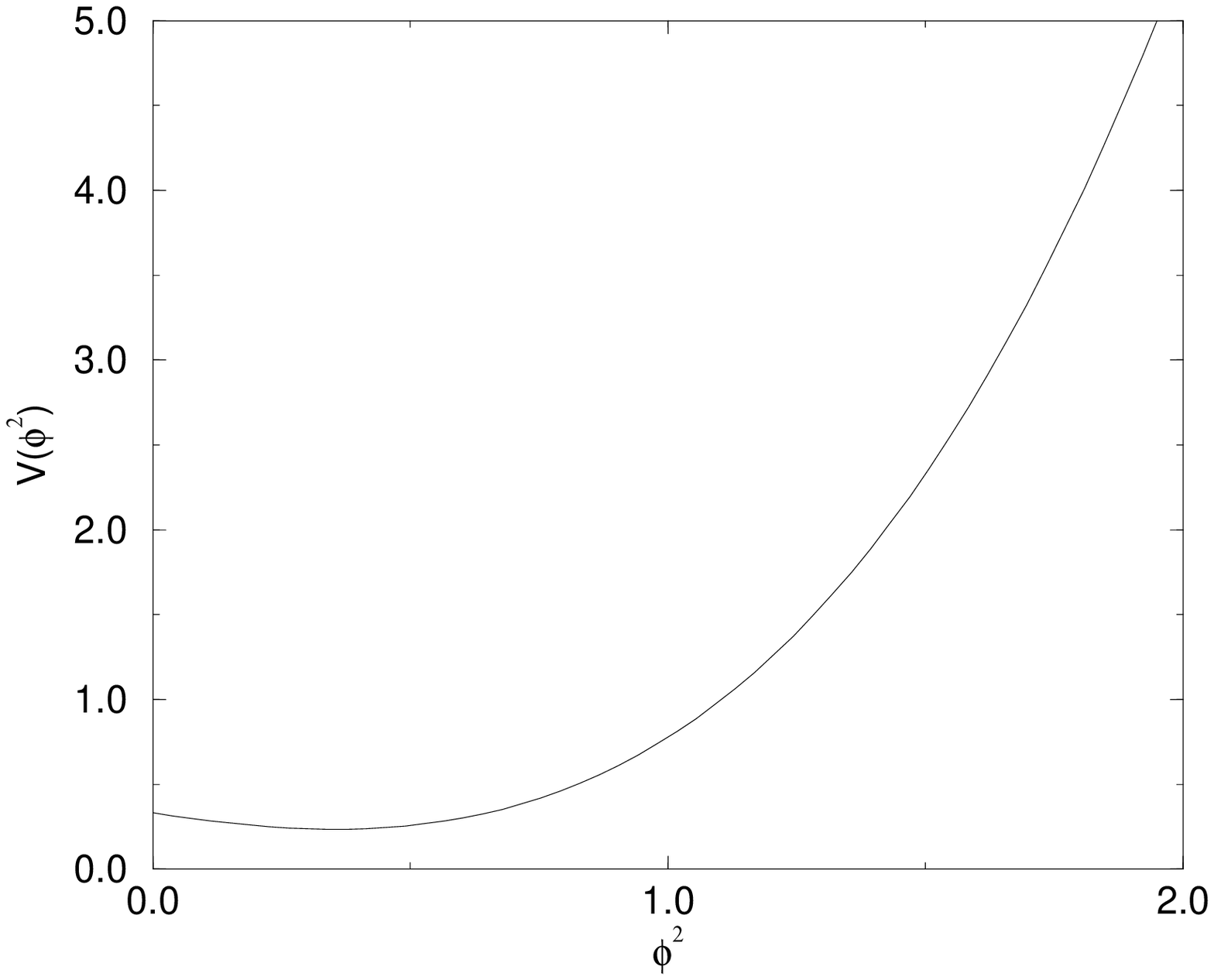,width=0.46
\textwidth , height=0.46 \textwidth}}
\subfigure[$Z(\phi^2)$]{\epsfig{figure=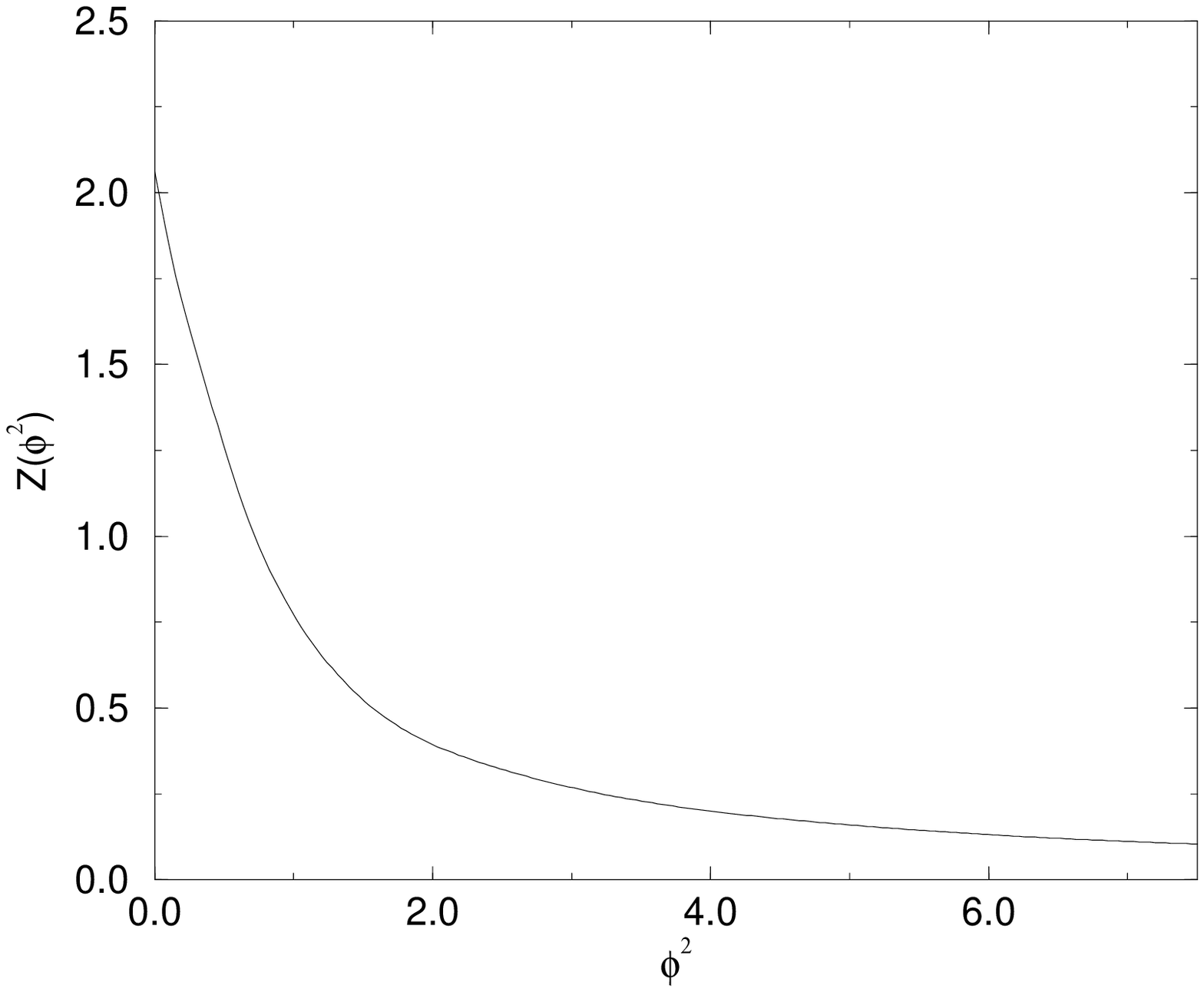,width=0.46
\textwidth , height=0.46 \textwidth}}
\caption{The potential, $V$, and the Z component of the wave function
renormalization of the 
Legendre effective action at second order in  
the derivative expansion, at $N=\infty$.}
\label{Zinfin}
\end{figure}

Of course this equation  can no longer be solved analytically,
as above, but the numerical solution is straightforward. Imposing the
constraint that $Z$ exists for all $\phi^2$ we get the solution shown
in figure~(\ref{Zinfin}b).

We have now calculated what the fixed point potentials look like,
but what we are really interested in is the critical exponents. To
calculate these we will again linearize about the fixed point
solutions by writing $V(\phi^2,t)=V^*(\phi^2)+ \varepsilon
\mathrm{e}^{\lambda t} 
v(\phi^2)$ and $K(\phi^2,t)=K^*(\phi^2)+ \varepsilon \mathrm{e}^{\lambda t}
k(\phi^2)$, where $V^*$ and $K^*$ denote the fixed point solutions
found above. Using these expressions in
equations~(\ref{e:Vinfin})~and~(\ref{e:Kinfin}) 
and expanding to first order in $\varepsilon$ yields the following
pair of differential equations,
\begin{eqnarray}
\lambda v +\phi^2 v' -3 v &=& \frac{1}{2} \frac{k + 2
v'}{(2+2V')^{3/2}} \label{e:vinfin} \\
\lambda k + \phi^2 k' & = & \frac{k'}{(2 V' + 2)^{3/2}} \label{e:kinfin}
\end{eqnarray}
We will consider equation~(\ref{e:kinfin}) first. Re-arranging the
equation we have,
\begin{equation}
\frac{k'}{k} =- \frac{\lambda (2 V' +2)^{3/2}}{\phi^2 (2 V' + 2)^{3/2}
- 1}
\end{equation}
Using  expression~(\ref{e:Vsol})   we then have,
\begin{equation}
\frac{k'}{k} =- \frac{ \lambda \left \{ 2 (\phi^2)^2 + 1 + 2 \phi^2
\sqrt{ (\phi^2)^2+1} \right \}^{3/2}}{ \phi^2 \left \{ 2 (\phi^2)^2 + 1 + 2 \phi^2
\sqrt{ (\phi^2)^2+1} \right \}^{3/2}  - 1 }
\end{equation}
This can easily integrated to give the solution for $k(\phi^2)$,
\begin{equation}
A_1 \left ( \mathrm{e}^{ 2 \sinh^{-1} \phi^2} -2 \right)^{-\lambda/2}
\label{e:ksol}
\end{equation} 
where $A_1$ is a  constant of integration.

We now consider the $v$ equation. This can be re-written as follows,
\begin{equation}
\frac{(\lambda-3)(2+2V')^{3/2} v}{\phi^2 ( 2+ 2V')^{3/2}-1} +v'
 = \frac{1}{2} \frac{k}{\phi^2 (2+2V')^{3/2} -1}
\end{equation}
Using an integrating factor of $(\exp(2 \sinh^{-1} \phi^2)
-2)^{\lambda/2 - 3/2}$, this equation becomes,
\begin{equation}
\frac{d}{d \phi^2} \left ( v(\phi^2) \, (\exp (2 \sinh^{-1} \phi^2)
-2)^{\lambda/2 - 3/2} \right ) = \frac{1}{2} \frac{ k \; (\exp(2
\sinh^{-1} \phi^2) 
-2)^{\lambda/2 - 3/2}}{\phi^2 (2 + 2 V')^{3/2} -1}
\end{equation}
Using the known results for $V(\phi^2)$ and $k(\phi^2)$ this equation
can be integrated to give,
\begin{eqnarray}
A_1 \frac{1}{12} \left (2 \mathrm{e}^{3 \sinh^{-1}\phi^2} -6 \mathrm{e}^{
\sinh^{-1}\phi^2} + 3 \mathrm{e}^{- \sinh^{-1}\phi^2} \right ) \left (
\mathrm{e}^{2 \sinh^{-1}\phi^2} - 2 \right )^{-\lambda/2}\nonumber  \\
+A_2 \left (
\mathrm{e}^{2 \sinh^{-1}\phi^2} - 2 \right )^{3/2-\lambda/2}
\label{e:vsol}
\end{eqnarray}

From equation~(\ref{e:ksol}) we note that for $k(0)$ to be real we
need $\lambda= - 2 m$, where $m$ is an integer, or have $A_1=0$. However,
we also note 
that to avoid a singularity near the origin we must have $m$ 
positive or zero ( as $\exp(2 \sinh^{-1} \phi^2) -2 < 0 $ for  $\phi^2 < \sinh(
\log \sqrt{2})$). On the other hand, considering equation~(\ref{e:vsol})
we see that we must also have $3/2-\lambda/2$ to be a positive integer,
or have $A_2=0$.
We see that we must have two mutually exclusive sets of solutions:-
\begin{itemize}
\item  a set of solutions  with eigenvalues $\lambda=3-2m$, for a positive
integer $m$, and $A_1=0$. These solutions have
$\lambda=3,1,-1,\ldots$ and satisfy
\begin{eqnarray}
v(\phi^2) & = & A_2 \left ( \mathrm{e}^{2 \sinh^{-1}\phi^2} - 2
\right )^{3/2-\lambda/2} \\
k(\phi^2) & = & 0
\end{eqnarray}
\item a set of solutions with $A_2=0$ and $\lambda=- 2 m$, for a
positive integer $m=0,1,2,3$. These solutions have
$\lambda=0,-2,-4,\ldots$ and satisfy,
\begin{eqnarray}
v(\phi^2) & = &A_1 \frac{1}{12} \left (2 \mathrm{e}^{3 \sinh^{-1}\phi^2} -6 \mathrm{e}^{\sinh^{-1}\phi^2} + 3 \mathrm{e}^{- \sinh^{-1}\phi^2} \right ) \nn\\ 
& & \ \ \ \ \ \times \left (\mathrm{e}^{2 \sinh^{-1}\phi^2} - 2 \right )^{-\lambda/2}  \\
k(\phi^2) & = & A_1 \left ( \mathrm{e}^{ 2 \sinh^{-1} \phi^2} -2
\right)^{-\lambda/2} 
\end{eqnarray}
\end{itemize}

\begin{figure}[H]
\centering
\subfigure[$v(\phi^2)$ for the first type of solution ]{\epsfig{figure=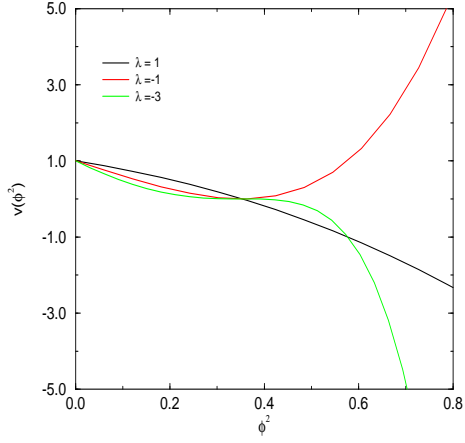,width=0.46
\textwidth,height=0.46\textwidth}}   
\subfigure[$v(\phi^2)$ for the second type of solution]{\epsfig{figure=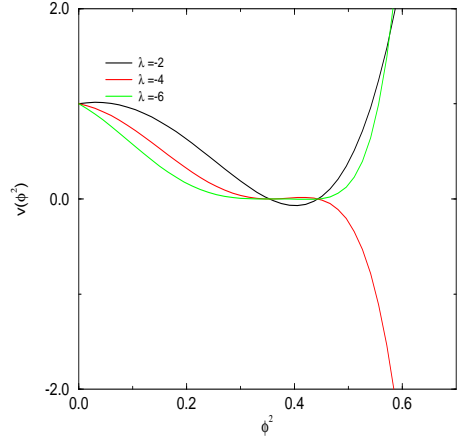,width=0.46
\textwidth,height=0.46\textwidth}}   
\subfigure[$k(\phi^2)$ for the second type of
solution]{\epsfig{figure=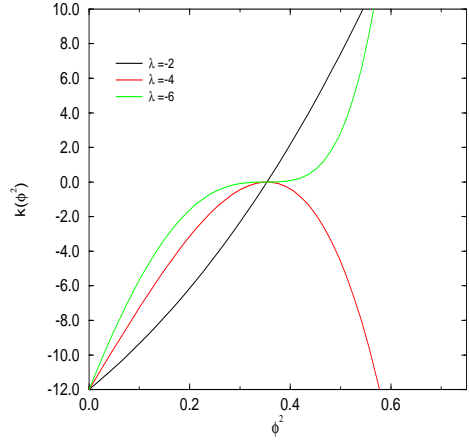,width=0.46\textwidth, 
height=0.46\textwidth}} 
\caption{The $v$ and $k$  components of the relevant, leading
irrelevant and sub-leading irrelevant
operators at $N=\infty$.}
\label{solinfin}
\end{figure}

The first set of results reproduce those first found by Wegner and
Houghton~\cite{a:WandH}, whilst the second set correspond to corrections
to scaling coming from the $(\partial_{\mu} \phi^a)^2$ part
of the action. We plot some of these functions for various values of
$\lambda$ in figure~(\ref{solinfin}). It
should be noted that all the solutions meet at the point
$(0,1/2\sqrt{2})$. This behaviour is related to the singularity
structure of the equations and is not of any particular importance.

\pagebreak

\section{Exact Results at $N=-2$}

The fact that the case of $N=-2$ is also an exactly solvable case was
noted long ago~\cite{a:balmin2,a:min2}. The initial hope was to be
able to combine the known 
results at $N=\infty$ with the new results at $N=-2$, and use
Pad\'{e} approximants to gain information about the physically
interesting case of $N=0,1,2,3$~\cite{a:balmin2,a:min2,a:fishmin2}. It
was found that when $N=-2$ the 
exponents reduce to those of the Gaussian model, $\eta=0$, $\nu=1/2$.
However there is one large difference: the $N=-2$ case describes the
\emph{critical} Gaussian model, as opposed to the \emph{tricritical}
Gaussian model~\cite{a:riedmin2}.  Therefore, we only expect to find one
relevant eigenvalue 
instead of the two relevant eigenvalues that we found  in
section~\ref{gaussian}. 

This phenomena can be understood as follows. As pointed out by
Fisher~\cite{a:min2}, 
at $N=-2$ only the quadratic part of the action plays a role, ie the
physics is totally independent of any quartic coupling etc. At the
tricritical fixed point we see two relevant operators corresponding to
a mass term with dimension two and a $\phi^4$ term with
dimension one. However, as we are now reduced to only having the
quadratic part of the  action playing a r\^ole, we only can expect  to
find one relevant operator.

We will show how the derivative expansion reproduces these results at
the leading order of the approximation before briefly discussing what
happens when we consider the second order terms.

\begin{figure}[th]
\centering
\subfigure[The fixed point potential, $V(\phi^2)$
]{\epsfig{figure=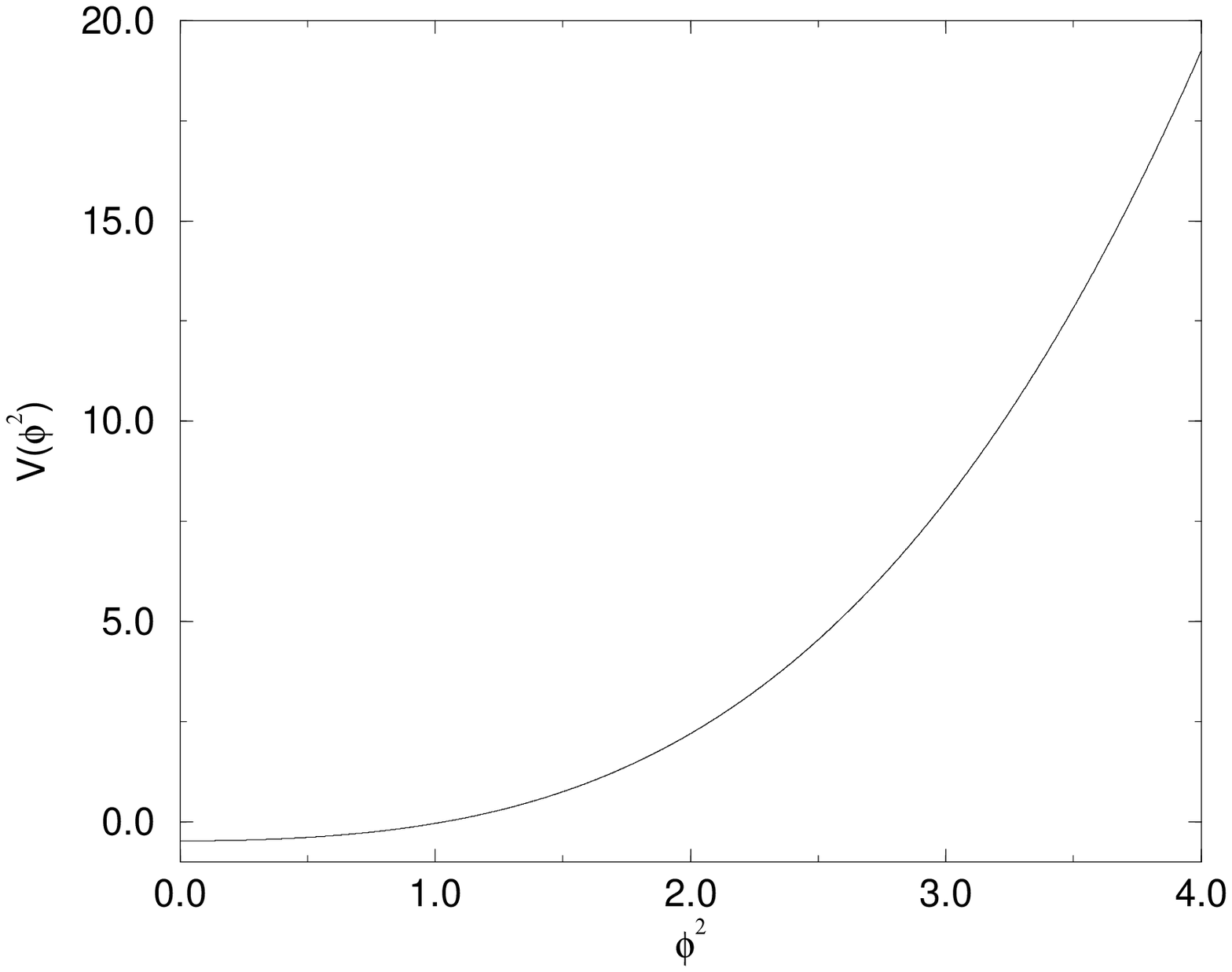,width=0.46 
\textwidth,height=0.46\textwidth}}   
\subfigure[The perturbation $v(\phi^2)$ for
$\lambda=2$.]{\epsfig{figure=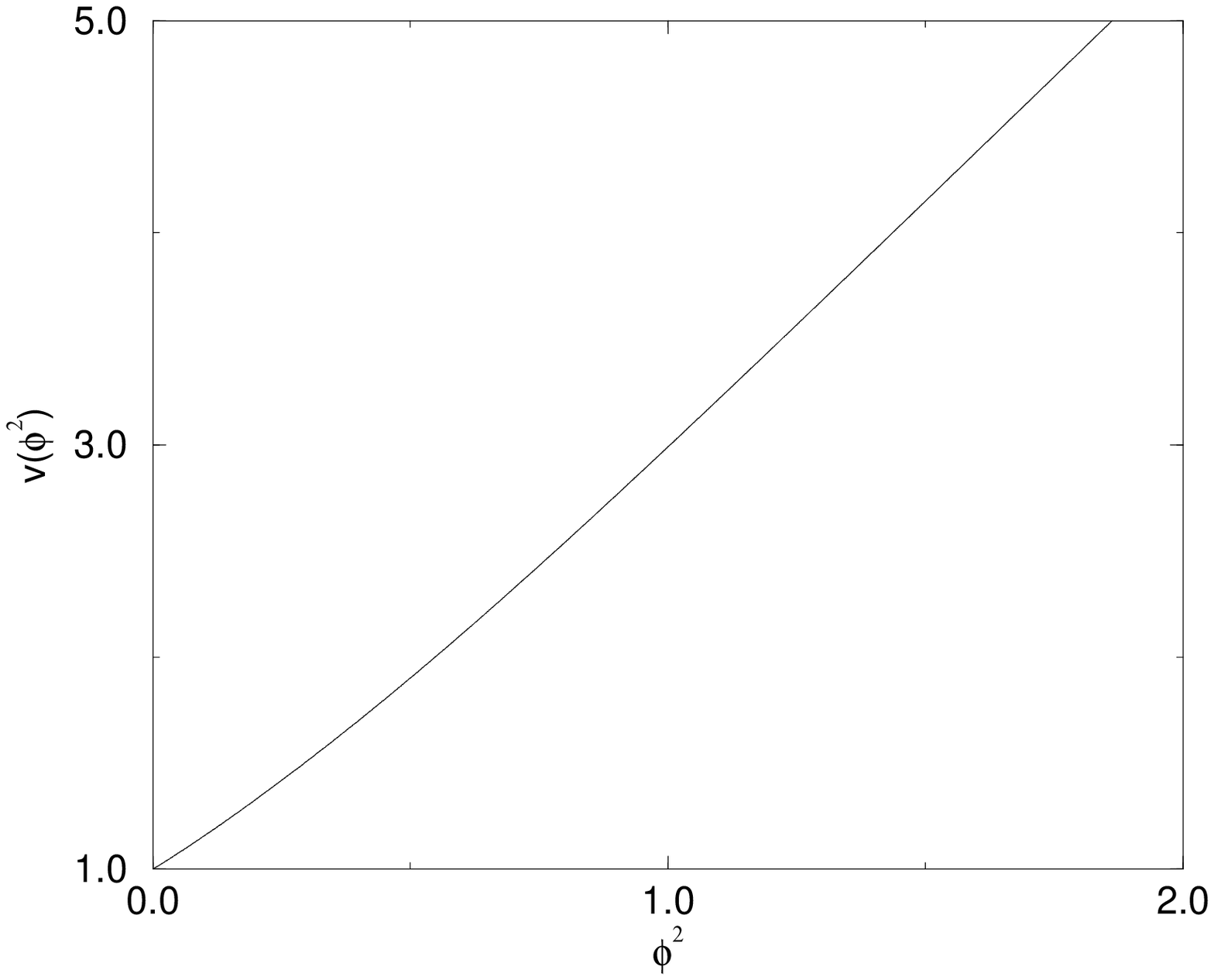,width=0.46  
\textwidth,height=0.46\textwidth}}   
\caption{The fixed point solution for the critical-Gaussian fixed
point, together with the only found perturbation, at $N=-2$.}
\label{minus2}
\end{figure}

Our starting point will be the equation for the potential at leading
order,
\begin{equation}
\phi^2 V'-3V=- \left \{ \frac{N-1}{\sqrt{2+2V'}} + \frac{1}{\sqrt{2+2V'+4
\phi^2 V''}} \right \}
\label{e:veqn-2}
\end{equation}
It turns out that we need only consider what happens at the origin. If
we differentiate the equation and set $\phi^2=0$ we get,
\begin{equation}
- \frac{N+2}{(2+2V'(0))^{3/2}}V''(0) - 2 V'(0)
\end{equation}
We see that at $N=-2$ this equation simplifies to $-2V'(0)=0$, so we
must have $V'(0)=0$. Considering equation~(\ref{e:veqn-2}) at $\phi^2=0$
we see that we must then have $V(0)=-\sqrt{2}/3$.
From this we see an interesting departure of the behaviour at the
origin from the cases $N=1,2,3,\ldots$, etc. Beforehand the boundary
condition at the origin used to leave $V(0)$ and $V'(0)$ as free
parameters, constrained by the given boundary condition. However we
now have $V(0)$ and $V'(0)$ exactly determined, with $V''(0)$
decoupling and becoming a free parameter. The question is whether we
can find   a non-trivial solution: after all if $V''(0)=0$ then series
methods show that we only reproduce the tricritical Gaussian solution,
$V(\phi^2)=N/(3\sqrt{2})$.  All analytic attempts to find such
a solution failed, but if we use $V(0)=-\sqrt{2}/3$ and $V'(0)=0$ as
boundary conditions then numerical methods quickly find a non-trivial
solution with $V''(0) \ne 0$ satisfying the required asymptotic
conditions at large $\phi^2$ (cf equation~(\ref{e:loasy})). The results
of this search are shown in 
figure~(\ref{minus2}a).

We now consider what happens to $v(\phi^2)$. We consider the linearized
equation,
\begin{equation}
(\lambda-3)v +\phi^2 v'= \frac{(N-1)v'}{(2+2V')^{3/2}} +
\frac{v'+\phi^2 v''}{(2 + 2 V' + 4 \phi^2 V'')^{3/2}}
\end{equation}
If we differentiate this equation and set $\phi^2=0$ we have,
\begin{equation}
- \frac{(N+2)v''}{(2 V'+2)^{3/2}} + 3 \frac{(N+2) v' V''}{(2 V'
+2)^{5/2}} + v' (\lambda-2)
\end{equation}
Hence we see that if we set $N=-2$ then the terms involving
$V''$ and $v''$ decouple and we are left with,
\begin{equation}
v' (\lambda-2)=0
\end{equation}

Therefore, we either have $\lambda=2$ or we must take $v'(0)=0$. Hence
we see that we find the first eigenvalue without re-course to any
analytic or numerical method, and so $\nu=1/2$.  Again we check
numerically that we find a solution that isn't just the tricritical
Gaussian solution. This is shown in figure~(\ref{minus2}b). If we take
$v'(0)=0$,
then setting $\phi^2=0$ in the linearized equation at $N=-2$,
shows that  $v(0)=0$. We now have a second order equation with two boundary
conditions set at the origin and hence we expect to find a unique
result. By inspection we see that this solution must be $v(\phi^2)=0$.

As pointed out by Fisher~\cite{a:min2} we should also expect to find
exact solutions for 
$N=-4,-6-8,\ldots$. We  find this to be the case. eg consider $N=-4$,
following a similar method to above, except this time  differentiating
twice, yields,
\begin{equation}
2 \sqrt{2} \frac{(\lambda-1)(\lambda-2) v'(0)}{N+2}
\end{equation}

From this we see that we must have $\lambda=2$ or $\lambda=1$ or have
$v'(0)=0$. As before the case of $v'(0)=0$ yields no further results.
This result should be expected  because, as Fisher pointed
out~\cite{a:min2}, at 
$N=-4$ interactions up to degree $\phi^4$ now play a r\^{o}le.
The program can be carried out for any negative even integer to yield
similar results, so for $N=-6$ we have $\lambda=2,1,0$ and
interactions up to degree $\phi^6$ become involved.

When we come to perform the analysis at the second order the equations
are vastly more complicated. Due to the fact that even modern symbolic
manipulation packages cannot handle the equations efficiently we
report no further results.

\include{conc} 
\appendix
\chapter{Numerical methods}

In this section we will outline the methods used to solve our
equations. We begin with the numerical methods used before discussing
particular problems that we encountered.

\section{Numerical Methods}

The equations form a two point boundary value problem. There are two
such methods of solving such equations: shooting~\cite{b:stoer,b:nr}
and relaxation~\cite{b:nr}. Both 
methods were used at some point in the research.

\subsection{Shooting}

Suppose we have a set $N$ of differential equations,
\begin{equation}
\frac{d y_i(x)}{dx}=g_i(x,y_1,\ldots,y_N) \ \ \ \ \ \ \ \ i=1,2,\ldots,N
\label{e:bcprob}
\end{equation}
with two boundary conditions. At the first boundary, $x=x_1$ the
solution is supposed to satisfy,
\begin{equation}
B_j(x_1,y_1,\ldots,y_N)=0   \ \ \ \ \ \ \ \ j=1,\ldots,n_1
\label{e:bc1}
\end{equation}
whilst at the second, $x=x_2$ it is supposed to satisfy,
\begin{equation}
\tilde{B}_j(x_2,y_1,\ldots,y_N)=0 \ \ \ \ \ \ \ \ j=1,\ldots,n_2
\label{e:bc2}
\end{equation}
with $n_1+n_2=N$.

In the shooting method we choose a set of values for all the dependent
values at the first boundary. These values  are chosen to be
consistent with the boundary 
condition~(\ref{e:bc1}), but are otherwise free to depend upon  a
set of free parameters, the values of which we guess. Using the values
of the dependent variables as initial conditions we integrate
equation~(\ref{e:bcprob}) out to the second boundary, using a standard
ordinary differential equation integrator. Of course we will usually
find that there is a discrepancy in how well the values at the second
boundary fit the boundary condition there. How well the  values at the
second boundary condition fit the boundary condition depend upon the
initial choice of the dependent variables, so we now have a problem of
finding what initial values of the dependent variables fit the second
boundary condition the best.  Hence, the problem has been reduced to a
multi-dimensional  root finding problem which can be solved using any
of the standard methods  of numerical analysis (eg Newton-Raphson
etc).

\subsection{Relaxation}

In the relaxation method the equations are replaced by a finite
difference mesh that spans the range that we are interested in. An
initial guess is then supplied: this guess need not satisfy the
equation or even the the boundary conditions. An iteration process
known as \emph{relaxation} is then performed to bring the equations
into closer agreement with the finite difference equations and the
boundary conditions. This again boils down to solving a
multi-dimensional set of non-linear equations using some numerical
method.

Relaxation is particularly suited to equations and boundary conditions
involving complicated expressions which can't be solved in closed form.
ie there is no need to explicitly write the equation in the form
$dy/dx=f(x,y)$. It is also the best method when we need to find
solutions that depend upon some parameter, such as the value of $N$ in
our $O(N)$ symmetry. Once we have found a solution for one value of
$N$ we can use this solution as an initial guess for a close value of
$N$. Given the nature of our problem it is not surprising that
relaxation turns out to be the principal method used during the research.

\section{The Boundary Conditions}

Before discussing the solutions of the equations  it is
necessary to discuss the equations and the boundary conditions
supplied to them.  

We wish to formulate the equations as a set of non-linear, coupled
second order differential equations. Noting that the equation relating
to the $Z$ part of the action contains powers of $V'''$ (cf appendix B) it is
necessary to differentiate the $V$ equation and find an expression for
$V'''$ in terms of $V,V',V'',K,K',Z$ and $Z'$. Similarly, when we come
to look at the boundary condition relating to the $Z$ equation we see
that it contains powers of $V''$.  Again we differentiated the $V$
equation  to find an expression for $V''(0)$ in terms of $V,V',K$ and
$Z$. Similar transformations where made when we determined the
equations for the perturbations.

The other set of boundary conditions seem to be  more problematic.  We
need the solutions to exist for all $\phi^2$. We know that if the
solutions exist for all $\phi^2$ then we can work out how the
solutions behave at large $\phi^2$. This is what provides the second
boundary condition. We choose a value of $\phi^2$ such that the
asymptotic expressions (shown in appendix B) become a good enough
approximation to the solutions. We then use the asymptotic expressions
for $V,K$ and $Z$ to provide boundary conditions for $V,V',K,K',Z$ and
$Z'$. The value of $\phi^2$ chosen, which we call $\phi^2_{\mathrm{asy}}$, must be
chosen large enough to ensure that the solutions satisfy the asymptotic
conditions sufficiently well, but not so large that numerical
instability prevents us from obtaining a solution. There is a certain
amount of trial and error in deciding where to set
$\phi^2_{\mathrm{asy}}$: we cannot 
really decide where to set it until we know something about the
solution. The solutions were also checked to be stable against
reasonable changes in the value of $\phi^2_{\mathrm{asy}}$.

In fact it is necessary to develop the asymptotic expressions to
beyond the leading order in order to find a solution. eg consider
equation~(\ref{e:lo}), if we just use
the leading order results, where the solutions behave according to
their scaling dimension, then we see that, as we set $\phi^2_{\mathrm{asy}}$ at a
finite value,  then $V''(\phi^2)$ would be left undetermined. Hence,
we are forced to expand the asymptotic expressions to beyond leading order. A
similar result holds for $K(\phi^2)$. However, when we consider
$Z(\phi^2)$, we see that we need to take the asymptotic expressions to next to next
to leading order. This is because when we calculate the correction
to the leading order asymptotic behaviour  it does not involve $Z''$,
hence leaving $Z''$ undetermined. In order to ensure that $Z''$ gets
determined, we need to extend the  asymptotic expression to one further order.

When we come to consider the perturbed equations it is sufficient to
use the leading order asymptotic expression.  As a check we
calculated the corrections to  the leading order behaviour and
also used these. This made no difference to the  values quoted.

\section{The Solution of the Equations}

Having determined the boundary conditions we can discuss how the
numerical solution of the equations was performed. The solution of the
equations is complicated by the fact that they are
stiff~\cite{b:stoer,b:nr}. This arises because small perturbations
from the true solutions lead to a singular behaviour. Relaxation is
particularly adept at handling stiff problems. Before discussing these
issues we briefly discuss the problems posed by the integrals in the
equations. 

\subsection{The Integrals $I_{a,b,c}$}

The integrals $I_{a,b,c}$ play an important part in the solution.
Defined by,
\begin{displaymath}
I_{a,b,c}=\int_0^{\infty} \frac{q^{2a} \, dq}{\left ( 1+ (2 V' + 4 \phi^2 V'')q^2
+(K+\phi^2 Z) q^4 \right )^b \left (1 + 2 V' q^2 + K q^4\right )^c}
\end{displaymath}
it is possible to find a general analytic result for them. 
This proves not to be a sensible thing to do. The integrals are
generally small numbers, but in the analytic expressions these small
numbers arise from the 
cancellation  of two large numbers. Roundoff errors start to play a
large part in determining their value using the analytic expression.
Even performing the calculations using higher precision Fortran
routines could not provide accuracy beyond four significant figures.
This level of accuracy was totally insufficient and so numerical
methods were used.

We used an adaptive integrator from the NAG libraries. The problem
was split  into two parts: a part from zero up to some finite
value, and the remaining bit. To avoid roundoff errors we used an
expression arising from asymptotic analysis to calculate the second
part of the integral, and only used the numerical routines in the part
of the real 
line
from the origin. The point were the crossover took place was set to
be as small as possible (to avoid numeric round off in the numerical
integration), whilst still allowing the asymptotic expression to be accurate.

\subsection{The  $Z$ equation at $N=\infty$}

The equation that determines $Z$ at $N=\infty$,
equation~(\ref{e:Zifin}) is a first order
non-linear equation. Being first order we only have one boundary
condition, which is that  the solution exists for all $\phi^2$ and has the
required asymptotic behaviour. For this problem we used the shooting
method. An initial value of $Z$ was used to integrate the equation out
towards the second boundary using an eight point Runge-Kutta
integrator. In general the initial value of $Z$ would be wrong: if it
was too large then the solution would tend to infinity at some finite
$\phi^2$; if it was too small the solution would tend to minus
infinity at some finite value of $\phi^2$. Using a binary chop we
could home into the correct value of $Z(0)$ and find the solution that
was defined for all $\phi^2$.

It was particularly difficult to get past the point
$\phi^2=1/2\sqrt{2}$. This is because at this point the right hand
side of the equations has a $0/0$ type behaviour. ie if
$Z'=n(\phi^2,Z)/d(\phi^2,Z)$, then as $\phi^2 \rightarrow 1/2/\sqrt{2}$
we see that $n(\phi^2,Z) \rightarrow 0$ and $d(\phi^2,Z) \rightarrow
0$, for the true solution. There are general methods of dealing with
such singular points~\cite{b:nr}, but our equations were  sufficiently well
behaved to avoid needing to use them.

\subsection{The Numerical Solution of the Flow Equations for
General $N$}

The numerical solution of the leading order equations proved to be
easy. It was only necessary to provide an initial guess and the
relaxation routines quickly found the solution. 

When we come to consider the second order equations we see that there
is vast increase in the complexity of  the equations. The numerical solution of the
second order equation is made difficult by the large number of
integrals that need to be calculated during each iteration. This places
a huge computational overhead on the numerical calculations.  It was
soon noted that the computational power required could not be provided
by normal machines. Parallel Fortran code,  making use of MPI~\cite{a:MPI}, was
developed to run on a 
sixteen node IBM SP2 system. The use of a parallel machine
considerably speeded the calculation up, with run times typically
between ten minutes to one hour (before developing parallel code the
programs could take up to one week to run).
Using the $N=\infty$ solutions as an initial guess allowed us to solve
the general $N$ cases using relaxation.

The shooting method was tried, but this proved to be  unsuitable. This
is because to use shooting effectively we need to be able to reach the
other boundary or 
have a particularly simple set of equations (eg the $N=\infty$ equations). The
singularity structure of the equations makes it impossible to shoot
from one side to the other and hence makes the more complicated
problem unsuited to shooting.

The main difficultly in solving the second order equation was removing
the instabilities in $Z$ near the origin. These manifested themselves
in the form of sharp spikes in $Z'$ near $\phi^2=0$. These were
removed by writing $Z(\phi^2)=s(\phi^2) \tilde{Z}(\phi^2)$, where
$s(\phi^2)$ is a known scaling function that makes $\tilde{Z}(\phi^2)$
as flat as possible. This greatly speeded up the calculations  and
removed the spikes in the solutions. 

When we came to consider the equations for the perturbations it was
realized that we could calculate the parts that relied upon the fixed
point solutions and store them in a file, thus saving a large
computational overhead. It was important that these functions, which
are the functions that 
multiplied $v,v',v'',k$ etc in each of the linearized equations, are
well determined. The size of the $z$
equation was such that small errors in the fixed point solutions made
it difficult to determine these functions accurately, for large $\phi^2$ .
This problem was solved by noting that we could determine these
functions by using the known asymptotic expressions for $V,K$ and $Z$
and then matching them onto the ones calculated from the fixed point
solutions.  It was checked that the eigenvalues were stable against
reasonable perturbations in where $\phi^2_{\mathrm{asy}}$ was set and the number of
points in the mesh.

\chapter{The Equations at Second Order}

In this appendix we show some of the expressions thought to too long
to appear in the main body of the text. We include the equations for
$K$ and $Z$ as well as the expressions for the asymptotics to next to
leading order.

\section{The Second Order Equations}

In the  the expression below $I_{a,b,c}$ represents the following
integral,
\begin{displaymath}
I_{a,b,c}=\int_0^{\infty} \frac{q^{2a} dq}{\left ( 1+ (2 V' + 4 \phi^2 V'')q^2
+(K+\phi^2 Z) q^4 \right )^b \left (1 + 2 V' q^2 + K q^4\right )^c}
\end{displaymath}

These integrals are not evaluated as the equations become very large. At 
present there are about three hundred terms in the equations. Evaluating
the integrals raises this to between twenty five and thirty thousand terms.
\newpage
The $K$ equation is given by
\begin{eqnarray*} 
\lefteqn{{\it (1+\eta)  \phi^2 K' + \eta K= } } \\ & & \\ && \frac{(\,4 - {
\eta}\,)}{\pi} \left[ {\vrule 
height0.80em width0em depth0.80em} \right. \! \! {\it I}_{4, 2,
1}\, \left( \! \,{\displaystyle \frac {4}{3}}\,
{\displaystyle \frac {{\it K'}^{2}\,{ \phi}}{{N}}} +
{\displaystyle \frac {4}{3}}\,{\displaystyle \frac {{\it K'}\,{Z}
\,{ \phi}}{{N}}}\, \!  \right) \mbox{} + {\it I}_{4, 1, 2}\, \left( \! \,
{\displaystyle \frac {4}{3}}\,{\displaystyle \frac {{\it K'}^{2}
\,{ \phi}}{{N}}} - 4\,{\displaystyle \frac {{\it K'}\,{Z}\,{ \phi
}}{{N}}} + {\displaystyle \frac {8}{3}}\,{\displaystyle \frac {{Z
}^{2}\,{ \phi}}{{N}}}\, \!  \right)  \\
 & & \mbox{} + {\displaystyle \frac {16}{3}}\,{\displaystyle
\frac {{\it I}_{3, 2, 1}\,{\it V''}\,{\it K'}\,{ \phi
}}{{N}}} + {\it I}_{3, 1, 2,}\, \left( \! \, - \,
{\displaystyle \frac {16}{3}}\,{\displaystyle \frac {{\it V''}\,
{\it K'}\,{ \phi}}{{N}}} + {\displaystyle \frac {16}{3}}\,
{\displaystyle \frac {{\it V''}\,{Z}\,{ \phi}}{{N}}}\, \! 
 \right)  \\
 & & \mbox{} + {\it I}_{2, 0, 2}\, \left( \! \,
{\displaystyle \frac {{\it K'}}{{N}}} - {\displaystyle \frac {{Z}
}{{N}}} - {\it K'}\, \!  \right)  + {\it I}_{2, 2, 0}
\, \left( \! \, - \,{\displaystyle \frac {{\it K'}}{{N}}} - 2\,  
{\displaystyle \frac {{\it K''}\,{ \phi}}{{N}}}\, \!  \right)    
 \! \! \left. {\vrule height0.80em width0em depth0.80em} \right] \\
\end{eqnarray*}

The longer $Z$ equation is given by,
\begin{eqnarray*}
\lefteqn{(\,1 + { \eta}\,)\,{ \phi^2}\,{\it Z'} + (\,1 + 2\,{ \eta}
\,)\,{Z} =} \\ & & \\ & &  {\displaystyle {\frac{(\,4 - { \eta}\,)}{\pi}}
\left[ {\vrule  
height0.86em width0em depth0.86em} \right. \! \!} 
  {{\it I}_{4, 0, 5}}\, \left( \! \,{\displaystyle \frac {
256}{3}}\,{\displaystyle \frac {{\it V''}^{2}\,{K}}{{N}}} - 
{\displaystyle \frac {32}{3}}\,{\displaystyle \frac {{\it K'}^{2}
}{{N}}} - {\displaystyle \frac {256}{3}}\,{\it V''}^{2}\,{K} + 
{\displaystyle \frac {32}{3}}\,{\it K'}^{2}\, \!  \right)  \\
 & & \\ & & + {{\it I}_{6, 1, 4}}\, \left( \! \,
{\displaystyle \frac {128}{3}}\,{\displaystyle \frac {{\it V''}^{
2}\,{K}^{2}}{{N}}} - {\displaystyle \frac {64}{3}}\,
{\displaystyle \frac {{K}\,{\it K'}^{2}}{{N}}} - {\displaystyle 
\frac {128}{3}}\,{\it V''}^{2}\,{K}^{2} + {\displaystyle \frac {
64}{3}}\,{K}\,{\it K'}^{2}\, \!  \right)    \\ 
 & & \\ && + {{\it I}_{4, 2, 3}} \left( {\vrule 
height0.80em width0em depth0.80em} \right. \! \! {\displaystyle 
\frac {256}{3}}\,{\displaystyle \frac {{\it V''}^{2}\,{K}}{{N}}}
 - {\displaystyle \frac {512}{3}}\,{\displaystyle \frac {{\it V''
}^{2}\,{\it K'}\,{ \phi^2}}{{N}}} - {\displaystyle \frac {256}{3}}
\,{\displaystyle \frac {{\it V''}^{2}\,{Z}\,{ \phi^2}}{{N}}} \\
 & &  + {\displaystyle \frac {256}{3}}\,{\displaystyle 
\frac {{\it V''}\,{\it V'''}\,{K}\,{ \phi^2}}{{N}}} - 
{\displaystyle \frac {16}{3}}\,{\displaystyle \frac {{\it K'}\,{Z
}}{{N}}} - {\displaystyle \frac {16}{3}}\,{\displaystyle \frac {
{\it K'}\,{\it Z'}\,{ \phi^2}}{{N}}} - {\displaystyle \frac {256}{3}}\,{\it V''}^{2}\,{K} \\
 & &  + {\displaystyle \frac {512}{3}}\,{\it V''}^{2}\,
{\it K'}\,{ \phi^2} + {\displaystyle \frac {256}{3}}\,{\it V''}^{2}
\,{Z}\,{ \phi^2} - {\displaystyle \frac {256}{3}}\,{\it V''}\,{\it 
V'''}\,{K}\,{ \phi^2} + {\displaystyle \frac {16}{3}}\,{\it K'}\,{Z
} \\
 & &  + {\displaystyle \frac {16}{3}}\,{\it K'}\,{\it Z'}
\,{ \phi^2} \! \! \left. {\vrule height0.80em width0em depth0.80em}
 \right) \\ & & \\ && + {{\it I}_{4, 3, 2}} \left( {\vrule 
height0.86em width0em depth0.86em} \right. \! \!  - \,
{\displaystyle \frac {1024}{3}}\,{\displaystyle \frac {{\it V''}
^{4}\,{ (\phi^2)}^{2}}{{N}}} + {\displaystyle \frac {128}{3}}\,
{\displaystyle \frac {{\it V''}^{2}\,{K}}{{N}}} \\
 & &  + {\displaystyle \frac {512}{3}}\,{\displaystyle 
\frac {{\it V''}^{2}\,{\it K'}\,{ \phi^2}}{{N}}} + {\displaystyle 
\frac {256}{3}}\,{\displaystyle \frac {{\it V''}^{2}\,{Z}\,{ \phi^2
}}{{N}}} - {\displaystyle \frac {16}{3}}\,{\displaystyle \frac {
{\it K'}^{2}}{{N}}} + {\displaystyle \frac {1024}{3}}\,{\it V''}
^{4}\,{ (\phi^2)}^{2} \\
 & &  - {\displaystyle \frac {128}{3}}\,{\it V''}^{2}\,{K}
 - {\displaystyle \frac {512}{3}}\,{\it V''}^{2}\,{\it K'}\,{ 
\phi^2} - {\displaystyle \frac {256}{3}}\,{\it V''}^{2}\,{Z}\,{ 
\phi^2} + {\displaystyle \frac {16}{3}}\,{\it K'}^{2} \! \! \left. 
{\vrule height0.86em width0em depth0.86em} \right)  \\
 & & \\ & & + {{\it I}_{5, 1, 4}}\, \left( \! \, - \,
{\displaystyle \frac {256}{3}}\,{\displaystyle \frac {{\it V''}\,
{K}\,{\it K'}}{{N}}} + {\displaystyle \frac {256}{3}}\,{\it V''}
\,{K}\,{\it K'}\, \!  \right) \\ & & \\ &&  + {{\it I}_{4, 2, 1}} 
   \left( \! \,{\displaystyle \frac {20}{3}}\,{\displaystyle 
\frac {{\it K'}^{2}}{{N}}} - {\displaystyle \frac {8}{3}}\,
{\displaystyle \frac {{\it K'}\,{Z}}{{N}}} - {\displaystyle 
\frac {4}{3}}\,{\displaystyle \frac {{\it K'}\,{\it Z'}\,{ \phi^2}
}{{N}}} + {\displaystyle \frac {{Z}^{2}}{{N}}} - 8\,{\it K'}^{2}
 + {\displaystyle \frac {4}{3}}\,{\it K'}\,{Z} + {\displaystyle 
\frac {4}{3}}\,{\it K'}\,{\it Z'}\,{ \phi^2}\, \!  \right)  \\
 & & \\ & & + {{\it I}_{6, 2, 2}} \left( {\vrule 
height0.80em width0em depth0.80em} \right. \! \! {\displaystyle 
\frac {56}{3}}\,{\displaystyle \frac {{K}\,{\it K'}^{2}}{{N}}} - 
{\displaystyle \frac {4}{3}}\,{\displaystyle \frac {{K}\,{\it K'}
\,{Z}}{{N}}} - {\displaystyle \frac {4}{3}}\,{\displaystyle 
\frac {{K}\,{\it K'}\,{\it Z'}\,{ \phi^2}}{{N}}} + {\displaystyle 
\frac {104}{3}}\,{\displaystyle \frac {{\it K'}^{2}\,{Z}\,{ \phi^2}
}{{N}}} \\
 & &  - {\displaystyle \frac {56}{3}}\,{K}\,{\it K'}^{2}
 + {\displaystyle \frac {4}{3}}\,{K}\,{\it K'}\,{Z} + 
{\displaystyle \frac {4}{3}}\,{K}\,{\it K'}\,{\it Z'}\,{ \phi^2} - 
{\displaystyle \frac {104}{3}}\,{\it K'}^{2}\,{Z}\,{ \phi^2} \! 
\! \left. {\vrule height0.80em width0em depth0.80em} \right) 
\\ & & + {{\it I}_{4, 4, 0}} \left( {\vrule 
height0.86em width0em depth0.86em} \right. \! \!  - 144\,{\displaystyle \frac {{\it V''}^{2}\,{K}}{{N}}} - 
144\,{\displaystyle \frac {{\it V''}^{2}\,{Z}\,{ \phi^2}}{{N}}} - 
192\,{\displaystyle \frac {{\it V''}\,{\it V'''}\,{K}\,{ \phi^2}}{{
N}}} - 192\,{\displaystyle \frac {{\it V''}\,{\it V'''}\,{Z}\,{ 
(\phi^2)}^{2}}{{N}}} \\
 & &  - 64\,{\displaystyle \frac {{\it V'''}^{2}\,{K}\,{ 
(\phi^2)}^{2}}{{N}}} - 64\,{\displaystyle \frac {{\it V'''}^{2}\,{Z}
\,{ (\phi^2)}^{3}}{{N}}} + 4\,{\displaystyle \frac {{\it K'}^{2}}{{N}
}} + 8\,{\displaystyle \frac {{\it K'}\,{Z}}{{N}}} + 8\,
{\displaystyle \frac {{\it K'}\,{\it Z'}\,{ \phi^2}}{{N}}} \\
 & &  + 4\,{\displaystyle \frac {{Z}^{2}}{{N}}} + 8\,
{\displaystyle \frac {{Z}\,{\it Z'}\,{ \phi^2}}{{N}}} + 4\,
{\displaystyle \frac {{\it Z'}^{2}\,{ (\phi^2)}^{2}}{{N}}} \! 
\! \left. {\vrule height0.86em width0em depth0.86em} \right)  \\
 & & \\ & & + {{\it I}_{2, 5, 0}}\, \left( \! \,192\,
{\displaystyle \frac {{\it V''}^{2}}{{N}}} + 256\,{\displaystyle 
\frac {{\it V''}\,{\it V'''}\,{ \phi^2}}{{N}}} + {\displaystyle 
\frac {256}{3}}\,{\displaystyle \frac {{\it V'''}^{2}\,{ \phi^2}^{2
}}{{N}}}\, \!  \right)  \\
 & & \\ & & + {{\it I}_{2, 3, 2}}\, \left( \! \, - \,
{\displaystyle \frac {64}{3}}\,{\displaystyle \frac {{\it V''}^{2
}}{{N}}} + {\displaystyle \frac {64}{3}}\,{\it V''}^{2}\, \! 
 \right) \\ & & \\ &&  + {{\it I}_{2, 3, 1}}\, \left( \! \,{\displaystyle 
\frac {128}{3}}\,{\displaystyle \frac {{\it V''}^{2}}{{N}}} - 
{\displaystyle \frac {128}{3}}\,{\it V''}^{2}\, \!  \right)  \\
 & & \\ & & + {{\it I}_{4, 1, 4}}\, \left( \! \, - \,
{\displaystyle \frac {256}{3}}\,{\displaystyle \frac {{\it V''}^{
2}\,{K}}{{N}}} + {\displaystyle \frac {32}{3}}\,{\displaystyle 
\frac {{\it K'}^{2}}{{N}}} + {\displaystyle \frac {256}{3}}\,
{\it V''}^{2}\,{K} - {\displaystyle \frac {32}{3}}\,{\it K'}^{2}
\, \!  \right)  \\
 & & \\ & & + {{\it I}_{2, 0, 3}}\, \left( \! \, - \,
{\displaystyle \frac {16}{3}}\,{\displaystyle \frac {{\it V''}^{2
}}{{N}}} + {\displaystyle \frac {16}{3}}\,{\it V''}^{2}\, \! 
 \right) \\ & & \\ && + {{\it I}_{5, 2, 3}} \left( {\vrule 
height0.80em width0em depth0.80em} \right. \! \! {\displaystyle 
\frac {512}{3}}\,{\displaystyle \frac {{\it V''}^{3}\,{K}\,{ \phi^2
}}{{N}}} 
  + {\displaystyle \frac {128}{3}}\,{\displaystyle 
\frac {{\it V''}\,{K}\,{\it K'}}{{N}}} + {\displaystyle \frac {64
}{3}}\,{\displaystyle \frac {{\it V''}\,{K}\,{Z}}{{N}}} + 
{\displaystyle \frac {64}{3}}\,{\displaystyle \frac {{\it V''}\,{
K}\,{\it Z'}\,{ \phi^2}}{{N}}} \\
 & &  - {\displaystyle \frac {256}{3}}\,{\displaystyle 
\frac {{\it V''}\,{\it K'}\,{Z}\,{ \phi^2}}{{N}}} + {\displaystyle 
\frac {128}{3}}\,{\displaystyle \frac {{\it V'''}\,{K}\,{\it K'}
\,{ \phi^2}}{{N}}} - {\displaystyle \frac {512}{3}}\,{\it V''}^{3}
\,{K}\,{ \phi^2} - {\displaystyle \frac {128}{3}}\,{\it V''}\,{K}\,
{\it K'} \\
 & &  - {\displaystyle \frac {64}{3}}\,{\it V''}\,{K}\,{Z}
 - {\displaystyle \frac {64}{3}}\,{\it V''}\,{K}\,{\it Z'}\,{ 
\phi^2} + {\displaystyle \frac {128}{3}}\,{\it V''}\,{\it K'}^{2}\,
{ \phi^2} + {\displaystyle \frac {256}{3}}\,{\it V''}\,{\it K'}\,{Z
}\,{ \phi^2} \\
 & &  - {\displaystyle \frac {128}{3}}\,{\it V'''}\,{K}\,
{\it K'}\,{ \phi^2}    - {\displaystyle \frac {128}{3}}\,
{\displaystyle \frac {{\it V''}\,{\it K'}^{2}\,{ \phi^2}}{{N}}} \! \! \left. {\vrule 
height0.80em width0em depth0.80em} \right) \\ & & \\ && + {{\it I}_{
8, 0, 5}}\, \left( \! \, - \,{\displaystyle \frac {32}{3}}\,
{\displaystyle \frac {{K}^{2}\,{\it K'}^{2}}{{N}}} + 
{\displaystyle \frac {32}{3}}\,{K}^{2}\,{\it K'}^{2}
\, \! 
 \right) \\ & & \\ && + {{\it I}_{8, 5, 0}} 
   \left( {\vrule height0.86em width0em depth0.86em}
 \right. \! \! {\displaystyle \frac {16}{3}}\,{\displaystyle 
\frac {{K}^{2}\,{\it K'}^{2}}{{N}}} + {\displaystyle \frac {32}{3
}}\,{\displaystyle \frac {{K}^{2}\,{\it K'}\,{Z}}{{N}}} + 
{\displaystyle \frac {32}{3}}\,{\displaystyle \frac {{K}^{2}\,
{\it K'}\,{\it Z'}\,{ \phi^2}}{{N}}} + {\displaystyle \frac {16}{3
}}\,{\displaystyle \frac {{K}^{2}\,{Z}^{2}}{{N}}} \\
 & &  + {\displaystyle \frac {32}{3}}\,{\displaystyle 
\frac {{K}^{2}\,{Z}\,{\it Z'}\,{ \phi^2}}{{N}}} + {\displaystyle 
\frac {16}{3}}\,{\displaystyle \frac {{K}^{2}\,{\it Z'}^{2}\,{ 
(\phi^2)}^{2}}{{N}}} + {\displaystyle \frac {32}{3}}\,{\displaystyle 
\frac {{K}\,{\it K'}^{2}\,{Z}\,{ \phi^2}}{{N}}} + {\displaystyle 
\frac {64}{3}}\,{\displaystyle \frac {{K}\,{\it K'}\,{Z}^{2}\,{ 
\phi^2}}{{N}}} \\
 & &  + {\displaystyle \frac {64}{3}}\,{\displaystyle 
\frac {{K}\,{\it K'}\,{Z}\,{\it Z'}\,{ (\phi^2)}^{2}}{{N}}} + 
{\displaystyle \frac {32}{3}}\,{\displaystyle \frac {{K}\,{Z}^{3}
\,{ \phi^2}}{{N}}} + {\displaystyle \frac {64}{3}}\,{\displaystyle 
\frac {{K}\,{Z}^{2}\,{\it Z'}\,{ (\phi^2)}^{2}}{{N}}} + 
{\displaystyle \frac {32}{3}}\,{\displaystyle \frac {{K}\,{Z}\,
{\it Z'}^{2}\,{ (\phi^2)}^{3}}{{N}}} \\
 & &  + {\displaystyle \frac {16}{3}}\,{\displaystyle 
\frac {{\it K'}^{2}\,{Z}^{2}\,{ (\phi^2)}^{2}}{{N}}} + 
{\displaystyle \frac {32}{3}}\,{\displaystyle \frac {{\it K'}\,{Z
}^{3}\,{ (\phi^2)}^{2}}{{N}}} + {\displaystyle \frac {32}{3}}\,
{\displaystyle \frac {{\it K'}\,{Z}^{2}\,{\it Z'}\,{ (\phi^2)}^{3}}{{
N}}} + {\displaystyle \frac {16}{3}}\,{\displaystyle \frac {{Z}^{
4}\,{ (\phi^2)}^{2}}{{N}}} \\
 & &  + {\displaystyle \frac {32}{3}}\,{\displaystyle 
\frac {{Z}^{3}\,{\it Z'}\,{ (\phi^2)}^{3}}{{N}}} + {\displaystyle 
\frac {16}{3}}\,{\displaystyle \frac {{Z}^{2}\,{\it Z'}^{2}\,{ 
(\phi^2)}^{4}}{{N}}} \! \! \left. {\vrule 
height0.86em width0em depth0.86em} \right) \\ & & \\ &&+ {{\it I}_{
8, 3, 2}} \left( {\vrule height0.86em width0em depth0.86em}
 \right. \! \!  - {\displaystyle \frac {64}{3}}\,{\displaystyle 
\frac {{K}\,{\it K'}^{2}\,{Z}\,{ \phi^2}}{{N}}} - {\displaystyle 
\frac {64}{3}}\,{\displaystyle \frac {{\it K'}^{2}\,{Z}^{2}\,{ 
(\phi^2)}^{2}}{{N}}} + {\displaystyle \frac {16}{3}}\,{K}^{2}\,{\it 
K'}^{2} + {\displaystyle \frac {64}{3}}\,{K}\,{\it K'}^{2}\,{Z}\,
{ \phi^2}   \\
 & &  + {\displaystyle \frac {64}{3}}\,{\it K'}^{2}\,{Z}^{
2}\,{ (\phi^2)}^{2}  - \,{\displaystyle \frac {16}{3}}\,
{\displaystyle \frac {{K}^{2}\,{\it K'}^{2}}{{N}}}  \! \! \left. {\vrule 
height0.86em width0em depth0.86em} \right) \\ & & \\ && + {{\it I}_{
8, 2, 3}} \left( {\vrule height0.80em width0em depth0.80em}
 \right. \! \!  - \,{\displaystyle \frac {16}{3}}\,
{\displaystyle \frac {{K}^{2}\,{\it K'}\,{Z}}{{N}}} - 
{\displaystyle \frac {16}{3}}\,{\displaystyle \frac {{K}^{2}\,
{\it K'}\,{\it Z'}\,{ \phi^2}}{{N}}} \\
 & &  + {\displaystyle \frac {64}{3}}\,{\displaystyle 
\frac {{K}\,{\it K'}^{2}\,{Z}\,{ \phi^2}}{{N}}} + {\displaystyle 
\frac {16}{3}}\,{K}^{2}\,{\it K'}\,{Z} + {\displaystyle \frac {16
}{3}}\,{K}^{2}\,{\it K'}\,{\it Z'}\,{ \phi^2} - {\displaystyle 
\frac {64}{3}}\,{K}\,{\it K'}^{2}\,{Z}\,{ \phi^2} \! \! \left. 
{\vrule height0.80em width0em depth0.80em} \right)  \\
 & & \\ & & + {{\it I}_{8, 1, 4}}\, \left( \! \,
{\displaystyle \frac {32}{3}}\,{\displaystyle \frac {{K}^{2}\,
{\it K'}^{2}}{{N}}} - {\displaystyle \frac {32}{3}}\,{K}^{2}\,
{\it K'}^{2}\, \!  \right)  \\
 & & \\ & & + {{\it I}_{7, 0, 5}}\, \left( \! \, - \,
{\displaystyle \frac {128}{3}}\,{\displaystyle \frac {{\it V''}\,
{K}^{2}\,{\it K'}}{{N}}} + {\displaystyle \frac {128}{3}}\,{\it 
V''}\,{K}^{2}\,{\it K'}\, \!  \right) \\& & \\ &&  + {{\it I}_{7, 5, 0}}
 \left( {\vrule height0.86em width0em depth0.86em} \right. \! \! 
  64\,{\displaystyle \frac {{\it V''}\,{K}^{2}\,{\it K'}}{{N}
}} + 64\,{\displaystyle \frac {{\it V''}\,{K}^{2}\,{Z}}{{N}}} + 
64\,{\displaystyle \frac {{\it V''}\,{K}^{2}\,{\it Z'}\,{ \phi^2}}{
{N}}} + 128\,{\displaystyle \frac {{\it V''}\,{K}\,{\it K'}\,{Z}
\,{ \phi^2}}{{N}}} \\
 & &  + 128\,{\displaystyle \frac {{\it V''}\,{K}\,{Z}^{2}
\,{ \phi^2}}{{N}}} + 128\,{\displaystyle \frac {{\it V''}\,{K}\,{Z}
\,{\it Z'}\,{ (\phi^2)}^{2}}{{N}}} + 64\,{\displaystyle \frac {{\it 
V''}\,{\it K'}\,{Z}^{2}\,{ (\phi^2)}^{2}}{{N}}} \\
 & &  + 64\,{\displaystyle \frac {{\it V''}\,{Z}^{3}\,{ 
(\phi^2)}^{2}}{{N}}} + 64\,{\displaystyle \frac {{\it V''}\,{Z}^{2}\,
{\it Z'}\,{ (\phi^2)}^{3}}{{N}}} + {\displaystyle \frac {128}{3}}\,
{\displaystyle \frac {{\it V'''}\,{K}^{2}\,{\it K'}\,{ \phi^2}}{{N}
}} \\
 & &  + {\displaystyle \frac {128}{3}}\,{\displaystyle 
\frac {{\it V'''}\,{K}^{2}\,{Z}\,{ \phi^2}}{{N}}} + {\displaystyle 
\frac {128}{3}}\,{\displaystyle \frac {{\it V'''}\,{K}^{2}\,{\it 
Z'}\,{ (\phi^2)}^{2}}{{N}}} + {\displaystyle \frac {256}{3}}\,
{\displaystyle \frac {{\it V'''}\,{K}\,{\it K'}\,{Z}\,{ (\phi^2)}^{2}
}{{N}}} \\
 & &  + {\displaystyle \frac {256}{3}}\,{\displaystyle 
\frac {{\it V'''}\,{K}\,{Z}^{2}\,{ (\phi^2)}^{2}}{{N}}} + 
{\displaystyle \frac {256}{3}}\,{\displaystyle \frac {{\it V'''}
\,{K}\,{Z}\,{\it Z'}\,{ (\phi^2)}^{3}}{{N}}} + {\displaystyle \frac {
128}{3}}\,{\displaystyle \frac {{\it V'''}\,{\it K'}\,{Z}^{2}\,{ 
(\phi^2)}^{3}}{{N}}} \\
 & &  + {\displaystyle \frac {128}{3}}\,{\displaystyle 
\frac {{\it V'''}\,{Z}^{3}\,{ (\phi^2)}^{3}}{{N}}} + {\displaystyle 
\frac {128}{3}}\,{\displaystyle \frac {{\it V'''}\,{Z}^{2}\,{\it 
Z'}\,{ (\phi^2)}^{4}}{{N}}} \! \! \left. {\vrule 
height0.86em width0em depth0.86em} \right) \\ & & \\ &&  + {{\it I}_{
7, 3, 2}} \left( {\vrule height0.86em width0em depth0.86em}
 \right. \! \!  - \,{\displaystyle \frac {64}{3}}\,
{\displaystyle \frac {{\it V''}\,{K}^{2}\,{\it K'}}{{N}}} - {\displaystyle \frac {128}{3}}\,{\displaystyle 
\frac {{\it V''}\,{K}\,{\it K'}^{2}\,{ \phi^2}}{{N}}} - 
{\displaystyle \frac {256}{3}}\,{\displaystyle \frac {{\it V''}\,
{K}\,{\it K'}\,{Z}\,{ \phi^2}}{{N}}}  \\
 & &  - {\displaystyle \frac {256}{3}}\,{\displaystyle 
\frac {{\it V''}\,{\it K'}\,{Z}^{2}\,{ (\phi^2)}^{2}}{{N}}} + 
{\displaystyle \frac {64}{3}}\,{\it V''}\,{K}^{2}\,{\it K'} + 
{\displaystyle \frac {128}{3}}\,{\it V''}\,{K}\,{\it K'}^{2}\,{ 
\phi^2} \\
 & &  + {\displaystyle \frac {256}{3}}\,{\it V''}\,{K}\,
{\it K'}\,{Z}\,{ \phi^2} + {\displaystyle \frac {256}{3}}\,{\it V''
}\,{\it K'}^{2}\,{Z}\,{ (\phi^2)}^{2} + {\displaystyle \frac {256}{3
}}\,{\it V''}\,{\it K'}\,{Z}^{2}\,{ (\phi^2)}^{2} \\ 
& &  - {\displaystyle
\frac {256}{3 
}}\,{\displaystyle \frac {{\it V''}\,{\it K'}^{2}\,{Z}\,{ \phi^2}^{
2}}{{N}}}\! \! \left. 
{\vrule height0.86em width0em depth0.86em} \right)    \\
 & & \\ && + {{\it I}_{7, 2, 3}} \left( {\vrule 
height0.80em width0em depth0.80em} \right. \! \!  - \,
{\displaystyle \frac {64}{3}}\,{\displaystyle \frac {{\it V''}\,{
K}^{2}\,{\it K'}}{{N}}} - {\displaystyle \frac {32}{3}}\,
{\displaystyle \frac {{\it V''}\,{K}^{2}\,{Z}}{{N}}} - 
{\displaystyle \frac {32}{3}}\,{\displaystyle \frac {{\it V''}\,{
K}^{2}\,{\it Z'}\,{ \phi^2}}{{N}}} \\
 & &  + {\displaystyle \frac {128}{3}}\,{\displaystyle 
\frac {{\it V''}\,{K}\,{\it K'}^{2}\,{ \phi^2}}{{N}}} + 
{\displaystyle \frac {256}{3}}\,{\displaystyle \frac {{\it V''}\,
{K}\,{\it K'}\,{Z}\,{ \phi^2}}{{N}}} - {\displaystyle \frac {64}{3
}}\,{\displaystyle \frac {{\it V'''}\,{K}^{2}\,{\it K'}\,{ \phi^2}
}{{N}}} \\
 & &  + {\displaystyle \frac {64}{3}}\,{\it V''}\,{K}^{2}
\,{\it K'} + {\displaystyle \frac {32}{3}}\,{\it V''}\,{K}^{2}\,{
Z} + {\displaystyle \frac {32}{3}}\,{\it V''}\,{K}^{2}\,{\it Z'}
\,{ \phi^2} - {\displaystyle \frac {128}{3}}\,{\it V''}\,{K}\,{\it 
K'}^{2}\,{ \phi^2} \\
 & &  - {\displaystyle \frac {256}{3}}\,{\it V''}\,{K}\,
{\it K'}\,{Z}\,{ \phi^2} + {\displaystyle \frac {64}{3}}\,{\it V'''
}\,{K}^{2}\,{\it K'}\,{ \phi^2} \! \! \left. {\vrule 
height0.80em width0em depth0.80em} \right)  \\
 & & \\ & & + {{\it I}_{7, 1, 4}}\, \left( \! \,
{\displaystyle \frac {128}{3}}\,{\displaystyle \frac {{\it V''}\,
{K}^{2}\,{\it K'}}{{N}}} - {\displaystyle \frac {128}{3}}\,{\it 
V''}\,{K}^{2}\,{\it K'}\, \!  \right)  \\
 & & \\ & & + {{\it I}_{6, 0, 5}}\, \left( \! \, - \,
{\displaystyle \frac {128}{3}}\,{\displaystyle \frac {{\it V''}^{
2}\,{K}^{2}}{{N}}} + {\displaystyle \frac {64}{3}}\,
{\displaystyle \frac {{K}\,{\it K'}^{2}}{{N}}} + {\displaystyle 
\frac {128}{3}}\,{\it V''}^{2}\,{K}^{2} - {\displaystyle \frac {
64}{3}}\,{K}\,{\it K'}^{2}\, \!  \right)  \\
 & & \\ & & + {{\it I}_{6, 0, 4}}\, \left( \! \,8\,
{\displaystyle \frac {{K}\,{\it K'}^{2}}{{N}}} - 8\,{K}\,{\it K'}
^{2}\, \!  \right) \\ & & + {{\it I}_{2, 0, 4}}\, \left( \! \,32\,
{\displaystyle \frac {{\it V''}^{2}}{{N}}} - 32\,{\it V''}^{2}\,
 \!  \right)  \\
 & & \\ & & + {{\it I}_{2, 0, 2}}\, \left( \! \,
{\displaystyle \frac {{\it Z'}}{{N}}} - {\it Z'}\, \!  \right) 
\\ & & + {{\it I}_{2, 1, 3}}\, \left( \! \, - \,{\displaystyle 
\frac {112}{3}}\,{\displaystyle \frac {{\it V''}^{2}}{{N}}} + 
{\displaystyle \frac {112}{3}}\,{\it V''}^{2}\, \!  \right)  \\
 & & \\ & & + {{\it I}_{2, 1, 2}}\, \left( \! \, - \,
{\displaystyle \frac {16}{3}}\,{\displaystyle \frac {{\it V''}^{2
}}{{N}}} + {\displaystyle \frac {16}{3}}\,{\it V''}^{2}\, \! 
 \right) \\ & & \\ && + {{\it I}_{3, 2, 3}} \left( {\vrule 
height0.80em width0em depth0.80em} \right. \! \!   - {\displaystyle
\frac {64}{3}}\,{\displaystyle  
\frac {{\it V''}\,{\it K'}}{{N}}} - {\displaystyle \frac {32}{3}}
\,{\displaystyle \frac {{\it V''}\,{Z}}{{N}}} - {\displaystyle 
\frac {32}{3}}\,{\displaystyle \frac {{\it V''}\,{\it Z'}\,{ \phi^2
}}{{N}}} - {\displaystyle \frac {64}{3}}\,{\displaystyle \frac {
{\it V'''}\,{\it K'}\,{ \phi^2}}{{N}}} + {\displaystyle \frac {512
}{3}}\,{\it V''}^{3}\,{ \phi^2} \\
 & &  + {\displaystyle \frac {64}{3}}\,{\it V''}\,{\it K'}
 + {\displaystyle \frac {32}{3}}\,{\it V''}\,{Z} + 
{\displaystyle \frac {32}{3}}\,{\it V''}\,{\it Z'}\,{ \phi^2}  - \,
{\displaystyle \frac {512}{3}}\,{\displaystyle \frac {{\it V''}^{
3}\,{ \phi^2}}{{N}}} +
{\displaystyle \frac {64}{3}}\,{\it V'''}\,{\it K'}\,{ \phi^2}  \! 
\! \left. {\vrule height0.80em width0em depth0.80em} \right) 
\\ & & \\ && + {{\it I}_{4, 2, 2}} \left( {\vrule 
height0.80em width0em depth0.80em} \right. \! \!  
 64\,{\displaystyle \frac {{\it V''}^{2}\,{K}}{{N}}} + 
{\displaystyle \frac {704}{3}}\,{\displaystyle \frac {{\it V''}^{
2}\,{\it K'}\,{ \phi^2}}{{N}}} + {\displaystyle \frac {416}{3}}\,
{\displaystyle \frac {{\it V''}^{2}\,{Z}\,{ \phi^2}}{{N}}} - 
{\displaystyle \frac {32}{3}}\,{\displaystyle \frac {{\it V''}\,
{\it V'''}\,{K}\,{ \phi^2}}{{N}}} \\
 & &  - 8\,{\displaystyle \frac {{\it K'}^{2}}{{N}}} + 
{\displaystyle \frac {20}{3}}\,{\displaystyle \frac {{\it K'}\,{Z
}}{{N}}} + {\displaystyle \frac {20}{3}}\,{\displaystyle \frac {
{\it K'}\,{\it Z'}\,{ \phi^2}}{{N}}} - 64\,{\it V''}^{2}\,{K} - 
{\displaystyle \frac {704}{3}}\,{\it V''}^{2}\,{\it K'}\,{ \phi^2}
 \\
 & &  - {\displaystyle \frac {416}{3}}\,{\it V''}^{2}\,{Z}
\,{ \phi^2} + {\displaystyle \frac {32}{3}}\,{\it V''}\,{\it V'''}
\,{K}\,{ \phi^2} + 8\,{\it K'}^{2} - {\displaystyle \frac {20}{3}}
\,{\it K'}\,{Z} - {\displaystyle \frac {20}{3}}\,{\it K'}\,{\it 
Z'}\,{ \phi^2} \! \! \left. {\vrule 
height0.80em width0em depth0.80em} \right)  \\
 & & \\ & & + {{\it I}_{2, 3, 0}}\, \left( \! \, - \,
{\displaystyle \frac {64}{3}}\,{\displaystyle \frac {{\it V''}^{2
}}{{N}}} + {\displaystyle \frac {64}{3}}\,{\it V''}^{2}\, \! 
 \right) \\ & & + {{\it I}_{2, 2, 0}}\, \left( \! \, - 5\,
{\displaystyle \frac {{\it Z'}}{{N}}} - 2\,{\displaystyle \frac {
{\it Z''}\,{ \phi^2}}{{N}}}\, \!  \right)  +  \\ && \\
 & & {{\it I}_{2, 2, 3}}\, \left( \! \, - \,{\displaystyle 
\frac {128}{3}}\,{\displaystyle \frac {{\it V''}^{2}}{{N}}} - 
{\displaystyle \frac {128}{3}}\,{\displaystyle \frac {{\it V''}\,
{\it V'''}\,{ \phi^2}}{{N}}} + {\displaystyle \frac {128}{3}}\,
{\it V''}^{2} + {\displaystyle \frac {128}{3}}\,{\it V''}\,{\it 
V'''}\,{ \phi^2}\, \!  \right)  \\
 & & \\ & & + {{\it I}_{2, 2, 2}}\, \left( \! \,
{\displaystyle \frac {64}{3}}\,{\displaystyle \frac {{\it V''}^{2
}}{{N}}} + {\displaystyle \frac {160}{3}}\,{\displaystyle \frac {
{\it V''}\,{\it V'''}\,{ \phi^2}}{{N}}} - {\displaystyle \frac {64
}{3}}\,{\it V''}^{2} - {\displaystyle \frac {160}{3}}\,{\it V''}
\,{\it V'''}\,{ \phi^2}\, \!  \right)  \\
 & & \\ & & + {{\it I}_{2, 2, 1}}\, \left( \! \,
{\displaystyle \frac {64}{3}}\,{\displaystyle \frac {{\it V''}^{2
}}{{N}}} - {\displaystyle \frac {32}{3}}\,{\displaystyle \frac {
{\it V''}\,{\it V'''}\,{ \phi^2}}{{N}}} - {\displaystyle \frac {64
}{3}}\,{\it V''}^{2} + {\displaystyle \frac {32}{3}}\,{\it V''}\,
{\it V'''}\,{ \phi^2}\, \!  \right)  \\
 & & \\ & & + {{\it I}_{2, 1, 4}}\, \left( \! \,
{\displaystyle \frac {128}{3}}\,{\displaystyle \frac {{\it V''}^{
2}}{{N}}} - {\displaystyle \frac {128}{3}}\,{\it V''}^{2}\, \! 
 \right) \\ & & \\ &&  + {{\it I}_{5, 4, 0}} \left( {\vrule 
height0.86em width0em depth0.86em} \right. \! \!  - 80\,{\displaystyle
\frac {{\it V''}\,{K}\,{Z}}{{N} 
}} - 80\,{\displaystyle \frac {{\it V''}\,{K}\,{\it Z'}\,{ \phi^2}
}{{N}}} - 80\,{\displaystyle \frac {{\it V''}\,{\it K'}\,{Z}\,{ 
\phi^2}}{{N}}} - 80\,{\displaystyle \frac {{\it V''}\,{Z}^{2}\,{ 
\phi^2}}{{N}}} \\
 & &  - 80\,{\displaystyle \frac {{\it V''}\,{Z}\,{\it Z'}
\,{ (\phi^2)}^{2}}{{N}}} - {\displaystyle \frac {160}{3}}\,
{\displaystyle \frac {{\it V'''}\,{K}\,{\it K'}\,{ \phi^2}}{{N}}}
 - {\displaystyle \frac {160}{3}}\,{\displaystyle \frac {{\it 
V'''}\,{K}\,{Z}\,{ \phi^2}}{{N}}} \\
 & &  - {\displaystyle \frac {160}{3}}\,{\displaystyle 
\frac {{\it V'''}\,{K}\,{\it Z'}\,{ (\phi^2)}^{2}}{{N}}} - 
{\displaystyle \frac {160}{3}}\,{\displaystyle \frac {{\it V'''}
\,{\it K'}\,{Z}\,{ (\phi^2)}^{2}}{{N}}} - {\displaystyle \frac {160}{
3}}\,{\displaystyle \frac {{\it V'''}\,{Z}^{2}\,{ (\phi^2)}^{2}}{{N}
}} \\
 & &  - {\displaystyle \frac {160}{3}}\,{\displaystyle 
\frac {{\it V'''}\,{Z}\,{\it Z'}\,{ (\phi^2)}^{3}}{{N}}} - 80\,
{\displaystyle \frac {{\it V''}\,{K}\,{\it K'}}{{N}}}   \! 
\! \left. {\vrule height0.86em width0em depth0.86em} \right) 
\\ & & \\ && + {{\it I}_{4, 3, 1}} \left( {\vrule 
height0.80em width0em depth0.80em} \right. \! \!  - \,
{\displaystyle \frac {128}{3}}\,{\displaystyle \frac {{\it V''}^{
2}\,{K}}{{N}}} - {\displaystyle \frac {512}{3}}\,{\displaystyle 
\frac {{\it V''}^{2}\,{\it K'}\,{ \phi^2}}{{N}}} \\
 & &  - {\displaystyle \frac {256}{3}}\,{\displaystyle 
\frac {{\it V''}^{2}\,{Z}\,{ \phi^2}}{{N}}} + {\displaystyle 
\frac {32}{3}}\,{\displaystyle \frac {{\it K'}^{2}}{{N}}} + 
{\displaystyle \frac {128}{3}}\,{\it V''}^{2}\,{K} + 
{\displaystyle \frac {512}{3}}\,{\it V''}^{2}\,{\it K'}\,{ \phi^2}
 \\
 & &  + {\displaystyle \frac {256}{3}}\,{\it V''}^{2}\,{Z}
\,{ \phi^2} - {\displaystyle \frac {32}{3}}\,{\it K'}^{2} \! 
\! \left. {\vrule height0.80em width0em depth0.80em} \right) 
\\ & & \\ && + {{\it I}_{3, 1, 4}}\, \left( \! \,{\displaystyle 
\frac {128}{3}}\,{\displaystyle \frac {{\it V''}\,{\it K'}}{{N}}}
 - {\displaystyle \frac {128}{3}}\,{\it V''}\,{\it K'}\, \! 
 \right)  \\
 & & \\ & & + {{\it I}_{3, 1, 2}}\, \left( \! \, - \,
{\displaystyle \frac {4}{3}}\,{\displaystyle \frac {{\it V''}\,{Z
}}{{N}}} + {\displaystyle \frac {16}{3}}\,{\it V''}\,{\it K'}\,
 \!  \right) \\ & & \\ &&  + {{\it I}_{6, 5, 0}} \left( {\vrule 
height0.86em width0em depth0.86em} \right. \! \! 192\,
{\displaystyle \frac {{\it V''}^{2}\,{K}^{2}}{{N}}} \\
 & &  + 384\,{\displaystyle \frac {{\it V''}^{2}\,{K}\,{Z}
\,{ \phi^2}}{{N}}} + 192\,{\displaystyle \frac {{\it V''}^{2}\,{Z}
^{2}\,{ (\phi^2)}^{2}}{{N}}} + 256\,{\displaystyle \frac {{\it V''}\,
{\it V'''}\,{K}^{2}\,{ \phi^2}}{{N}}} \\
 & &  + 512\,{\displaystyle \frac {{\it V''}\,{\it V'''}\,
{K}\,{Z}\,{ (\phi^2)}^{2}}{{N}}} + 256\,{\displaystyle \frac {{\it 
V''}\,{\it V'''}\,{Z}^{2}\,{ (\phi^2)}^{3}}{{N}}} + {\displaystyle 
\frac {256}{3}}\,{\displaystyle \frac {{\it V'''}^{2}\,{K}^{2}\,{
 (\phi^2)}^{2}}{{N}}} \\
 & &  + {\displaystyle \frac {512}{3}}\,{\displaystyle 
\frac {{\it V'''}^{2}\,{K}\,{Z}\,{ (\phi^2)}^{3}}{{N}}} + 
{\displaystyle \frac {256}{3}}\,{\displaystyle \frac {{\it V'''}
^{2}\,{Z}^{2}\,{ (\phi^2)}^{4}}{{N}}} - {\displaystyle \frac {32}{3}}
\,{\displaystyle \frac {{K}\,{\it K'}^{2}}{{N}}} - 
{\displaystyle \frac {64}{3}}\,{\displaystyle \frac {{K}\,{\it K'
}\,{Z}}{{N}}} \\
 & &  - {\displaystyle \frac {64}{3}}\,{\displaystyle 
\frac {{K}\,{\it K'}\,{\it Z'}\,{ \phi^2}}{{N}}} - {\displaystyle 
\frac {32}{3}}\,{\displaystyle \frac {{K}\,{Z}^{2}}{{N}}} - 
{\displaystyle \frac {64}{3}}\,{\displaystyle \frac {{K}\,{Z}\,
{\it Z'}\,{ \phi^2}}{{N}}} - {\displaystyle \frac {32}{3}}\,
{\displaystyle \frac {{K}\,{\it Z'}^{2}\,{ (\phi^2)}^{2}}{{N}}} \\
 & &  - {\displaystyle \frac {32}{3}}\,{\displaystyle 
\frac {{\it K'}^{2}\,{Z}\,{ \phi^2}}{{N}}} - {\displaystyle \frac {
64}{3}}\,{\displaystyle \frac {{\it K'}\,{Z}^{2}\,{ \phi^2}}{{N}}}
 - {\displaystyle \frac {64}{3}}\,{\displaystyle \frac {{\it K'}
\,{Z}\,{\it Z'}\,{ (\phi^2)}^{2}}{{N}}} - {\displaystyle \frac {32}{3
}}\,{\displaystyle \frac {{Z}^{3}\,{ \phi^2}}{{N}}} - 
{\displaystyle \frac {64}{3}}\,{\displaystyle \frac {{Z}^{2}\,
{\it Z'}\,{ (\phi^2)}^{2}}{{N}}} \\
 & &  - {\displaystyle \frac {32}{3}}\,{\displaystyle 
\frac {{Z}\,{\it Z'}^{2}\,{ (\phi^2)}^{3}}{{N}}} \! \! \left. 
{\vrule height0.86em width0em depth0.86em} \right)   \\ && \\ &&+ {
{\it I}_{6, 4, 0}} \left( {\vrule 
height0.86em width0em depth0.86em} \right. \! \!  - \,
{\displaystyle \frac {28}{3}}\,{\displaystyle \frac {{K}\,{\it K'
}^{2}}{{N}}} - {\displaystyle \frac {56}{3}}\,{\displaystyle 
\frac {{K}\,{\it K'}\,{Z}}{{N}}} - {\displaystyle \frac {56}{3}}
\,{\displaystyle \frac {{K}\,{\it K'}\,{\it Z'}\,{ \phi^2}}{{N}}}
 \\
 & &  - {\displaystyle \frac {28}{3}}\,{\displaystyle 
\frac {{K}\,{Z}^{2}}{{N}}} - {\displaystyle \frac {56}{3}}\,
{\displaystyle \frac {{K}\,{Z}\,{\it Z'}\,{ \phi^2}}{{N}}} - 
{\displaystyle \frac {28}{3}}\,{\displaystyle \frac {{K}\,{\it Z'
}^{2}\,{ (\phi^2)}^{2}}{{N}}} - {\displaystyle \frac {28}{3}}\,
{\displaystyle \frac {{\it K'}^{2}\,{Z}\,{ \phi^2}}{{N}}} - 
{\displaystyle \frac {56}{3}}\,{\displaystyle \frac {{\it K'}\,{Z
}^{2}\,{ \phi^2}}{{N}}} \\
 & &  - {\displaystyle \frac {56}{3}}\,{\displaystyle 
\frac {{\it K'}\,{Z}\,{\it Z'}\,{ (\phi^2)}^{2}}{{N}}} - 
{\displaystyle \frac {28}{3}}\,{\displaystyle \frac {{Z}^{3}\,{ 
\phi^2}}{{N}}} - {\displaystyle \frac {56}{3}}\,{\displaystyle 
\frac {{Z}^{2}\,{\it Z'}\,{ (\phi^2)}^{2}}{{N}}} - {\displaystyle 
\frac {28}{3}}\,{\displaystyle \frac {{Z}\,{\it Z'}^{2}\,{ \phi^2}
^{3}}{{N}}} \! \! \left. {\vrule 
height0.86em width0em depth0.86em} \right) \\ & & \\ &&  + {{\it I}_{
6, 3, 2}} \left( {\vrule height0.86em width0em depth0.86em}
 \right. \! \!   - \,{\displaystyle \frac {64}{3}}\,{\displaystyle \frac {
{\it V''}^{2}\,{K}^{2}}{{N}}} - {\displaystyle \frac {512}{3}}\,
{\displaystyle \frac {{\it V''}^{2}\,{K}\,{\it K'}\,{ \phi^2}}{{N}
}} - {\displaystyle \frac {256}{3}}\,{\displaystyle \frac {{\it 
V''}^{2}\,{K}\,{Z}\,{ \phi^2}}{{N}}} - {\displaystyle \frac {256}{3
}}\,{\displaystyle \frac {{\it V''}^{2}\,{\it K'}^{2}\,{ \phi^2}^{2
}}{{N}}} \\
 & &  - {\displaystyle \frac {1024}{3}}\,{\displaystyle 
\frac {{\it V''}^{2}\,{\it K'}\,{Z}\,{ (\phi^2)}^{2}}{{N}}} - 
{\displaystyle \frac {256}{3}}\,{\displaystyle \frac {{\it V''}^{
2}\,{Z}^{2}\,{ (\phi^2)}^{2}}{{N}}} + {\displaystyle \frac {32}{3}}\,
{\displaystyle \frac {{K}\,{\it K'}^{2}}{{N}}} + {\displaystyle 
\frac {64}{3}}\,{\displaystyle \frac {{\it K'}^{2}\,{Z}\,{ \phi^2}
}{{N}}} \\
 & &  + {\displaystyle \frac {64}{3}}\,{\it V''}^{2}\,{K}
^{2} + {\displaystyle \frac {512}{3}}\,{\it V''}^{2}\,{K}\,{\it 
K'}\,{ \phi^2} + {\displaystyle \frac {256}{3}}\,{\it V''}^{2}\,{K}
\,{Z}\,{ \phi^2} + {\displaystyle \frac {256}{3}}\,{\it V''}^{2}\,
{\it K'}^{2}\,{ (\phi^2)}^{2} \\
 & &  + {\displaystyle \frac {1024}{3}}\,{\it V''}^{2}\,
{\it K'}\,{Z}\,{ (\phi^2)}^{2} + {\displaystyle \frac {256}{3}}\,
{\it V''}^{2}\,{Z}^{2}\,{ (\phi^2)}^{2} - {\displaystyle \frac {32}{3
}}\,{K}\,{\it K'}^{2} - {\displaystyle \frac {64}{3}}\,{\it K'}^{
2}\,{Z}\,{ \phi^2} \! \! \left. {\vrule 
height0.86em width0em depth0.86em} \right)  \\
 & & \\ & & + {{\it I}_{6, 3, 1}}\, \left( \! \, - \,
{\displaystyle \frac {32}{3}}\,{\displaystyle \frac {{K}\,{\it K'
}^{2}}{{N}}} - {\displaystyle \frac {64}{3}}\,{\displaystyle 
\frac {{\it K'}^{2}\,{Z}\,{ \phi^2}}{{N}}} + {\displaystyle \frac {
32}{3}}\,{K}\,{\it K'}^{2} + {\displaystyle \frac {64}{3}}\,{\it 
K'}^{2}\,{Z}\,{ \phi^2}\, \!  \right)  + \\ &&  \\
 & & {{\it I}_{6, 2, 3}} \left( {\vrule 
height0.80em width0em depth0.80em} \right. \! \!  - \,
{\displaystyle \frac {128}{3}}\,{\displaystyle \frac {{\it V''}^{
2}\,{K}^{2}}{{N}}} + {\displaystyle \frac {512}{3}}\,
{\displaystyle \frac {{\it V''}^{2}\,{K}\,{\it K'}\,{ \phi^2}}{{N}
}} + {\displaystyle \frac {256}{3}}\,{\displaystyle \frac {{\it 
V''}^{2}\,{K}\,{Z}\,{ \phi^2}}{{N}}} \\
 & &  - {\displaystyle \frac {128}{3}}\,{\displaystyle 
\frac {{\it V''}\,{\it V'''}\,{K}^{2}\,{ \phi^2}}{{N}}} + 
{\displaystyle \frac {32}{3}}\,{\displaystyle \frac {{K}\,{\it K'
}\,{Z}}{{N}}} + {\displaystyle \frac {32}{3}}\,{\displaystyle 
\frac {{K}\,{\it K'}\,{\it Z'}\,{ \phi^2}}{{N}}} - {\displaystyle 
\frac {64}{3}}\,{\displaystyle \frac {{\it K'}^{2}\,{Z}\,{ \phi^2}
}{{N}}} \\
 & &  + {\displaystyle \frac {128}{3}}\,{\it V''}^{2}\,{K}
^{2} - {\displaystyle \frac {512}{3}}\,{\it V''}^{2}\,{K}\,{\it 
K'}\,{ \phi^2} - {\displaystyle \frac {256}{3}}\,{\it V''}^{2}\,{K}
\,{Z}\,{ \phi^2} \\
 & &  + {\displaystyle \frac {128}{3}}\,{\it V''}\,{\it 
V'''}\,{K}^{2}\,{ \phi^2} - {\displaystyle \frac {32}{3}}\,{K}\,
{\it K'}\,{Z} - {\displaystyle \frac {32}{3}}\,{K}\,{\it K'}\,
{\it Z'}\,{ \phi^2} + {\displaystyle \frac {64}{3}}\,{\it K'}^{2}\,
{Z}\,{ \phi^2} \! \! \left. {\vrule 
height0.80em width0em depth0.80em} \right)  \\
 & & \\ & & + {{\it I}_{6, 1, 3}}\, \left( \! \, - \,
{\displaystyle \frac {20}{3}}\,{\displaystyle \frac {{K}\,{\it K'
}^{2}}{{N}}} + {\displaystyle \frac {20}{3}}\,{K}\,{\it K'}^{2}\,
 \!  \right)  \\
 & & \\ & & + {{\it I}_{5, 0, 5}}\, \left( \! \,
{\displaystyle \frac {256}{3}}\,{\displaystyle \frac {{\it V''}\,
{K}\,{\it K'}}{{N}}} - {\displaystyle \frac {256}{3}}\,{\it V''}
\,{K}\,{\it K'}\, \!  \right)  \\
 & & \\ & & + {{\it I}_{5, 0, 4}}\, \left( \! \,
{\displaystyle \frac {64}{3}}\,{\displaystyle \frac {{\it V''}\,{
K}\,{\it K'}}{{N}}} - {\displaystyle \frac {64}{3}}\,{\it V''}\,{
K}\,{\it K'}\, \!  \right) \\ & & \\ &&  + {{\it I}_{5, 3, 2}} \left( 
{\vrule height0.86em width0em depth0.86em} \right. \! \!  - \,
{\displaystyle \frac {512}{3}}\,{\displaystyle \frac {{\it V''}^{
3}\,{K}\,{ \phi^2}}{{N}}} \\
 & &  - {\displaystyle \frac {1024}{3}}\,{\displaystyle 
\frac {{\it V''}^{3}\,{\it K'}\,{ (\phi^2)}^{2}}{{N}}} - 
{\displaystyle \frac {1024}{3}}\,{\displaystyle \frac {{\it V''}
^{3}\,{Z}\,{ (\phi^2)}^{2}}{{N}}} + {\displaystyle \frac {128}{3}}\,
{\displaystyle \frac {{\it V''}\,{K}\,{\it K'}}{{N}}} \\
 & &  + {\displaystyle \frac {128}{3}}\,{\displaystyle 
\frac {{\it V''}\,{\it K'}^{2}\,{ \phi^2}}{{N}}} + {\displaystyle 
\frac {256}{3}}\,{\displaystyle \frac {{\it V''}\,{\it K'}\,{Z}\,
{ \phi^2}}{{N}}} + {\displaystyle \frac {512}{3}}\,{\it V''}^{3}\,{
K}\,{ \phi^2} \\
 & &  + {\displaystyle \frac {1024}{3}}\,{\it V''}^{3}\,
{\it K'}\,{ (\phi^2)}^{2} + {\displaystyle \frac {1024}{3}}\,{\it V''
}^{3}\,{Z}\,{ (\phi^2)}^{2} - {\displaystyle \frac {128}{3}}\,{\it 
V''}\,{K}\,{\it K'} - {\displaystyle \frac {128}{3}}\,{\it V''}\,
{\it K'}^{2}\,{ \phi^2} \\
 & &  - {\displaystyle \frac {256}{3}}\,{\it V''}\,{\it K'
}\,{Z}\,{ \phi^2} \! \! \left. {\vrule 
height0.86em width0em depth0.86em} \right) \\ & & \\ && + {{\it I}_{
5, 3, 1}} \left( {\vrule height0.80em width0em depth0.80em}
 \right. \! \!  - \,{\displaystyle \frac {128}{3}}\,
{\displaystyle \frac {{\it V''}\,{K}\,{\it K'}}{{N}}} - 
{\displaystyle \frac {128}{3}}\,{\displaystyle \frac {{\it V''}\,
{\it K'}^{2}\,{ \phi^2}}{{N}}} \\
 & &  - {\displaystyle \frac {256}{3}}\,{\displaystyle 
\frac {{\it V''}\,{\it K'}\,{Z}\,{ \phi^2}}{{N}}} + {\displaystyle 
\frac {128}{3}}\,{\it V''}\,{K}\,{\it K'} + {\displaystyle 
\frac {128}{3}}\,{\it V''}\,{\it K'}^{2}\,{ \phi^2} + 
{\displaystyle \frac {256}{3}}\,{\it V''}\,{\it K'}\,{Z}\,{ \phi^2}
   \! \! \left. {\vrule height0.80em width0em depth0.80em}
 \right) \\ & & \\ &&  + {{\it I}_{4, 0, 4}}\, \left( \! \,
{\displaystyle \frac {32}{3}}\,{\displaystyle \frac {{\it V''}^{2
}\,{K}}{{N}}} + {\displaystyle \frac {8}{3}}\,{\displaystyle 
\frac {{\it K'}^{2}}{{N}}} - {\displaystyle \frac {32}{3}}\,{\it 
V''}^{2}\,{K} - {\displaystyle \frac {8}{3}}\,{\it K'}^{2}\, \! 
 \right)  \\
 & & \\ & & + {{\it I}_{4, 0, 3}}\, \left( \! \, - \,
{\displaystyle \frac {8}{3}}\,{\displaystyle \frac {{\it K'}^{2}
}{{N}}} - 4\,{\displaystyle \frac {{\it K'}\,{Z}}{{N}}} + 
{\displaystyle \frac {8}{3}}\,{\it K'}^{2} + 4\,{\it K'}\,{Z}\,
 \!  \right) \\ & & \\ &&  + {{\it I}_{4, 5, 0}} \left( {\vrule 
height0.86em width0em depth0.86em} \right. \! \!  -
384\,{\displaystyle \frac {{\it V''}^{2}\,{K}}{{N}}} -  
384\,{\displaystyle \frac {{\it V''}^{2}\,{Z}\,{ \phi^2}}{{N}}} - 
512\,{\displaystyle \frac {{\it V''}\,{\it V'''}\,{K}\,{ \phi^2}}{{
N}}} - 512\,{\displaystyle \frac {{\it V''}\,{\it V'''}\,{Z}\,{ 
(\phi^2)}^{2}}{{N}}} \\
 & &  - {\displaystyle \frac {512}{3}}\,{\displaystyle 
\frac {{\it V'''}^{2}\,{K}\,{ (\phi^2)}^{2}}{{N}}} - {\displaystyle 
\frac {512}{3}}\,{\displaystyle \frac {{\it V'''}^{2}\,{Z}\,{ 
(\phi^2)}^{3}}{{N}}} + {\displaystyle \frac {16}{3}}\,{\displaystyle 
\frac {{\it K'}^{2}}{{N}}} + {\displaystyle \frac {32}{3}}\,
{\displaystyle \frac {{\it K'}\,{Z}}{{N}}} \\
 & &  + {\displaystyle \frac {32}{3}}\,{\displaystyle 
\frac {{\it K'}\,{\it Z'}\,{ \phi^2}}{{N}}} + {\displaystyle 
\frac {16}{3}}\,{\displaystyle \frac {{Z}^{2}}{{N}}} + 
{\displaystyle \frac {32}{3}}\,{\displaystyle \frac {{Z}\,{\it Z'
}\,{ \phi^2}}{{N}}} + {\displaystyle \frac {16}{3}}\,
{\displaystyle \frac {{\it Z'}^{2}\,{ (\phi^2)}^{2}}{{N}}} \! 
\! \left. {\vrule height0.86em width0em depth0.86em} \right) 
\\ & & \\ && + {{\it I}_{4, 3, 0}} \left( {\vrule 
height0.86em width0em depth0.86em} \right. \! \! 4\,
{\displaystyle \frac {{\it K'}^{2}}{{N}}}
  + {\displaystyle \frac {56}{3}}\,{\displaystyle 
\frac {{\it K'}\,{Z}}{{N}}} + {\displaystyle \frac {56}{3}}\,
{\displaystyle \frac {{\it K'}\,{\it Z'}\,{ \phi^2}}{{N}}} + 
{\displaystyle \frac {28}{3}}\,{\displaystyle \frac {{Z}^{2}}{{N}
}} + {\displaystyle \frac {56}{3}}\,{\displaystyle \frac {{Z}\,
{\it Z'}\,{ \phi^2}}{{N}}} \\ & &  + {\displaystyle \frac {28}{3}}\,
{\displaystyle \frac {{\it Z'}^{2}\,{ (\phi^2)}^{2}}{{N}}} 
  + {\displaystyle \frac {16}{3}}\,{\it K'}^{2} \! 
\! \left. {\vrule height0.86em width0em depth0.86em} \right) 
\\ & & \\ && + {{\it I}_{4, 1, 3}}\, \left( \! \, - \,
{\displaystyle \frac {80}{3}}\,{\displaystyle \frac {{\it V''}^{2
}\,{K}}{{N}}} - {\displaystyle \frac {28}{3}}\,{\displaystyle 
\frac {{\it K'}^{2}}{{N}}} + {\displaystyle \frac {80}{3}}\,{\it 
V''}^{2}\,{K} + {\displaystyle \frac {28}{3}}\,{\it K'}^{2}\, \! 
 \right)  \\
 & & \\ & & + {{\it I}_{4, 1, 2}}\, \left( \! \, - \,
{\displaystyle \frac {8}{3}}\,{\displaystyle \frac {{\it K'}^{2}
}{{N}}} + 4\,{\displaystyle \frac {{\it K'}\,{Z}}{{N}}} - 
{\displaystyle \frac {5}{3}}\,{\displaystyle \frac {{Z}^{2}}{{N}
}} + {\displaystyle \frac {4}{3}}\,{\it K'}^{2}\, \!  \right)  \\
 & & \\ & & + {{\it I}_{3, 0, 5}}\, \left( \! \, - \,
{\displaystyle \frac {128}{3}}\,{\displaystyle \frac {{\it V''}\,
{\it K'}}{{N}}} + {\displaystyle \frac {128}{3}}\,{\it V''}\,
{\it K'}\, \!  \right)  \\
 & & \\ & & + {{\it I}_{3, 0, 4}}\, \left( \! \,
{\displaystyle \frac {64}{3}}\,{\displaystyle \frac {{\it V''}\,
{\it K'}}{{N}}} - {\displaystyle \frac {64}{3}}\,{\it V''}\,{\it 
K'}\, \!  \right)  \\
 & & \\ & & + {{\it I}_{3, 0, 3}}\, \left( \! \, - \,
{\displaystyle \frac {16}{3}}\,{\displaystyle \frac {{\it V''}\,
{\it K'}}{{N}}} - 8\,{\displaystyle \frac {{\it V''}\,{Z}}{{N}}}
 + {\displaystyle \frac {16}{3}}\,{\it V''}\,{\it K'} + 8\,{\it 
V''}\,{Z}\, \!  \right) \\ & & \\ &&  + {{\it I}_{3, 5, 0}}  \left( {\vrule height0.86em width0em depth0.86em}
 \right. \! \! 64\,{\displaystyle \frac {{\it V''}\,{\it K'}}{{N}
}} + 64\,{\displaystyle \frac {{\it V''}\,{Z}}{{N}}} + 64\,
{\displaystyle \frac {{\it V''}\,{\it Z'}\,{ \phi^2}}{{N}}} + 
{\displaystyle \frac {128}{3}}\,{\displaystyle \frac {{\it V'''}
\,{\it K'}\,{ \phi^2}}{{N}}} \\
 & &  + {\displaystyle \frac {128}{3}}\,{\displaystyle 
\frac {{\it V'''}\,{Z}\,{ \phi^2}}{{N}}} + {\displaystyle \frac {
128}{3}}\,{\displaystyle \frac {{\it V'''}\,{\it Z'}\,{ (\phi^2)}^{2}
}{{N}}} \! \! \left. {\vrule height0.86em width0em depth0.86em}
 \right) \\ & & \\ &&  + {{\it I}_{3, 4, 0}} \left( {\vrule 
height0.86em width0em depth0.86em} \right. \! \! 16\,
{\displaystyle \frac {{\it V''}\,{\it K'}}{{N}}} 
  + 16\,{\displaystyle \frac {{\it V''}\,{Z}}{{N}}} +
16\,{\displaystyle \frac {{\it V''}\,{\it Z'}\,{ \phi^2}}{{N}}} +
{\displaystyle \frac {32}{3}}\,{\displaystyle \frac {{\it  
V'''}\,{\it K'}\,{ \phi^2}}{{N}}} \\ & &  + {\displaystyle \frac
{32}{3}}\,{\displaystyle \frac {{\it  
V'''}\,{Z}\,{ \phi^2}}{{N}}}  + {\displaystyle \frac {32}{3}}\,{\displaystyle 
\frac {{\it V'''}\,{\it Z'}\,{ (\phi^2)}^{2}}{{N}}} \! \! \left. 
{\vrule height0.86em width0em depth0.86em} \right) \\ & & \\ && + {
{\it I}_{3, 3, 0}} \left( {\vrule 
height0.86em width0em depth0.86em} \right. \! \! {\displaystyle 
\frac {80}{3}}\,{\displaystyle \frac {{\it V''}\,{\it K'}}{{N}}}
 + 48\,{\displaystyle \frac {{\it V''}\,{Z}}{{N}}} + 48\,
{\displaystyle \frac {{\it V''}\,{\it Z'}\,{ \phi^2}}{{N}}} \\
 & &  + 32\,{\displaystyle \frac {{\it V'''}\,{\it K'}\,{ 
\phi^2}}{{N}}} + 32\,{\displaystyle \frac {{\it V'''}\,{Z}\,{ \phi^2}
}{{N}}} + 32\,{\displaystyle \frac {{\it V'''}\,{\it Z'}\,{ \phi^2}
^{2}}{{N}}} + {\displaystyle \frac {64}{3}}\,{\it V''}\,{\it K'}
 \! \! \left. {\vrule height0.86em width0em depth0.86em} \right) 
 \\
 & & \\ & & + {{\it I}_{3, 3, 2}}\, \left( \! \,
{\displaystyle \frac {512}{3}}\,{\displaystyle \frac {{\it V''}^{
3}\,{ \phi^2}}{{N}}} - {\displaystyle \frac {64}{3}}\,
{\displaystyle \frac {{\it V''}\,{\it K'}}{{N}}} - 
{\displaystyle \frac {512}{3}}\,{\it V''}^{3}\,{ \phi^2} + 
{\displaystyle \frac {64}{3}}\,{\it V''}\,{\it K'}\, \!  \right) 
 \\
 & & \\ & & + {{\it I}_{3, 3, 1}}\, \left( \! \, - \,
{\displaystyle \frac {512}{3}}\,{\displaystyle \frac {{\it V''}^{
3}\,{ \phi^2}}{{N}}} + {\displaystyle \frac {128}{3}}\,
{\displaystyle \frac {{\it V''}\,{\it K'}}{{N}}} + 
{\displaystyle \frac {512}{3}}\,{\it V''}^{3}\,{ \phi^2} - 
{\displaystyle \frac {128}{3}}\,{\it V''}\,{\it K'}\, \! 
 \right)  \\
 & & \\ & & + {{\it I}_{3, 2, 2}} \left( {\vrule 
height0.80em width0em depth0.80em} \right. \! \! {\displaystyle 
\frac {704}{3}}\,{\displaystyle \frac {{\it V''}^{3}\,{ \phi^2}}{{N
}}} - {\displaystyle \frac {16}{3}}\,{\displaystyle \frac {{\it 
V''}\,{\it K'}}{{N}}} + {\displaystyle \frac {40}{3}}\,
{\displaystyle \frac {{\it V''}\,{Z}}{{N}}} + {\displaystyle 
\frac {40}{3}}\,{\displaystyle \frac {{\it V''}\,{\it Z'}\,{ \phi^2
}}{{N}}} \\
 & &  + {\displaystyle \frac {80}{3}}\,{\displaystyle 
\frac {{\it V'''}\,{\it K'}\,{ \phi^2}}{{N}}} - {\displaystyle 
\frac {704}{3}}\,{\it V''}^{3}\,{ \phi^2} + {\displaystyle \frac {
16}{3}}\,{\it V''}\,{\it K'} - {\displaystyle \frac {40}{3}}\,
{\it V''}\,{Z} - {\displaystyle \frac {40}{3}}\,{\it V''}\,{\it 
Z'}\,{ \phi^2} \\
 & &  - {\displaystyle \frac {80}{3}}\,{\it V'''}\,{\it K'
}\,{ \phi^2} \! \! \left. {\vrule height0.80em width0em depth0.80em
} \right) \\ & & \\ && + {{\it I}_{3, 2, 1}} \left( {\vrule 
height0.80em width0em depth0.80em} \right. \! \! {\displaystyle 
\frac {64}{3}}\,{\displaystyle \frac {{\it V''}\,{\it K'}}{{N}}}
 + {\displaystyle \frac {4}{3}}\,{\displaystyle \frac {{\it V''}
\,{Z}}{{N}}} - {\displaystyle \frac {8}{3}}\,{\displaystyle 
\frac {{\it V''}\,{\it Z'}\,{ \phi^2}}{{N}}} \\
 & &  - {\displaystyle \frac {16}{3}}\,{\displaystyle 
\frac {{\it V'''}\,{\it K'}\,{ \phi^2}}{{N}}} - {\displaystyle 
\frac {80}{3}}\,{\it V''}\,{\it K'} + {\displaystyle \frac {8}{3
}}\,{\it V''}\,{Z} + {\displaystyle \frac {8}{3}}\,{\it V''}\,
{\it Z'}\,{ \phi^2} + {\displaystyle \frac {16}{3}}\,{\it V'''}\,
{\it K'}\,{ \phi^2} \! \! \left. {\vrule 
height0.80em width0em depth0.80em} \right)  \\
 & & \\ & & + {{\it I}_{3, 1, 3}}\, \left( \! \, - \,
{\displaystyle \frac {112}{3}}\,{\displaystyle \frac {{\it V''}\,
{\it K'}}{{N}}} + {\displaystyle \frac {112}{3}}\,{\it V''}\,
{\it K'}\, \!  \right)  \\
 & & \\ & & + {{\it I}_{2, 0, 5}}\, \left( \! \, - \,
{\displaystyle \frac {128}{3}}\,{\displaystyle \frac {{\it V''}^{
2}}{{N}}} + {\displaystyle \frac {128}{3}}\,{\it V''}^{2}\, \! 
 \right) \\ & & \\ && + {{\it I}_{5, 5, 0}} \left( {\vrule 
height0.86em width0em depth0.86em} \right. \! \!  - 128\,
{\displaystyle \frac {{\it V''}\,{K}\,{\it K'}}{{N}}} \\
 & &  - 128\,{\displaystyle \frac {{\it V''}\,{K}\,{Z}}{{N
}}} - 128\,{\displaystyle \frac {{\it V''}\,{K}\,{\it Z'}\,{ \phi^2
}}{{N}}} - 128\,{\displaystyle \frac {{\it V''}\,{\it K'}\,{Z}\,{
 \phi^2}}{{N}}} - 128\,{\displaystyle \frac {{\it V''}\,{Z}^{2}\,{ 
\phi^2}}{{N}}} \\
 & &  - 128\,{\displaystyle \frac {{\it V''}\,{Z}\,{\it Z'
}\,{ (\phi^2)}^{2}}{{N}}} - {\displaystyle \frac {256}{3}}\,
{\displaystyle \frac {{\it V'''}\,{K}\,{\it K'}\,{ \phi^2}}{{N}}}
 - {\displaystyle \frac {256}{3}}\,{\displaystyle \frac {{\it 
V'''}\,{K}\,{Z}\,{ \phi^2}}{{N}}} \\
 & &  - {\displaystyle \frac {256}{3}}\,{\displaystyle 
\frac {{\it V'''}\,{K}\,{\it Z'}\,{ (\phi^2)}^{2}}{{N}}} - 
{\displaystyle \frac {256}{3}}\,{\displaystyle \frac {{\it V'''}
\,{\it K'}\,{Z}\,{ (\phi^2)}^{2}}{{N}}} - {\displaystyle \frac {256}{
3}}\,{\displaystyle \frac {{\it V'''}\,{Z}^{2}\,{ (\phi^2)}^{2}}{{N}
}} \\
 & &  - {\displaystyle \frac {256}{3}}\,{\displaystyle 
\frac {{\it V'''}\,{Z}\,{\it Z'}\,{ (\phi^2)}^{3}}{{N}}} \! 
\! \left. {\vrule height0.86em width0em depth0.86em} \right) 
\\ & & \\ && + {{\it I}_{5, 2, 2}} \left( {\vrule 
height0.80em width0em depth0.80em} \right. \! \! {\displaystyle 
\frac {208}{3}}\,{\displaystyle \frac {{\it V''}\,{K}\,{\it K'}}{
{N}}} - {\displaystyle \frac {8}{3}}\,{\displaystyle \frac {{\it 
V''}\,{K}\,{Z}}{{N}}} \\
 & &  - {\displaystyle \frac {8}{3}}\,{\displaystyle 
\frac {{\it V''}\,{K}\,{\it Z'}\,{ \phi^2}}{{N}}} + {\displaystyle \frac {176}{3}}\,{\displaystyle \frac {
{\it V''}\,{\it K'}^{2}\,{ \phi^2}}{{N}}} + {\displaystyle \frac {416}{3}}\,{\displaystyle \frac {
{\it V''}\,{\it K'}\,{Z}\,{ \phi^2}}{{N}}} \\
 & &  - {\displaystyle \frac {16}{3}}\,{\displaystyle 
\frac {{\it V'''}\,{K}\,{\it K'}\,{ \phi^2}}{{N}}} - {\displaystyle \frac {208}{3}}\,{\it V''}\,{K}\,{\it K'}
 + {\displaystyle \frac {8}{3}}\,{\it V''}\,{K}\,{Z} + 
{\displaystyle \frac {8}{3}}\,{\it V''}\,{K}\,{\it Z'}\,{ \phi^2}
 \\
 & &  - {\displaystyle \frac {176}{3}}\,{\it V''}\,{\it K'
}^{2}\,{ \phi^2} - {\displaystyle \frac {416}{3}}\,{\it V''}\,{\it 
K'}\,{Z}\,{ \phi^2} + {\displaystyle \frac {16}{3}}\,{\it V'''}\,{K
}\,{\it K'}\,{ \phi^2} \! \! \left. {\vrule 
height0.80em width0em depth0.80em} \right)  \\
 & & \\ & & + {{\it I}_{2, 4, 0}}\, \left( \! \, - 48\,
{\displaystyle \frac {{\it V''}^{2}}{{N}}} - 64\,{\displaystyle 
\frac {{\it V''}\,{\it V'''}\,{ \phi^2}}{{N}}} - {\displaystyle 
\frac {64}{3}}\,{\displaystyle \frac {{\it V'''}^{2}\,{ (\phi^2)}^{2}
}{{N}}}\, \!  \right)  \\
 & & \\ & & + {{\it I}_{5, 1, 3}}\, \left( \! \, - \,
{\displaystyle \frac {80}{3}}\,{\displaystyle \frac {{\it V''}\,{
K}\,{\it K'}}{{N}}} + {\displaystyle \frac {80}{3}}\,{\it V''}\,{
K}\,{\it K'}\, \!  \right)  \! \! \left. {\vrule 
height0.86em width0em depth0.86em} \right]  
\end{eqnarray*}

\newpage

\section{Asymptotic expressions for $V,K$ and $Z$}

We can also calculate expressions for $V,K$ and $Z$ for large $\phi^2$.
These were imposed as boundary conditions at some suitable endpoint.
$A_v$, $A_k$ and $A_z$ denote constants.

We have $V(\phi^2)$ behaving as,
\begin{eqnarray*}
\lefteqn{{\it A_v}\,{ (\phi^2)}^{ \left( \! \,3\,\frac {1}{1 + { \eta}
}\, \!  \right) } + } \\
 & &  \left( \! \,{\displaystyle \frac {1}{24}}\,{\displaystyle 
\frac {\sqrt {6}\,(\,{N} - 1\,)\,\sqrt {1 + { \eta}}}{\sqrt {
{\it A_v}}\,\sqrt {{\it A_k}}\,{N}}} + {\displaystyle \frac {1}{24
}}\,{\displaystyle \frac {\sqrt {6}\,(\,1 + { \eta}\,)}{\sqrt {
{\it A_v}}\,{N}\,\sqrt {{\it A_k} + {\it A_z}}\,\sqrt {5 - { \eta}}
}}\, \!  \right) \,{ (\phi^2)}^{ \left( \! \,\frac { - 1 + { \eta}}{1
 + { \eta}}\, \!  \right) }
\end{eqnarray*}

$K(\phi^2)$ behaves as,

\begin{eqnarray*}
\lefteqn{{\it A_k}\,{ (\phi^2)}^{ \left( \! \, - \,\frac {{ \eta}}{1
 + { \eta}}\, \!  \right) } + 2 \left( {\vrule 
height0.79em width0em depth0.79em} \right. \! \! {\displaystyle 
\frac {1}{864}}\sqrt {6}\,(\,1 + { \eta}\,) \left( {\vrule 
height0.43em width0em depth0.43em} \right. \! \! 4\,{ \eta}^{4}\,
{\it A_k}^{3} + 16\,{\it A_k}^{2}\,{ \eta}^{4}\,{\it A_z}} \\
 & & \mbox{} + 11\,{\it A_k}\,{ \eta}^{4}\,{\it A_z}^{2} + 48\,{ 
\eta}^{3}\,{\it A_z}^{3} + 55\,{\it A_k}\,{ \eta}^{3}\,{\it A_z}^{2}
 - 52\,{\it A_k}^{3}\,{ \eta}^{3} \\
 & & \mbox{} - 56\,{\it A_k}^{2}\,{ \eta}^{3}\,{\it A_z} - 579\,{ 
\eta}^{2}\,{\it A_k}\,{\it A_z}^{2} - 480\,{\it A_k}^{2}\,{ \eta}^{2
}\,{\it A_z} - 80\,{\it A_k}^{3}\,{ \eta}^{2} \\
 & & \mbox{} - 144\,{ \eta}^{2}\,{\it A_z}^{3} - 703\,{ \eta}\,
{\it A_k}\,{\it A_z}^{2} + 336\,{ \eta}\,{\it A_k}^{3} + 40\,{\it A_k
}^{2}\,{ \eta}\,{\it A_z} \\
 & & \mbox{} - 432\,{ \eta}\,{\it A_z}^{3} - 80\,{\it A_k}\,{\it A_z
}^{2} - 240\,{\it A_z}^{3} + 160\,{\it A_k}^{2}\,{\it A_z} \! 
\! \left. {\vrule height0.43em width0em depth0.43em} \right) 
 \left/ {\vrule height0.43em width0em depth0.43em} \right. \! \! 
 \left( {\vrule height0.44em width0em depth0.44em} \right. \! \! 
\,\sqrt {{\it A_k} + {\it A_z}} \\
 & & (\, - 4\,{\it A_k} + 2\,{\it A_k}\,{ \eta} + {\it A_z} + { \eta
}\,{\it A_z}\,)^{2}\,(\,5 - { \eta}\,)^{3/2}\,{N}\,{\it A_v}^{3/2}
\, \! \! \left. {\vrule height0.44em width0em depth0.44em}
 \right) \mbox{} + {\displaystyle \frac {1}{864}}\sqrt {6} \\
 & & \sqrt {1 + { \eta}} \left( {\vrule 
height0.43em width0em depth0.43em} \right. \! \!  \left( \! \,12
\,{ \eta}^{3}\,{N} + 28\,{ \eta}^{3} + 48\,{ \eta}\,{N} - 48\,{ 
\eta}^{2}\,{N} - 32\,{ \eta}^{2} - 48\,{ \eta}\, \!  \right) \,
{\it A_k}^{3} \\
 & & \mbox{} +  \left( {\vrule height0.43em width0em depth0.43em}
 \right. \! \! 68\,{ \eta}^{2}\,{\it A_z} - 24\,{ \eta}\,{N}\,
{\it A_z} - 12\,{ \eta}^{2}\,{N}\,{\it A_z} + 44\,{ \eta}^{3}\,
{\it A_z} + 8\,{ \eta}\,{\it A_z} \\
 & & \mbox{} + 12\,{ \eta}^{3}\,{N}\,{\it A_z} - 16\,{\it A_z} \! 
\! \left. {\vrule height0.43em width0em depth0.43em} \right) 
{\it A_k}^{2}\mbox{} +  \left( {\vrule 
height0.43em width0em depth0.43em} \right. \! \! 3\,{ \eta}^{3}\,
{N}\,{\it A_z}^{2} + 3\,{ \eta}\,{N}\,{\it A_z}^{2} + 21\,{ \eta}^{
3}\,{\it A_z}^{2} \\
 & & \mbox{} + 69\,{ \eta}\,{\it A_z}^{2} + 66\,{ \eta}^{2}\,{\it 
A_z}^{2} + 24\,{\it A_z}^{2} + 6\,{ \eta}^{2}\,{N}\,{\it A_z}^{2}
 \! \! \left. {\vrule height0.43em width0em depth0.43em} \right) 
{\it A_k}\mbox{} + 5\,{ \eta}^{3}\,{\it A_z}^{3} \\
 & & \mbox{} + 15\,{ \eta}\,{\it A_z}^{3} + 15\,{ \eta}^{2}\,{\it 
A_z}^{3} + 5\,{\it A_z}^{3} \! \! \left. {\vrule 
height0.43em width0em depth0.43em} \right)  \left/ {\vrule 
height0.43em width0em depth0.43em} \right. \! \!  \left( {\vrule 
height0.44em width0em depth0.44em} \right. \! \! \,{\it A_v}^{3/2}
\,\sqrt {{\it A_k}}\,{N} \\
 & & (\,(\,2\,{ \eta} - 4\,)\,{\it A_k} + { \eta}\,{\it A_z} + 
{\it A_z}\,)^{2}\, \! \! \left. {\vrule 
height0.44em width0em depth0.44em} \right)  \! \! \left. {\vrule 
height0.79em width0em depth0.79em} \right)   \! \,{ (\phi^2)}^{ \left(
\! \, - \,\frac {4}{1 + { \eta}}\, \! \right) } 
\end{eqnarray*}

Finally $Z(\phi^2)$ behaves as,
\begin{eqnarray*}
\lefteqn{{\it A_z}\,{ (\phi^2)}^{ \left( \! \, - \,\frac {1 + 2\,{ 
\eta}}{1 + { \eta}}\, \!  \right) } +  \left( {\vrule 
height0.93em width0em depth0.93em} \right. \! \! {\displaystyle 
\frac {1}{144}} \left( {\vrule height0.43em width0em depth0.43em}
 \right. \! \! 64\,{ \eta}^{5}\,{N} - 65\,{ \eta}^{5} + 185\,{ 
\eta}^{4} - 172\,{ \eta}^{4}\,{N} - 716\,{ \eta}^{3}\,{N}} \\
 & & \mbox{} + 652\,{ \eta}^{3} + 604\,{ \eta}^{2}\,{N} - 452\,{ 
\eta}^{2} - 2524\,{ \eta} + 2348\,{ \eta}\,{N} - 1184 \\
 & & \mbox{} + 1264\,{N} \! \! \left. {\vrule 
height0.43em width0em depth0.43em} \right) \sqrt {6} \left/ 
{\vrule height0.43em width0em depth0.43em} \right. \! \!  \left( 
\! \,\sqrt {{\it A_v}}\,{N}\,(\, - 2 + { \eta}\,)^{2}\,(\,1 + { 
\eta}\,)\,(\,5 - { \eta}\,)^{3/2}\, \!  \right)  \\
 & & \mbox{} + {\displaystyle \frac {1}{24}}\,{\displaystyle 
\frac {\sqrt {6}\, \left( \! \,20\,{ \eta}^{4} - 33\,{ \eta}^{3}
 - 6\,{ \eta}^{2} - 58\,{ \eta} - 24\, \!  \right) \,(\,{N} - 1\,
)}{(\,1 + { \eta}\,)^{3/2}\,(\, - 2 + { \eta}\,)^{2}\,{N}\,
\sqrt {{\it A_v}}}} \! \! \left. {\vrule 
height0.93em width0em depth0.93em} \right) { (\phi^2)}^{ \left( \! \,
 - \,3/2\,\frac {{ \eta} + 2}{1 + { \eta}}\, \!  \right) }
\end{eqnarray*}

% The end bit:
\newpage
\bibliographystyle{phaip}
\bibliography{thesis}

\begin{thebibliography}{10}

\bibitem{a:topcond}
W.~Bardeen, C.~Hill, and M.Lindner,
\newblock Phys. Rev. {\bf 62}, 2793 (1989).

\bibitem{a:techni1}
S.~Weinberg,
\newblock Phys. Rev. {\bf D13}, 974 (1976).

\bibitem{a:techni2}
S.~Weinberg,
\newblock Phys. Rev. {\bf D19}, 1277 (1979).

\bibitem{a:georgiGUTS}
H.~Georgi and S.~Glashow,
\newblock Phys. Rev. Lett. {\bf 32}, 438 (1974).

\bibitem{a:supsym}
Wess and Zumino,
\newblock Nucl. Phys. {\bf B78}, 1 (1974).

\bibitem{hierarch}
A.~Natale and R.~Shellard,
\newblock J. Phys {\bf G8}, 635 (1980).

\bibitem{a:patisalam}
J.~Pati and A.~Salam,
\newblock Phys. Rev. {\bf D10}, 275 (1974).

\bibitem{a:weineff}
S.Weinberg,
\newblock Phys. Lett. {\bf 91B}, 51 (1980).

\bibitem{a:halleff}
L.~Hall,
\newblock Nucl. Phys. {\bf B178}, 75 (1978).

\bibitem{a:apple}
T.~Applequist and J.~Carrazone,
\newblock  {\bf D11}, 2856 (1975).

\bibitem{a:effect}
A.~Manohar,
\newblock hep-ph/9508245.

\bibitem{b:georgi}
H.~Georgi,
\newblock {\em Weak Interactions and Modern Particle Physics},
\newblock Benjamin/Cummings, 1984.

\bibitem{a:pol}
J.~Polchinski,
\newblock Nucl. Phys. {\bf B231}, 269 (1984).

\bibitem{a:wk}
K.~Wilson and J.~Kogut,
\newblock Phys. Rep. {\bf 12C}, 75 (1974).

\bibitem{a:ball/thorne}
R.~Ball and R.~Thorne,
\newblock Ann. Phys. {\bf 236}, 117 (1994).

\bibitem{a:timappr}
T.~Morris,
\newblock Int. J. Mod. Phys. {\bf A9}, 2411 (1994).

\bibitem{a:warr}
B.~Warr,
\newblock Ann. Phys {\bf 183}, 1 and 59 (1983).

\bibitem{a:weinunitary}
S.~Weinberg,
\newblock Physica {\bf 96A}, 327 (1979).

\bibitem{p:wein}
S.~Weinberg,
\newblock Conference summary,
\newblock 1992,
\newblock Talk presented at the XXVI International Conference on High Energy
  Physics.

\bibitem{a:hh}
A.~Hasenfratz and P.~Hasenfratz,
\newblock Nucl. Phys. {\bf B270}, 685 (1986).

\bibitem{a:Seiberg}
N.~Seiberg and E.~Witten,
\newblock Nucl. Phys {\bf B462}, 19 (1994).

\bibitem{b:zinn}
J.~Zinn-Justin,
\newblock {\em Quantum Field Theory and Critical Phenomena},
\newblock Clarendon Press, Oxford, 1989.

\bibitem{b:amit}
D.~Amit,
\newblock {\em Field Theory, the Renormalization Group abd Critical Phenomena},
\newblock World Scientific, 2nd edition edition, 1984.

\bibitem{a:marco}
M.~Bonini, M.~D'Attanasio, and G.~Marchensini,
\newblock Nucl. Phys. {\bf B409}, 441 (1993).

\bibitem{a:salm}
M.~Salmhofer,
\newblock Nucl. Phys. {\bf B (Proc. Suppl) 30}, 81 (1993).

\bibitem{a:timmom}
T.~Morris,
\newblock Nucl. Phys. {\bf B458}, 477 (1996).

\bibitem{a:golreparam}
G.~Golner,
\newblock Phys. Rev. {\bf B33}, 7863 (1986).

\bibitem{a:reparam1}
T.~Bell and K.~Wilson,
\newblock Phys. Rev. {\bf B11}, 3431 (1975).

\bibitem{a:reparam2}
E.~Riedel, G.~Golner, and K.~Newman,
\newblock Ann. Phys. {\bf 161}, 178 (1985).

\bibitem{a:timderiv}
T.~Morris,
\newblock Phys. Lett. {\bf B329}, 241 (1994).

\bibitem{timpriv}
T.~Morris,
\newblock Private communication.

\bibitem{a:ncs2}
J.~Nicoll, T.~Chang, and H.~Stanley,
\newblock Phys.Rev. Lett. {\bf 33}, 541 (1974).

\bibitem{a:HazenNager}
P.~Hazenfratz and J.~Nager,
\newblock Z. Phys. {\bf 37C}, 476 (1988).

\bibitem{a:wet}
C.~Wetterich,
\newblock Phys. Lett. {\bf B301}, 90 (1993).

\bibitem{a:ellwanger}
U.~Ellwanger and L.~Vergara,
\newblock Nucl. Phys. {\bf B398}, 52 (1993).

\bibitem{a:timtrunc}
T.~Morris,
\newblock Phys. Lett. {\bf B334}, 355 (1994).

\bibitem{a:t+wett}
N.~Tetradis and C.~Wetterich,
\newblock Nucl. Phys. {\bf B422}, 541 (1994).

\bibitem{a:alford}
M.~Alford,
\newblock Phys. Lett. {\bf B336}, 237 (1994).

\bibitem{a:timlat}
T.~Morris,
\newblock {\it Lattice '94}, Nucl. Phys. {\bf B(Proc. Suppl.)42}, 811 (1995).

\bibitem{a:timhalp}
T.~Morris,
\newblock hep-ph/9601128, to be published in Phys. Rev D.

\bibitem{a:WandH}
F.~Wegner and A.~Houghton,
\newblock Phys. Rev. {\bf A8}, 401 (1973).

\bibitem{a:ncs1}
J.~Nicoll, T.~Chang, and H.~Stanley,
\newblock Phys. Lett. {\bf 57A}, 7 (1976).

\bibitem{a:wein}
S.~Weinberg,
\newblock Critical phenomena for field theorists,
\newblock in {\em Lectures, Erice Subnucl. Phys.}, page~1, 1976.

\bibitem{a:widom}
B.~Widom,
\newblock Journal of Chemical Physics {\bf 43}, 3898 (1965).

\bibitem{b:rothe}
H.~Rothe,
\newblock {\em Lattice Gauge Theories, An Introduction},
\newblock World Scientific, 1992.

\bibitem{a:deGennes}
P.~de~Gennes,
\newblock Phys. Lett. {\bf 38A}, 339 (1972).

\bibitem{a:Weigel}
F.~Weigel,
\newblock Phys. Rep. {\bf 16C}, 57 (1975).

\bibitem{a:YangLee}
C.~Yang and T.~Lee,
\newblock Phys. Rev. {\bf 87}, 404 (1952).

\bibitem{a:wilzcek}
F.~Wilzcek and R.~Pizarki,
\newblock Phys. Rev. {\bf D29}, 338 (1984).

\bibitem{a:critbin}
D.~Beysens,
\newblock Status of the experimental situation in critical binary fluids,
\newblock in {\em Phase Transitions}, edited by M.~Levy, J.~Le~Guillou, and
  J.~Zinn-Justin, Plenum Press, 1980,
\newblock Cargese, 1980.

\bibitem{a:wilsciam}
K.~Wilson,
\newblock Scientific American {\bf August 1979}.

\bibitem{b:realspace}
T.~Burkhardt and J.~van Leeuwen,
\newblock {\em Real-Space Renormalization},
\newblock Springer,Berlin, 1982.

\bibitem{a:mcrg1}
S.~Ma,
\newblock Phys. Rev. {\bf 37}, 461 (1976).

\bibitem{a:mcrg2}
L.~Kadanoff,
\newblock Rev. Mod. Phys. {\bf 49}, 267 (1977).

\bibitem{b:mcrg3}
K.~Wilson,
\newblock Recent developments in gauge theories,
\newblock 1979,
\newblock Cargese.

\bibitem{a:nu:1overN2}
Y.~Okabe and M.~Oku,
\newblock Prog. Theor. Phys. {\bf 60}, 1277 (1978).

\bibitem{a:KanN=4}
K.~Kanaya and S.~Kaya,
\newblock Phys. Rev. {\bf D51}, 2404 (1995).

\bibitem{a:om:1overN}
S.~Ma,
\newblock Phys. Rev {\bf A10}, 1818 (1974).

\bibitem{a:wet2}
J.~Adams et~al.,
\newblock Mod. Phys. Lett. {\bf A10}, 2367 (1995).

\bibitem{a:wet3}
J.~Berges, N.~Tetradis, and C.~Wetterich,
\newblock hep-th/9507159.

\bibitem{a:ballrubbish}
R.~Ball, P.~Haagensen, J.~Latorre, and E.~Moreno,
\newblock Phys. Lett. {\bf B347}, 80 (1995).

\bibitem{a:Form}
J.~Vermaseren,
\newblock The symbolic manipulation program form,
\newblock KEK-preprint-92-1, 1992.

\bibitem{a:redun}
F.~Wegner,
\newblock J. Phys. {\bf C7}, 2098 (1974).

\bibitem{a:tim2d}
T.~Morris,
\newblock Phys. Lett. {\bf B345}, 139 (1995).

\bibitem{a:stanley}
H.~Stanley,
\newblock Phys. Rev. {\bf 176}, 718 (1968).

\bibitem{a:joyce}
G.~Joyce,
\newblock Phys. Rev {\bf 146}, 349 (1966).

\bibitem{a:balmin2}
R.~Balian and G.~Toulouse,
\newblock Phys. Rev. Lett. {\bf 30}, 544 (1973).

\bibitem{a:min2}
M.~Fisher,
\newblock Phys. Rev. Lett. {\bf 30}, 679 (1973).

\bibitem{a:fishmin2}
M.~Fisher,
\newblock Rev. Mod. Phys , 597 (1974).

\bibitem{a:riedmin2}
K.~Newman and E.~Riedel,
\newblock Phys. Rev. {\bf 30B}, 6615 (1984).

\bibitem{a:tetlit}
N.~Tetradis and D.~Litim,
\newblock Nucl. Phys. {\bf B464}, 492 (1996).

\bibitem{a:Marcogauge}
M.~Bonini, M.~D'Attanasio, and G.~Marchesini,
\newblock Nucl. Phys. {\bf B437}, 163 (1995).

\bibitem{a:ellwanger2}
U.~Ellwanger,
\newblock Phys. Lett {\bf B335}, 364 (1994).

\bibitem{a:u1com}
T.~Morris,
\newblock Phys. Rev. {\bf D53}, 7250 (1996).

\bibitem{a:u1noncom}
T.~Morris,
\newblock Phys. Lett. {\bf B357}, 225 (1995).

\bibitem{b:stoer}
J.~Stoer and R.~Bulirsch,
\newblock {\em Introduction to Numerical Analysis},
\newblock Springer, 2nd edition edition, 1993.

\bibitem{b:nr}
W.~Press, S.~Teukolsky, W.~Vetterling, and B.~Flannery,
\newblock {\em Numerical Recipes: the art of scientific computing},
\newblock Cambridge University Press, 1986.

\bibitem{a:MPI}
M.~P.~I. Forum,
\newblock Mpi: A message passing interface standard,
\newblock University of Tennesse, Report No. CS-94-230, 1994,
\newblock See also International Journal of Supercomputing Applications volume
  8, number 3/4, 1994 and
  http://www.mcs.anl.gov/mpi/mpi-report/mpi-report.html.

\end{thebibliography}
\newpage

\pagestyle{empty} 
\begin{center} 
\vspace*{10cm} 
\emph{Drink Beer, Play Darts}
\end{center}

\end{document}